\documentclass[11pt]{article}

\usepackage[utf8]{inputenc}
\usepackage[left=25mm,right=25mm,top=25mm,bottom=25mm]{geometry}

\usepackage{multirow}
\usepackage{booktabs}
\usepackage{float}
\usepackage{psfrag}

\usepackage{dsfont}
\usepackage{amssymb}
\usepackage{amsmath}
\usepackage{amsthm}
\usepackage{nicefrac}
\usepackage{bbm}
\theoremstyle{definition}   
\usepackage{chngcntr}
\usepackage{apptools}

\usepackage[usenames,dvipsnames,svgnames]{xcolor}
\usepackage{tikz}
\ifdefined\figH\else
	\newlength\figH
\fi
\ifdefined\figW\else
	\newlength\figW
\fi

\usepackage{pgfplots}
\usepgfplotslibrary{groupplots}
\usepackage{pgfplotstable}
\usepackage{subcaption}
\usepackage[font=small]{caption}
\usepackage{wrapfig}
\pgfplotsset{compat=newest}
\usepgfplotslibrary{fillbetween}
\usetikzlibrary{patterns}
\usepgfplotslibrary{colormaps}
\usetikzlibrary{positioning}

\usepackage[maxbibnames=20, maxcitenames=3, 
    natbib=true, style=numeric,
    isbn=false, url=false,giveninits=true,sortcites]{biblatex}   
\bibliography{bibliography}

\usepackage[german, main=english]{babel}
\usepackage[autostyle]{csquotes}
\usepackage[T1]{fontenc}
\usepackage{hyperref}
\usepackage[capitalise]{cleveref}
\crefname{section}{Sect.}{Sect.}
\crefname{appendix}{Appx.}{Appx.}
\usepackage[affil-it]{authblk}
\usepackage{color}
\usepackage{colortbl}

\newtheoremstyle{mystyle}{3pt}{3pt}{\itshape}{}{\bfseries}{.}{ }{}
\theoremstyle{mystyle}
\newtheorem{theorem}{Theorem}

\newtheorem{remark}[theorem]{Remark}

\usepackage{pgf-pie}

\newcommand{\ba}{\vec{a}}
\newcommand{\bb}{\vec{b}}

\newcommand{\bA}{\tens{A}}
\newcommand{\bB}{\tens{B}}

\newcommand{\bI}{\tens{I}}

\newcommand{\bP}{\tens{P}}
\newcommand{\bQ}{\tens{Q}}

\newcommand{\bX}{\tens{X}}

\newcommand{\SO}{\text{SO}}
\newcommand{\norm}[1]{\left\lVert#1\right\rVert}

\newcommand{\bbR}{{\mathbb{R}}}

\newcommand{\tr}{\operatorname{tr}}

\renewcommand{\wedge}{\hhash}

\newcommand{\R}{\mathbb{R}}
\newcommand{\T}{\mathbb{T}}
\renewcommand{\S}{\mathbb{S}}

\newcommand{\voigt}{\textrm{V}}

\newcommand{\subVarphi}{\ensuremath{\boldsymbol{\varphi}}}

\newcommand{\subT}{\ensuremath{\boldsymbol{T}}}
\newcommand{\subX}{\ensuremath{\boldsymbol{\mathrm{X}}}}

\newcommand{\subv}{\ensuremath{\boldsymbol{\mathrm{v}}}}
\newcommand{\subA}{\ensuremath{\boldsymbol{\mathrm{A}}}}

\newcommand{\subC}{\ensuremath{\boldsymbol{\mathrm{C}}}}
\newcommand{\subG}{\ensuremath{\boldsymbol{\mathrm{G}}}}
\newcommand{\subI}{\ensuremath{\boldsymbol{\mathrm{\mathcal{I}}}}}

\newcommand{\LambdaC}{\tens{\Lambda}^{\subC}}
\newcommand{\LambdaG}{\tens{\Lambda}^{\subG}}
\newcommand{\Lambdac}{\Lambda^{C}}

\newcommand{\subF}{\ensuremath{\boldsymbol{\mathrm{F}}}}
\newcommand{\subH}{\ensuremath{\boldsymbol{\mathrm{H}}}}

\newcommand{\spaceV}{\mathds{V}}

\renewcommand{\nmid}{n+\frac{1}{2}}

\newcommand{\W}{W^{\text{I}}}
\newcommand{\WF}{W^{\text{F}}}
\newcommand{\WFt}{\hat W^{\text{F}}}
\newcommand{\WC}{W^{\text{C}}}
\newcommand{\WG}{W^{\text{G}}}
\newcommand{\WCt}{\hat W^{\text{C}}}
\newcommand{\WI}{W^{\text{I}}}
\newcommand{\WIt}{\hat{W}^{\text{I}}}
\newcommand{\F}{\tens{F}}
\renewcommand{\H}{\tens{H}}
\newcommand{\FPhi}{\tens{F}_{\subVarphi}}

\newcommand{\JPhi}{J_{\subVarphi}}
\newcommand{\C}{\tens{C}}
\newcommand{\G}{\tens{G}}
\renewcommand{\c}{C}
\newcommand{\CPhi}{\tens{C}_{\subVarphi}}
\newcommand{\GPhi}{\tens{G}_{\subVarphi}}

\newcommand{\I}{I_{\subC}}
\newcommand{\II}{II_{\subC}}
\newcommand{\III}{III_{\subC}}

\newcommand{\IPhiN}{I_{\subC_{n}}}
\newcommand{\IIPhiN}{II_{\subC_{n}}}
\newcommand{\JPhiN}{J_{n}}
\newcommand{\IPhiNOne}{I_{\subC_{{n+1}}}}
\newcommand{\IIPhiNOne}{II_{\subC_{{n+1}}}}
\newcommand{\JPhiNOne}{J_{{n+1}}}
\newcommand{\CPhiN}{\C_{n}}
\newcommand{\GPhiN}{\G_{n}}
\newcommand{\cPhiN}{C_{n}}
\newcommand{\CPhiNOne}{\C_{{n+1}}}
\newcommand{\GPhiNOne}{\G_{{n+1}}}
\newcommand{\cPhiNOne}{C_{{n+1}}}

\newcommand{\IPhit}{I_{\subC_{t}}}
\newcommand{\IIPhit}{II_{\subC_{t}}}
\newcommand{\IIIPhit}{III_{\subC_{t}}}
\newcommand{\CPhit}{\tens{C}_{t}}
\newcommand{\GPhit}{\tens{G}_{t}}
\newcommand{\JPhit}{J_{t}}
\newcommand{\FPhit}{\tens{F}_{t}}

\DeclareMathOperator*{\assemtext}{\text{{\ensuremath{\mathbf{\textsf A}}}}}
\makeatletter
\DeclareRobustCommand\bigop[1]{%
  \mathop{\vphantom{\sum}\mathpalette\bigop@{#1}}\slimits@
}
\newcommand{\bigop@}[2]{%
  \vcenter{%
    \sbox\z@{$#1\sum$}%
    \hbox{\resizebox{\ifx#1\displaystyle.9\fi\dimexpr\ht\z@+\dp\z@}{!}{$\m@th#2$}}%
  }%
}
\makeatother
\newcommand{\assem}{\DOTSB\bigop{\text{\ensuremath{\mathbf{\textsf A}}}}}

\renewcommand{\vec}[1]{\ensuremath{\mbox{\boldmath $\mathrm{#1}$}}}
\newcommand{\tens}[1]{\vec{#1}}
\newcommand{\cof}{\ensuremath{\operatorname{cof}}}

\renewcommand{\tr}{\ensuremath{\operatorname{tr}}}

\renewcommand{\d}[1][]{\ensuremath{\,\mathrm{d}#1}}
\newcommand{\D}[1][]{\ensuremath{\mathrm{D}#1}}

\newcommand{\B}{\ensuremath{\mathcal{B}_0}}

\newcommand{\h}{{\text{h}}}
\newcommand{\transp}{^\mathrm{T}}
\newcommand{\ntransp}{^\mathrm{-T}}

\usepackage{bbding}

\usepackage{stackengine}

\stackMath
\def\newhash{\mathrel{\abovebaseline[-.9ex]{\rotatebox{45}{\stackon[0pt]{%
					\stackon[0pt]{\rule[.28em]{.8em}{.045em}}{\rule[.67em]{.8em}{.045em}}%
				}{\rule[.1em]{.045em}{.8em}\rule{.3em}{0pt}\rule[.1em]{.045em}{.8em}}}}}}
\newcommand{\hhash}{\scalebox{0.75}{\raisebox{.7ex}{$\newhash$}}}

\definecolor{color11}{RGB}{236, 188, 0}
\definecolor{color12}{RGB}{50, 67, 121}
\definecolor{color13}{RGB}{222, 222, 222}
\definecolor{color21}{RGB}{92, 134, 196}
\definecolor{color22}{RGB}{249, 156, 0}
\definecolor{color23}{RGB}{222, 222, 222}
\definecolor{color31}{RGB}{222, 222, 222}
\definecolor{color32}{RGB}{222, 222, 222}
\definecolor{color33}{RGB}{191, 60, 60}

\pgfplotscreateplotcyclelist{mycolorlist}{%
only marks,solid, mark options={solid,fill=color11,fill opacity=1,mark size=4pt},mark=square*,color11\\
only marks,solid, mark options={solid,fill=color12,fill opacity=1,mark size=4pt},mark=square*,color12\\
only marks,solid, mark options={solid,fill=color13,fill opacity=1,mark size=4pt},mark=square*,color13\\
only marks,solid, mark options={solid,fill=color21,fill opacity=1,mark size=4pt},mark=square*,color21\\
only marks,solid, mark options={solid,fill=color22,fill opacity=1,mark size=4pt},mark=square*,color22\\
only marks,solid, mark options={solid,fill=color23,fill opacity=1,mark size=4pt},mark=square*,color23\\
only marks,solid, mark options={solid,fill=color31,fill opacity=1,mark size=4pt},mark=square*,color31\\
only marks,solid, mark options={solid,fill=color32,fill opacity=1,mark size=4pt},mark=square*,color32\\
only marks,solid, mark options={solid,fill=color33,fill opacity=1,mark size=4pt},mark=square*,color33\\
color11, dashed, very thick,mark=*, mark options={solid}, mark repeat = 15\\
color12, dashed, very thick ,mark=*, mark options={solid}, mark repeat = 15\\
color13, dashed, very thick ,mark=*, mark options={solid}, mark repeat = 15\\
color21, dashed, very thick ,mark=*, mark options={solid}, mark repeat = 15\\
color22, dashed, very thick ,mark=*, mark options={solid}, mark repeat = 15\\
color23, dashed, very thick ,mark=*, mark options={solid}, mark repeat = 15\\
color31, dashed, very thick,mark=*, mark options={solid} , mark repeat = 15\\
color32, dashed, very thick ,mark=*, mark options={solid}, mark repeat = 15\\
color33, dashed, very thick,mark=*, mark options={solid} , mark repeat = 15\\
color11, very thick \\
color12, very thick \\
color13, very thick \\
color21, very thick \\
color22, very thick \\
color23, very thick \\
color31, very thick \\
color32, very thick \\ 
color33, very thick \\
}

\pgfplotscreateplotcyclelist{mycolorlistWFXcell}{%
only marks,solid, mark options={solid,fill=color11,fill opacity=1,mark size=4pt},mark=square*,color11\\
only marks,solid, mark options={solid,fill=color22,fill opacity=1,mark size=4pt},mark=square*,color22\\
only marks,solid, mark options={solid,fill=color33,fill opacity=1,mark size=4pt},mark=square*,color33\\
color11, dashed, very thick,mark=*, mark options={solid}, mark repeat = 15\\
color22, dashed, very thick,mark=*, mark options={solid}, mark repeat = 15\\
color33, dashed, very thick,mark=*, mark options={solid}, mark repeat = 15\\
color11, very thick \\
color22, very thick \\
color33, very thick \\
}

\pgfplotscreateplotcyclelist{mycolorlistvol}{%
color11, dashed, very thick,mark=*, mark options={solid}, mark repeat = 15\\
color21, very thick \\
color33, very thick \\
}

\pgfplotscreateplotcyclelist{mycolorlist2}{%
only marks,solid, mark options={solid,fill=color11,fill opacity=1,mark size=4pt},mark=square*,color11\\
only marks,solid, mark options={solid,fill=color12,fill opacity=1,mark size=4pt},mark=square*,color12\\
only marks,solid, mark options={solid,fill=color13,fill opacity=1,mark size=4pt},mark=square*,color13\\
only marks,solid, mark options={solid,fill=color21,fill opacity=1,mark size=4pt},mark=square*,color21\\
only marks,solid, mark options={solid,fill=color22,fill opacity=1,mark size=4pt},mark=square*,color22\\
only marks,solid, mark options={solid,fill=color23,fill opacity=1,mark size=4pt},mark=square*,color23\\
only marks,solid, mark options={solid,fill=color31,fill opacity=1,mark size=4pt},mark=square*,color31\\
only marks,solid, mark options={solid,fill=color32,fill opacity=1,mark size=4pt},mark=square*,color32\\
only marks,solid, mark options={solid,fill=color33,fill opacity=1,mark size=4pt},mark=square*,color33\\
color11 \\
color11 \\
color11, dashed, very thick,mark=*, mark options={solid}, mark repeat = 15\\
color12, dashed, very thick ,mark=*, mark options={solid}, mark repeat = 15\\
color13, dashed, very thick ,mark=*, mark options={solid}, mark repeat = 15\\
color21, dashed, very thick ,mark=*, mark options={solid}, mark repeat = 15\\
color22, dashed, very thick ,mark=*, mark options={solid}, mark repeat = 15\\
color23, dashed, very thick ,mark=*, mark options={solid}, mark repeat = 15\\
color31, dashed, very thick,mark=*, mark options={solid} , mark repeat = 15\\
color32, dashed, very thick ,mark=*, mark options={solid}, mark repeat = 15\\
color33, dashed, very thick,mark=*, mark options={solid} , mark repeat = 15\\
color11, very thick \\
color12, very thick \\
color13, very thick \\
color21, very thick \\
color22, very thick \\
color23, very thick \\
color31, very thick \\
color32, very thick \\ 
color33, very thick \\
color11 \\
color11 \\
}

\pgfplotscreateplotcyclelist{mycolorlist123}{%
only marks,solid, mark options={solid,fill=color11,fill opacity=1,mark size=4pt},mark=square*,color11\\
only marks,solid, mark options={solid,fill=color22,fill opacity=1,mark size=4pt},mark=square*,color22\\
only marks,dashed, mark options={solid,fill=color33,fill opacity=1,mark size=4pt},mark=square*,color33\\
color11, very thick,dashed, mark options={solid}, mark repeat = 15\\
color22, very thick , mark options={solid}, mark repeat = 15\\
color33, very thick,dashed , mark options={solid}, mark repeat = 15\\
}

\pgfplotscreateplotcyclelist{mycolorlistSNE}{%
only marks,solid, mark options={solid,fill=color11,fill opacity=1,mark size=4pt},mark=square*,color11\\
only marks,solid, mark options={solid,fill=color22,fill opacity=1,mark size=4pt},mark=square*,color22\\
only marks,dashed, mark options={solid,fill=color33,fill opacity=1,mark size=4pt},mark=square*,color33\\
color11, very thick, dashed,  mark options={solid}, mark repeat = 15,mark=*\\
color22, very thick, dashed , mark options={solid}, mark repeat = 15, mark=*\\
color33, very thick,dashed , mark options={solid}, mark repeat = 15, mark=*\\
color11, very thick \\
color22, very thick \\
color33, very thick \\
}

\usetikzlibrary{patterns}
\usetikzlibrary{shapes}
\usetikzlibrary{arrows.meta} 

\tikzset{>=latex} 
\tikzstyle{node}=[thick,circle,draw=black,minimum size=22,inner sep=0.5,outer sep=0.6]
\tikzstyle{node icnn}=[color=color22!10!black,node,draw=black,fill=color22!25, text = black]
\tikzstyle{node in_out}=[node,color21!10!black,draw=black,fill=color21!25, text=black]
\tikzstyle{node inv_pot}=[rectangle, node,color33!10!black,draw=black,fill=color33!25, text=black]
\tikzstyle{connect}=[thick,black] 
\tikzstyle{connect arrow}=[-{Latex[length=4,width=3.5]},thick,black,shorten <=0.5,shorten >=1]

\definecolor{CPSgreen}{RGB}{22,164,138}
\definecolor{CPSlightblue}{RGB}{104,143,198}
\definecolor{CPSdarkblue}{RGB}{67,83,132}
\definecolor{CPSgrey}{RGB}{204, 204, 204}
\definecolor{CPSorange}{RGB}{246,163,21}
\definecolor{CPSred}{RGB}{194,76,76}

\usepackage{multirow}
\usepackage{listofitems} 
\usetikzlibrary{arrows.meta} 
\usepackage[outline]{contour} 
\contourlength{1.4pt}

\theoremstyle{definition}

\usepackage{pgfplotstable}
\usepackage{longtable}
\usepackage{multirow}
\usepackage{tabularx}
\usepackage{booktabs}
\pgfplotstableset{
	every head row/.style={
		before row={\toprule},
		after row={\midrule\midrule},
	},
	every last row/.style={after row={\bottomrule}},
}

\pgfplotstableset{every table/.append style={outfile={tables/#1.tex}}}
\pgfplotstableset{include outfiles,force remake=false}

\title{Advanced discretization techniques for hyperelastic\\ physics-augmented neural networks\vspace{1ex}
}

\author[1,*]{Marlon~Franke}
\author[2]{Dominik~K.~Klein}
\author[2]{Oliver Weeger}
\author[1]{Peter Betsch}

\affil[1]{\footnotesize Institute of Mechanics, Karlsruhe Institute of Technology, 76131 Karlsruhe, Germany}
\affil[2]{\footnotesize Cyber-Physical Simulation, 
Department of Mechanical Engineering,\protect\\Technical University of Darmstadt, 64293 Darmstadt, Germany}
\affil[*]{\footnotesize Corresponding author, email: marlon.franke@kit.edu}

\date{June 15, 2023}

\begin{document}
\maketitle
\par\noindent\rule{\textwidth}{0.4pt}
\begin{abstract}
\noindent
In the present work, advanced spatial and temporal discretization techniques are tailored to hyperelastic physics-augmented neural networks, i.e., neural network based constitutive models which fulfill all relevant mechanical conditions of hyperelasticity by construction. 
The framework takes into account the structure of neural network-based constitutive models, in particular, that their derivatives are more complex compared to analytical models.
The proposed framework allows for convenient mixed Hu-Washizu like finite element formulations applicable to nearly incompressible material behavior. 
The key feature of this work is a tailored energy-momentum scheme for time discretization, which allows for energy and momentum preserving dynamical simulations.
Both the mixed formulation and the energy-momentum discretization are applied in finite element analysis. 
For this, a hyperelastic physics-augmented neural network model is calibrated to data generated with an analytical potential.
In all finite element simulations, the proposed discretization techniques show excellent performance.
All of this demonstrates that, from a formal point of view, neural networks are essentially mathematical functions. 
As such, they can be applied in numerical methods as straightforwardly as analytical constitutive models. 
Nevertheless, their special structure suggests to tailor advanced discretization methods, to arrive at compact mathematical formulations and convenient implementations.

\end{abstract}
\vspace*{2ex}
\par\noindent\rule{\textwidth}{0.4pt}

\section{Introduction}\label{sec:intro}
In continuum mechanics, the motion of solid bodies is described by field equations, 
comprised of kinematic relations and balance laws which both are valid in general, and material specific constitutive models \cite{TruesdellNoll}. 
These field equations reflect our understanding of how material bodies behave and carefully fulfill important physical conditions. 
However, when applying standard numerical methods 
such as the widely used Newmark scheme or midpoint rule, 
which are well-established for linear problems, the discretization usually influences the physics in the nonlinear regime and is inclined to numerical instabilities. 
This is particularly the case in dynamic simulations, where the temporal discretization does not preserve necessary physical properties of their continuous counterparts \cite{greenspan1984,gonzalez2000}. 

To address this problem, an abundand number of structure-preserving methods, also known as geometric integrators, were proposed. These methods are designed to preserve the physics of the continuous system during the time discretization process and thereby enhance the numerical stability. 
There is a great variety of structure-preserving methods in the literature, which roughly can be divided into symplectic methods and energy-momentum schemes (EMS) \cite{hairer2006,sharma2020}. 
These methods can be distinguished by preserving different invariants of the system, where for instance the simultaneous preservation of the invariants, i.e.\ total energy, momentum maps and simplecticity, is in general not possible \cite{lew2004,zhong1988}. 
Here we restrict our consideration to the established EMS, in particular to the discrete gradient versions thereof, which preserve both linear and angular momentum maps, and the total energy of the system. 
The mechanism behind the EMS can be understood as an explicit \cite{simo1992} or, in case of the discrete gradients, implicit \cite{gonzalez1996,gonzalez2000} momentum-preserving projection of the midpoint rule onto the surface of constant energy \cite{simo1994,sharma2020}.
EMS are promising since they fulfill most design criteria for a practical time-stepping scheme as suggested by \cite{bathe2007}.
In that regard, the EMS is applicable for elastic and inelastic problems, it does not use additional variables (e.g.\ Lagrange multipliers) or parameters to be provided by the analyst, it is second-order accurate, and exhibits similar stability properties in large deformations and long term simulations for coarse time step sizes as energy-conserving integrators in linear analysis. 
The only drawbacks associated with the EMS are that the method leads to unsymmetric tangent stiffness matrices and that the implementation is more involved by different time point evaluations of stresses and remaining parts of the equations. 

The development of EMS can be traced back to the early works of \cite{laBudde1974,labudde1976a,laBudde1976b,greenspan1984}, which provide energy and momentum conserving methods for particle dynamics. 
Especially the work of \cite{greenspan1984} provided a formula which conserves both momentum maps and total energy for energy densities with only scalar dependencies. 
In the pioneering work \cite{simo1992} the so-called discrete energy-momentum method has been proposed, which has been improved and properly reformulated in \cite{laursen2001}. 
A milestone in the development of EMS, which circumvents the use of the mean value algorithm and its associated solution of a nonlinear equation on quadrature point level used in \cite{simo1992}, is related to the work \cite{gonzalez1996,gonzalez2000} where the discrete gradient EMS has been proposed. 
It is applicable to arbitrary nonlinear hyperelastic material models, leading to a projection-based algorithmic stress formula, and as a special case contains both the midpoint averaging of the strains for quadratic potentials \cite{simo1992} and the Greenspan formula \cite{greenspan1984} for scalar dependencies of the energy density. 
Other areas of successful developments of EMS are located in the fields of structural mechanics for nonlinear beams and shells \cite{romero2002,betsch2003,simo1994,betsch2016b}, inelasticity \cite{mohr2008,gross2010,martin2014,hesch2014}, elastodynamical contact problems \cite{laursen1997,chawla1998,hesch2011}, and systems with (non-)holonomic constraints \cite{gonzalez1999,celledoni2019}, to name but a few. 
An extension of EMS to dissipative systems is based on the general equation for non-equilibrium reversible irreversible coupling (GENERIC) formulation, which provides a systematic framework facilitating the design of energy-momentum and even entropy consistent schemes \cite{grmela1997,oettinger1997,romero2009,krueger2016,martin2016,betsch2019}. 

Recently, in \cite{betsch2018} a new family of partioned discrete gradients has been proposed which is inspired by polyconvex strain energies \cite{ball1976,Ball1977}, leading to a new algorithmic stress formula based on three partioned discrete gradients which represent the algorithmic counterparts of the work conjugates of the right Cauchy-Green strain tensor, its cofactor, and its determinant. 
The formulation in \cite{betsch2018} benefits from the so-called tensor cross-product, first published in \cite{deBoer1982} and kind of rediscovered in the field of computational nonlinear elasticity in \cite{bonet2015}, which is based on a Hu-Washizu type mixed variational formulation and leads to a remarkably simple structure for the algorithmic stress. 
This work has given rise to several polyconvexity-inspired works in the field of coupled problems, in particular for finite thermo-elastodynamics \cite{franke2018,ortigosa2020}, nonlinear electro-elastodynamics \cite{ortigosa2018,franke2019}, and nonlinear thermo-electro-elastodynamics \cite{franke2022,franke2023}. 

In addition to the above mentioned sophisticated temporal discretization methods, significant progress has been made in the development of advanced spatial discretization techniques over the past few decades. 
In partiuclar, an abundand number of mixed finite element (FE) formulations in elasticity based on Hu-Washizu variational functionals \cite{washizu1982} have been developed \cite{simo1990,armero2000,schroeder2011,bonet2015,betsch2018}. 
These formulations aim to address the issue of locking which standard displacement-based formulations are prone to.

\medskip

Shifting the focus to constitutive modeling, the mechanical conditions underlying hyperelasticity were extensively discussed in the last decades \cite{TruesdellNoll,ciarlet1988}, and sophisticated analytical models were formulated to fulfill all constitutive conditions by construction \cite{schroeder2003,Ebbing2010}. 
While the constitutive conditions that these models fulfill have a sound mechanical motivation, their explicit choice of functional relationship usually does not. Rather, it is heuristically motivated, such as the polynomial form of the Ogden model \cite{ogden2004}. This human choice of functional relationship easily restricts the function space that a model can represent, thus introducing an undesired \emph{model uncertainty} \cite{huellermeier2021}. This is a major drawback of analytical constitutive models.
On the other side, artificial neural networks (NN) are highly flexible functions, in fact, they are universal approximators \cite{Hornik1991}. However, when applied to mechanical problems, standard NNs easily violate physical conditions and require large amounts of calibration data. Hyperelastic physics-augmented neural networks (PANNs)\footnote{Following \cite{linden2023}, we denote physics-augmented neural networks (PANNs) as NN-based constitutive models which fulfill all relevant mechanical conditions by construction.} aim to combine the best of both worlds. Here, NN-based models are formulated which fulfill all mechanical conditions of hyperelasticity by construction. This yields highly flexible yet reliable constitutive models. Furthermore, the constitutive conditions included in the model formulation serve as an \emph{inductive bias} \cite{haussler1988}, which highly improves the models generalization properties and allows for a calibration with moderately sized datasets that could stem from experimental material characterization.

Basically, NNs are applied to represent hyperelastic potentials, which benefit from the excellent flexibility of NNs. In the same manner as analytical constitutive models \cite{schroeder2003,Ebbing2010}, NN-based potentials can be formulated in terms of invariants \cite{kalina2023,klein2022a,klein2022b,Linka2020,tac2022}, thus ensuring several mechanical conditions by construction.
Other approaches formulate potentials directly in terms of the components of strain tensors \cite{stpierre2023,klein2022a,asad2022,Fernandez2020,Vlassis2022a}.
For the formulation of polyconvex potentials, several approaches exist \cite{klein2022a,Chen2022,linka2023,tac2022,tac2023}, where the most noteworthy approaches are based on input-convex neural networks (ICNNs). Proposed in a pioneering work by \cite{Amos2016}, this special network architecture is attractive for, e.g., physical applications which require convexity \cite{huang2022a} and convex optimization \cite{calafiore2020a,calafiore2020b}.
While the aforementioned NN-based models fulfill some constitutive conditions in an exact way, other conditions are only approximated. E.g., the model proposed by \cite{klein2022a} fulfills several conditions including polyconvexity in an exact way, but only approximates the stress-free reference configuration. 
Indeed, in the framework of NN-based constitutive models, it is possible to fulfill constitutive conditions only in an approximate fashion \cite{weber2021,weber2023,masi2022}, e.g., the NN can learn to approximate a condition through data augmentation \cite{Ling2016}, or the loss function can include physical conditions \cite{karniadakis2021}. For some applications this is even necessary, e.g., when a formulation in invariants leads to a loss of information, which is the case for some anisotropy classes \cite{klein2022a}.
However, since fulfilling a mechanical condition by construction is the only way to be absolutely sure about its fulfillment, if possible, this should be preferred over approximations.

Finally, in the works of \cite{linka2023} and \cite{linden2023}, NN-based models were proposed which fulfill all common mechanical conditions of hyperelasticity by construction. While the work of \cite{linka2023} is restricted to incompressible material behavior, the work of \cite{linden2023} is formulated for general, compressible material behavior. 
In the present work, the model proposed by \cite{linden2023} is applied, which is an extension of the polyconvex model proposed by \cite{klein2022a}. 
The model fulfills mechanical conditions in several ways, first of all, by a cautious \emph{choice of inputs} for the NN (i.e., polyconvex invariants). Furthermore, the \emph{special choice of network architecture} (i.e., ICNNs) preserves the polyconvexity of the invariants. In addition, \emph{analytical growth terms} and \emph{invariant-based normalization terms} ensure a physically sensible energy and stress behavior. In that light, one could argue that the PANN model proposed by \cite{linden2023} is actually quite close to analytical constitutive models, with the only difference being the choice of ansatz function for the constitutive model. By using ICNNs as ansatz functions, the flexibility of the model can be increased to an arbitrary amount, while even for large number of model parameters, their calibration is still straightforward.

The most obvious reason to apply PANN models is their extraordinary flexibility in conjunction with their physically well motivated basis. In particular, PANN constitutive models can represent the highly challenging homogenized behavior of microstructured materials, for which analytical models may not be flexible enough \cite{klein2022a}. This allows for very efficient simulation of microstructured materials \cite{gaertner2021,kalina2023,masi2022}.
The flexibility has another implication, namely, PANNs can represent a variety of analytical constitutive models. In fact, for a wide range of applications, analytical constitutive models are flexible enough. However, there is a vast number of analytical constitutive models, and finding a suitable one can be very demanding and requires a lot of expert knowledge in both the choice and calibration of the model \cite{Ricker2023}. This can be bypassed by NN-based constitutive models, which automatically find the best function out of a very rich function space. Another noteworthy approach for this is the data-driven framework proposed by \cite{flaschel2021,flaschel2023}, where an approach for the automated discovery of analytical constitutive models is introduced.
On the downside, the flexibility of PANNs goes along with a lack of traceability. Indeed, some analytical constitutive models base their choice of functional relationship on physical reasoning, such as the hyperelastic Hencky model \cite{Hencky1928,Hencky1929,martin2017a}. For such models, the material parameters have a clear physical interpretation. This is in contrast to NN-based models, which have a large amount of parameters which are generally inaccessible to a clear interpretation \cite[Sect.~6]{klein2022a}. However, it should be noted that this is also the case for moderately flexible analytical models such as the Ogden model \cite{ogden2004}. 
Finally, with data-driven approaches it is also possible to circumvent the formulation of an explicit, analytical constitutive model \cite{kirchdoerfer2016,carrara2020,Prume_Reese_Ortiz_2023}.
To close, in the era of {artificial intelligence} and {big data}, it is indeed possible to largely abandon the formulation of models and let ML methods purely work on data \cite{pietsch_2021}.
However, it is broadly accepted that including scientific knowledge in ML methods has a multitude of advantages in many scientific fields \cite{rueden2021,peng2021,karniadakis2021,Kumar2022,liu2021e}. Overall, the scientific knowledge developed in the last centuries is not to be omitted, instead, the use of ML methods allows to apply them in a new light, and thereby, to exploit the advantages of up-to-date techniques.

\medskip

From a formal point of view, NNs are nothing more but mathematical functions. As such, they can be applied in standard numerical methods as straightforwardly as analytical constitutive models. However, compared to analytical constitutive models, NN-based constitutive models comprise far more complicated functional relationships. Thus, while applying standard methods is indeed possible, it potentially leads to demanding mathematical formulations and implementations. This can be circumvented by tailoring numerical methods to the special structure of NN-based constitutive models.

In the present work, advanced spatial and temporal discretization techniques are tailored to the special structure of hyperelastic PANN constitutive models. In particular, inspired by \cite{bonet2015,betsch2018,kraus2019}, the polyconvex structure of PANN models is taken into account.
Furthermore, the numerical methods are formulated to conveniently handle the - possibly very complex - functional relationships of NN-based constitutive models. This is done by setting the focus on invariant-based formulations.
Based on these ideas, a new Hu-Washizu like mixed formulation is proposed which uses scalar-valued strain invariants as additional unknowns. Furthermore, a novel discrete gradient EMS is designed, which mostly relies on scalar-valued formulations in terms of strain invariants.
To the best knowledge of the authors, both mixed spatial methods and discrete gradient integration schemes have not yet been tailored to PANN constitutive models so far.

\medskip

The outline of the manuscript is as follows. In \cref{sec:hyperel}, the basics of hyperelasticity are introduced, which are relevant for the PANN constitutive model introduced in \cref{sec:PANN}. In \cref{sec:cont_form}, continuum formulation and variation tailored to PANN models are introduced. This is followed by the energy- and momentum consistent time discretization introduced in \cref{sec:time_disc} and the spatial FE discretization discussed in \cref{sec:spat_disc}. Stability and robustness of the proposed framework are examined in representative numerical examples in \cref{sec:num}. After the conclusion in \cref{sec:conc}, the appendices provide some additional information on the proposed EMS.

\paragraph{Notation}
Throughout this work, tensor composition and contraction are denoted by $\left(\tens{A}\,\bB\right)_{ij}=A_{ik}B_{kj}$ and $\bA:\bB=A_{ij}B_{ij}$, respectively, with vectors $\ba$, $\bb$ and second order tensors $\bA$, $\bB$. 
Note that the Einstein summation convention is applied if not otherwise stated. 
The tensor product is denoted by~$\otimes$, the second order identity tensor by $\bI$. Transpose and inverse of a second order tensor $\bA$ are denoted as $\bA\transp$ and $\bA^{-1}$, respectively.
The tensor cross product as introduced by \cite{deBoer1982} is defined as $  \tens{A}\wedge\tens{B} = \varepsilon_{ijk}\,\varepsilon_{\alpha\beta\gamma}\,A_{j\beta}\,B_{k\gamma}\,\vec{e}_i\otimes\vec{e}_{\alpha}$, where \(\varepsilon_{ijk}\) denotes the third-order permutation tensor.
For a discussion of the properties of the tensor cross product, see \cite{bonet2015}.
Furthermore, trace, determinant and cofactor are denoted by $\tr\bA$, $\det\bA$ and $\cof\bA:=\det\left(\bA\right)\,\bA\ntransp$, respectively. 
The space of second order tensors is denoted as \(\T^3 := \{\tens{A}\in\mathcal{L}(\R^3,\R^3)\}\), the set of invertible second order tensors with positive determinant is denoted by \(\T^3_+ := \{\tens{A}\in\T^3\,\rvert\,\det(\tens{A})>0\}\), and the set of symmetric invertible second order tensors with positive determinant is denoted by \(\S^3_+:=\{\tens{A}\in\T^3_+\,\rvert\,\tens{A}=\tens{A}\transp\}\).
The orthogonal and special orthogonal group in $\mathbb{R}^3$ are denoted by $\text{O}(3):=\big\{\bA \in\allowbreak \T^3\;\rvert\allowbreak \;\bA\transp\bA=\bI\big\}$ and $\SO(3):=\big\{\bA \in\allowbreak \T^3\;\rvert\allowbreak \;\bA\transp\bA=\bI,\;\det \bA =1\big\}$, respectively.
The space of positive and negative real numbers are denoted by $\bbR_+:=\big\{x\in\bbR\;\rvert\allowbreak \;x>0\big\}$ and $\bbR_-:=\big\{x\in\bbR\;\rvert\allowbreak \;x<0\big\}$, respectively.
The first Fr\'echet derivative of a function $f$ w.r.t.\ $\bX$ is denoted by $\partial_{\subX}f$, while discrete gradients are denoted by $D_{\subX}f$.

\section{Basics of hyperelasticity}\label{sec:hyperel}
In this section, the kinematic relationships of nonlinear solid mechanics are introduced in \cref{sec:kinem}, followed by the constitutive conditions of hyperelasticity in \cref{sec:hyperel_cond}. In \cref{sec:hyperel_inv}, the formulation of hyperelastic potentials in terms of strain invariants is discussed.
\subsection{Kinematics}\label{sec:kinem}
Let us consider the motion of a solid body. The reference configuration \(\B\subset\R^3\) and the current, deformed configuration \(\mathcal{B}\subset\R^3\) are connected via the bijective deformation mapping \(\vec{\varphi}:\B\rightarrow\bbR^3\), linking material particles \(\vec{X}\in\B\) to $\vec{x} = \vec{\varphi}(\vec{X})\in\mathcal{B}$.
The material gradient of the deformation field \(\vec{\varphi}\), which is denoted as the deformation gradient \(\F:\B\rightarrow\T^3_+\), is defined as
\begin{equation}
  \F = \partial_{\subX}\vec{\varphi}(\vec{X})\,.
\end{equation}
Then, its determinant \(J:\B\rightarrow\R_+\) and its cofactor \(\H:\B\rightarrow\T^3_+\) follow as
\begin{equation}\label{eq:JH}
\begin{aligned}
  J &= \det(\F) = \frac{1}{3}\,\H:\F = \frac{1}{6}\,(\F\wedge\F):\F\,,
  \\
  \H& = \cof(\F) = \frac{1}{2}\,\F\wedge\F\,.
  \end{aligned}
\end{equation}
Let us furthermore introduce the symmetric Cauchy-Green strain tensor \(\C:\B\rightarrow\S^3_+\) as
\begin{equation}
  \C = \F\transp\,\F\,,
\end{equation}
with the corresponding determinant \(C:\B\rightarrow\R_+\) and cofactor \(\G:\B\rightarrow\S^3_+\) defined as
\begin{equation}
    \begin{aligned}
    C &= \det(\C) = \frac{1}{3}\,\G:\C = \frac{1}{6}\,(\C\wedge\C):\C\,,
    \\
      \G& = \cof(\C) = \frac{1}{2}\,\C\wedge\C\,.
    \end{aligned}
\end{equation}
Based on the above symmetric kinematic set, we introduce the isotropic invariants \(\I:\B\rightarrow\R_+\), \(\II:\B\rightarrow\R_+\) and \(\III:\B\rightarrow\R_+\) as
\begin{equation}\label{eq:invs}
  \I = \operatorname{tr}(\C)\,,\qquad \II = \operatorname{tr}(\G)\,,\qquad \III = C\,.
\end{equation}
\subsection{Constitutive conditions}\label{sec:hyperel_cond}
The specific mechanical behavior of a hyperelastic material can be expressed by a potential
\begin{equation}\label{eq:pot}
    \WF\colon\T^3_+\rightarrow\bbR\,,\qquad \F\mapsto \WF\left(\F\right)\,,
\end{equation}
which corresponds to the strain energy density stored in the body due to deformation \cite{holzapfel2000,Haupt2002,ciarlet1988}. 
With the first Piola-Kirchhoff (PK1) stress given as the gradient field
\begin{equation}
    \tens{P}=\partial_{\subF}\WF\left(\F\right)\,,
\end{equation}
the \textbf{(i) second law of thermodynamics} is fulfilled by construction. The second Piola-Kirchhoff (PK2) stress is given by
\begin{equation}
\tens{S}=\tens{F}^{-1}\tens{P}\,.
\end{equation}
The \textbf{(ii) balance of angular momentum} implies that $\tens{S}$ is symmetric. 
Furthermore, the principle of \textbf{(iii) objectivity} states that a model should be independent on the choice of observer, which is formalized as
\begin{equation}
    \WF\left(\bQ\,\F\right)=\WF\left(\F\right)\quad\forall\,\bQ\in\SO(3)\,.
\end{equation}
Throughout this work, isotropic material behavior is considered, which is taken into account by the \textbf{(iv) material symmetry condition}
\begin{equation}
    \WF(\F\,\bQ\transp)=\WF\left(\F\right)\quad\forall\,\bQ\in\text{O}(3)\,.
\end{equation}
Furthermore, we consider \textbf{(v) polyconvex} potentials which allow for a representation
\begin{equation}\label{eq:pc}
    \WF\left(\F\right)=\WFt\left(\F,\,\H,\,J\right)\,,
\end{equation}
with the kinematical quantities as introduced in \cref{eq:JH} and the function $\WFt$ which is convex in the components of $\F,\,\H$, and $J$. The notion of polyconvexity as introduced by Ball \cite{ball1976,Ball1977} stems from a quite theoretical context and is linked to the existence of solutions in finite elasticity. However, it is also the most straightforward way of fulfilling the ellipticity condition \cite{Zee1983,neff2015}
\begin{equation}
    \left(\ba\otimes\bb\right)\colon\frac{\partial^2 \WF\left(\F\right)}{\partial \F \partial \F}\colon \left(\ba\otimes\bb\right)\geq 0\quad\forall\,\ba,\bb\in\bbR^3\,.
\end{equation}
Also known as material stability, this condition leads to a favorable behavior in numerical applications. Furthermore, a physically sensible energy and stress behavior requires fulfillment of the \textbf{(vi) growth condition}
\begin{equation}
    \WF\rightarrow\infty \quad \text{as}\quad J \rightarrow 0^+\,,
\end{equation}
as well as that the reference configuration $\F=\bI$ is energy- and stress-free, i.e.,  
\begin{equation}
  \WF(\bI)=0 \,,\qquad  \bP(\bI)=\boldsymbol{0} \,,
\end{equation}
also referred to as \textbf{(vii) energy normalisation} and \textbf{(viii) stress normalisation}.
\subsection{Invariant-based modeling}\label{sec:hyperel_inv}
To conclude on the previous section, while in hyperelasticity the \textbf{(i) second law of thermodynamics} 
is fulfilled by construction, the remaining conditions require further considerations in the model formulation.
In a first step, by formulating the potential in terms of invariants of the right Cauchy-Green tensor $\C$, cf.~\cref{eq:invs}, conditions \textbf{(ii--iv)} can be fulfilled. In particular, as isotropic material behavior is assumed, the hyperelastic potential is formulated in terms of isotropic invariants, allowing for the representation
\begin{equation}
        \WI\colon\bbR_+\times\bbR_+\times\bbR_+\rightarrow\bbR\,,\qquad \vec{\mathcal{I}}^0\mapsto \WI\left(\vec{\mathcal{I}}^0\right)
\end{equation}
with
\begin{equation}
        \vec{\mathcal{I}}^0:=\left(\I,\,\II,\,J\right)\,.
\end{equation}
The isotropic invariants introduced in \cref{eq:invs} are polyconvex, in particular, $\I$ is convex in $\F$ and $\II$ is convex in $\H$. 
To preserve the polyconvexity of the invariants, the potential $\WI$ is chosen as a convex function in $\vec{\mathcal{I}}^0$ and furthermore as \emph{non-decreasing} in $\I,\,\II$, which overall ensures fulfillment of the \textbf{(v) polyconvexity condition}. The non-decreasing condition is owed to the fact that $\I,\,\II$ are already non-linear functions of the arguments of the polyconvexity condition, cf.~\cref{eq:pc} and \cref{eq:invs},  see also \cite[Rem.~A.10]{klein2022a} for an explicit proof and \cite[Sect.~2]{klein2023a} for a simple 1D example. In contrast to that, $J$ is the only invariant quantity in the arguments of the polyconvexity condition, meaning that the potential must be convex in $J$, but may indeed be non-decreasing in $J$.
This additional flexibility of the model in $J$ is also the reason to not use $\III$, which would again be a non-linear function in $J$, and is furthermore essential for the model to represent negative stress values at all, cf.~\cite{klein2022a}.
The subsequent NN-based model formulation is simplified with the slightly adapted version 
\begin{equation}\label{eq:W_inv}
        \WIt\colon\bbR_+\times\bbR_+\times\bbR_+\times\bbR_-\rightarrow\bbR\,,\qquad \vec{\mathcal{I}}\mapsto \WIt\left(\vec{\mathcal{I}}\right)\,,
\end{equation}
with
\begin{equation}
        \vec{\mathcal{I}}:=\left(\I,\,\II,\,J,\,J^*\right)\,,\qquad J^*:=-J\,.
\end{equation}
By this adaption, the function $\WIt$ can be chosen as a convex function in $\vec{\mathcal{I}}$ and a non-decreasing function in all of its arguments.
Note that this general form of the potential does not yet fulfill conditions \textbf{(vi--viii)}, which ensure a physically sensible energy and stress behavior of the model. 
For a formulation of the potential in isotropic invariants, the second Piola-Kirchhoff stress follows as
\begin{equation}\label{eq:PK2}
        \tens{S}=2
      \partial_{\subC}\,   \WI
      =2\partial_{\I} \WI\,
      \bI +2\partial_{\II}\WI\,
      \tens{I}\wedge\C +\partial_{J} \WI\,
      J^{-1}\G\,,
\end{equation}
where we made use of the tensor generators \cite{kalina2022a}
\begin{equation}\label{eq:generators}
    \begin{aligned}
    \partial_{\subC}\I=\bI\,,\qquad
       \partial_{\subC}\II=\tens{I}\wedge\C\,, \qquad
       \partial_{\subC}J=\frac{1}{2}J^{-1}\G\,.
    \end{aligned}
\end{equation}
An analytical example for hyperelastic potentials is the Mooney-Rivlin model, given by
\begin{equation}\label{equation:MR}
  W^{\text{MR}} = a\,(\I-3) + b\,(\II-3) + \frac{c}{2}\,(J-1)^2 - d\,\log(J)\,,\qquad a,b,c\geq 0\,,
\end{equation}
which is formulated such that it fulfills the remaining conditions \textbf{(vi--viii)}. 
In particular, the stress normalization condition is ensured by the additional constraint $d=2(a+2b)$.

\section{Physics-augmented neural network constitutive model}\label{sec:PANN}
In this section, the constitutive equations of the neural network (NN) constitutive model applied throughout this work are introduced in \cref{sec:PANN_con_eq}, followed by a description of the model calibration process in \cref{sec:PANN_calibration}. Afterwards, in \cref{sec:derivs}, the derivatives of the NN-based constitutive model required throughout this work are introduced.
\subsection{Constitutive equations}\label{sec:PANN_con_eq}
\begin{figure}[t!]
\centering
\resizebox{0.9\textwidth}{!}{
\begin{tikzpicture}[x=1.6cm,y=1.1cm]
  \large
  \def\NC{6} 
  \def\nstyle{int(\lay<\Nnodlen?(\lay<\NC?min(2,\lay):3):4)} 
  \tikzset{ 
    node 1/.style={node in_out},
    node 2/.style={node inv_pot},
    node 3/.style={node icnn},
  }
  
\draw[color33!40,fill=color33,fill opacity=0.02,rounded corners=4]
    (1.3,-1.5) --++ (0,4.0) --++ (0.8,0) --++ (0,-4.0) -- cycle;
     
\draw[color22!40,fill=color22,fill opacity=0.02,rounded corners=4]
    (3.1,-2) rectangle++ (0.8,5.0);

\draw[color33!40,fill=color33,fill opacity=0.02,rounded corners=4]
    (4.4,-0.3) rectangle++ (5.8,1.6);
  
 \node[node 1, outer sep=0.6] (2-1) at (0.5,0.5) {$\C$};

\node[anchor=west] (3-1) at (1.45,2) {$\I$};
\node[anchor=west] (3-2) at (1.45,1) {$\II$};
\node[anchor=west] (3-3) at (1.45,0) {$J$};
\node[anchor=west] (3-4) at (1.45,-1) {$J^*$};

\draw[connect arrow] (2-1) -- (3-1);
\draw[connect arrow] (2-1) -- (3-2);
\draw[connect arrow] (2-1) -- (3-3);
\draw[connect arrow] (2-1) -- (3-4);

\def\N{5}
\foreach \i [evaluate={\y=\N/2-\i+1.0;}] in {1,...,3}
{ 
    \node[node 3,outer sep=0.6] (4-\i) at (3.5,\y) {};
    
    \foreach \j in {1,2,3,4}{
        \draw[connect arrow]  (3-\j) -- (4-\i);
    }
}

\node[scale=1.2] (4-4) at (3.5,-0.5) {$\vdots$};

\foreach \i [evaluate={\y=\N/2-\i+1.0;}] in {5,...,5}
{ 
    \node[node 3,outer sep=0.6] (4-\i) at (3.5,\y) {};
    
    \foreach \j in {1,2,3,4}{
        \draw[connect arrow]  (3-\j) -- (4-\i);
    }
}


\node[node 2, outer sep=0.6] (6-1) at (5,0.5) {$W^{\text{NN}}$};

\foreach \j in {1,...,3}{
    \draw[connect arrow]  (4-\j) -- (6-1);
}

\foreach \j in {5,...,\N}{
    \draw[connect arrow]  (4-\j) -- (6-1);
}

\node[align = center] (7-1) at (7.75,0.5) {$+\:W^{\text{stress}}+W^{\text{energy}}+W^{\text{growth}}=:W^{\text{PANN}}$};

\node (8-1) at (10.8,0.5) {$\partial_{\subC} $};
\draw[connect arrow]  (7-1) -- (8-1);

\node[node 1, outer sep=0.6] (9-1) at (11.8,0.5) {$ \tens{S}$};
\draw[connect arrow]  (8-1) -- (9-1);

  \end{tikzpicture}
}
\caption{Illustration of the PANN based constitutive model. Note that the hidden layer (yellow) of the NN may be multilayered.}
\label{fig:PANN}
\end{figure}
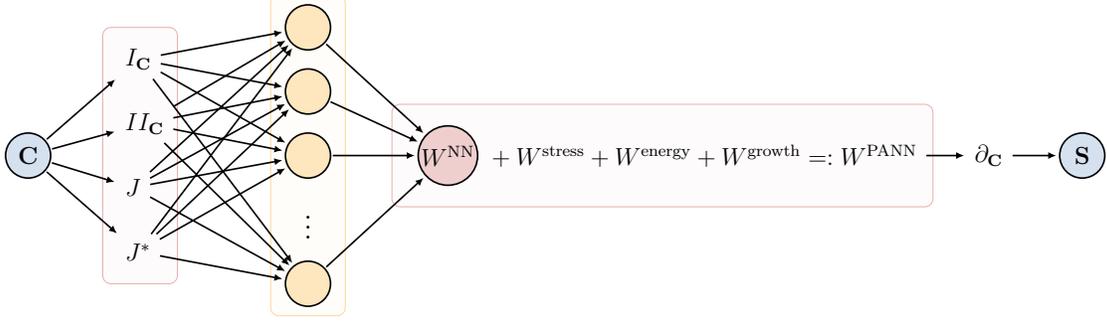

As discussed in \cref{sec:hyperel_inv}, hyperelastic potentials are usually formulated in terms of invariants of the right Cauchy-Green tensor. When polyconvex invariants are considered and the hyperelastic potential is a convex and non-decreasing function of the invariants, the overall potential is polyconvex. 
The simple structure and recursive definition of feed-forward neural networks (FFNNs) \cite{kollmannsberger2021,aggarwal2018} make them a very natural choice for the construction of convex and non-decreasing functions \cite{Amos2016}, which suggests to use them for the formulation of polyconvex hyperelastic potentials \cite{klein2022a}. 
Still, the NN is only a part of the overall {physics-augmented neural network} (PANN) constitutive model proposed by \cite{linden2023}, which is given by
\begin{equation}\label{eq:PANN}
    W^{\text{PANN}}\left(\vec{\mathcal{I}}^0\right)=W^{\text{NN}}\left( \vec{\mathcal{I}}\right)
    +W^{\text{stress}}\left(J\right)+W^{\text{energy}}+W^{\text{growth}}\left(J\right)\,,
\end{equation}
with $\vec{\mathcal{I}}^0,\,\vec{\mathcal{I}}$ as defined in \cref{sec:hyperel_inv}.
Here, $W^{\text{NN}}$ denotes the NN part of the model, which provides the model with its excellent flexibility, while the remaining terms ensure a physically sensible stress and energy behavior of the model, cf.~\cref{sec:hyperel_cond}. The overall flow and structure of the PANN model is visualized in \cref{fig:PANN}.
In the present work, without loss of generality, neural networks with only one hidden layer are applied, which has shown to be sufficient for the representation of most isotropic hyperelastic potentials \cite{klein2022b,linden2023}. For an introduction to multilayered PANN architectures, the reader is referred to \cite{linden2023}. In the present work, the NN takes the form
\begin{equation}\label{eq:WNN}
      W^{\text{NN}}(\vec{\mathcal{I}}) = \vec{w}^2\cdot\mathcal{SP}\left(\vec{w}^1\cdot\vec{\mathcal{I}}+\vec{b}\right)\,,
\end{equation}
where $\mathcal{SP}(x) = \log(1+e^x)$ denotes the \emph{Softplus} function, which is applied component-wise. This non-linear function is referred to as {activation function}, while $\vec{w}^1\in\bbR^{n\times 4}_{\geq 0},\,\vec{w}^2\in\bbR^n_{\geq 0}$ are referred to as {weights}, and $\vec{b}\in\bbR^n$ is referred to as {bias}. 
When all weights are greater or equal to zero, and the activation function is a convex and non-decreasing function, the overall function \cref{eq:WNN} is convex and non-decreasing in $\vec{\mathcal{I}}$, thus leading to a polyconvex potential \cite{klein2022a}. This is the motivation for using the \emph{Softplus} activation function, which is convex and non-decreasing, where it should be noted that also other suitable activation functions exist \cite{asad2022,Amos2016}. In general, NNs which are convex in the input are referred to as input-convex neural networks (ICNNs) \cite{Amos2016}.   

Together, weights and bias form the set of model parameters, here denoted as $\vec{\mathcal{P}}$, which are optimized in the calibration process to fit the model to a given dataset. For a given input dimension, the size of the weights and bias matrices is determined by $n$, which specifies the number of {nodes} / {neurons} in the NN. This is a {hyperparameter} of the NN, which is fixed before the model calibration. By increasing $n$, the model flexibility is increased.

The polyconvex stress normalization term as proposed by \cite{linden2023} is given by
\begin{equation}
    W^\text{stress}(J):=-\mathfrak{n}\,(J-1)\;,
\end{equation}
where the constant
\begin{equation}
    \mathfrak{n}:=2\left(\partial_{\I}\, W^\text{NN}
    + 2 \partial_{\II}\,W^\text{NN}   +\frac{1}{2}\partial_{J}\,W^\text{NN}-\frac{1}{2}\partial_{J^*}\,W^\text{NN}
    \right)\bigg\rvert_{\footnotesize \F=\bI}\in\bbR
\end{equation}
is a weighted sum of derivatives of the NN potential with respect to the invariants for the undeformed state $\F=\bI$. The stress correction term preserves polyconvexity as it is linear in $J$. 
The energy normalization term is given by
\begin{equation}
    W^\text{energy}:=-W^\text{NN}\Big\rvert_{\footnotesize \F=\bI}\in\bbR\,.
\end{equation}
Finally, the volumetric growth term is chosen as the analytical, polyconvex function
\begin{equation}
  W^{\text{growth}}(J) = (J + J^{-1} - 2)^2\,.
\end{equation}

Setting aside the nomenclature of machine learning, one could consider \cref{eq:PANN} as a classical constitutive model, which uses linear transformations in combination with the \emph{Softplus} function as an ansatz for hyperelastic potentials, cf.~\cref{eq:WNN}. Then, two benefits compared to other hyperelastic constitutive models become obvious. 
First of all, the function $W^{\text{NN}}$ allows for a strong interrelation between the different invariants, which becomes even more clear when considering the derivatives of the potential, cf.~\cref{sec:derivs}. This is in contrast to most analytical polyconvex constitutive models. For such models, the strain energy function is usually additively decomposed in the invariants \cite{Ebbing2010}, see e.g.~\cref{equation:MR}. This is due to the fact that multiplication of polyconvex invariants does not preserve polyconvexity \cite{Hartmann2003}. The NN-potential in \cref{eq:WNN}, however, allows for a strong interrelation between the invariants, while also preserving their polyconvexity.  
The second benefit is the immediate possibility to increase the model flexibility, basically to an arbitrary amount \cite{Hornik1991}. While analytical constitutive models generally can become also very flexible, e.g., by considering enough terms in the polynomial ansatz of the Ogden model \cite{ogden2004}, their calibration quickly becomes infeasible. In contrast, the NN ansatz in \cref{eq:WNN} has proven to be very stable in the calibration, even for large number of parameters \cite{klein2022a}. In this regard, it should be mentioned that the PANN model does not suffer from phenomena such as overfitting, since the inclusion of physics provides the model with a pronounced mathematical structure \cite{linden2023}.

\subsection{Model calibration}\label{sec:PANN_calibration}

After fixing the model architecture, here, the number of nodes $n$, the model parameters $\vec{\mathcal{P}}$ can be optimized to fit the model to a given dataset. Throughout this work, datasets of the form
\begin{equation}
    \mathcal{D}=\left\{\left(^1\tens{C},\,^1\tens{S}\right),\dotsc,\left(^m\tens{C},\,^m\tens{S}\right)\right\}\,,
\end{equation}
are considered, consisting of $m$ strain-stress tuples in the right Cauchy-Green deformation tensor $\tens{C}$ and the second Piola-Kirchhoff stress $\tens{S}$. Thereby, two distinct datasets are created. A calibration dataset $\mathcal{D}_c$, which is used to calibrate the model, and a test dataset $\mathcal{D}_t$, which is used to examine the models generalization, i.e., its capability to predict load cases not included in the calibration process. To this end, the test dataset $\mathcal{D}_t$ should consist of fairly general load cases. 
Then, the loss function given as the mean squared error
\begin{equation}\label{eq:MSE}
    \mathcal{MSE}\left(\vec{\mathcal{P}}\right)=\frac{1}{9\text{Pa}^2}\frac{1}{m_c}\sum_{i=1}^{m_c}\norm{^i\tens{S}-\tens{S}^{\text{PANN}}\left(^i\tens{C};\,\vec{\mathcal{P}}\right)}^2\,,
\end{equation}
is minimized, where $m_c=\lvert\mathcal{D}_c\rvert$ denotes the number of tuples in $\mathcal{D}_c$ and $\norm{\cdot}$ denotes the Frobenius norm. Note that the model proposed by \cite{linden2023} is independent of the choice of loss function and also other loss functions could be applied for calibration.
As can be seen from \cref{eq:MSE}, the PANN model is calibrated only through its gradients $\tens{S}^{\text{PANN}}=\partial_{\subC}W^{\text{PANN}}$, which is referred to as Sobolev training \cite{Vlassis2020,Vlassis2022a}. In general, also values of the potential itself could be included in the loss function. However, while synthetic datasets can yield information about energies, this is generally not the case for real-world experiments. Thus, including energy values in the loss function would lead to a less general formulation and not even lead to significant benefits in the calibrated models \cite{Fernandez2020}.
\subsection{Derivatives of the neural network potential}\label{sec:derivs}
For the FE implementation of the PANN model, both first and second derivatives of the potential w.r.t.\ the invariants are required. 
For FFNN architectures with a large number of hidden layers, calculating derivatives in an explicit way quickly leads to huge expressions. However, for the single-layered FFNN used in this work, the explicit derivatives are very manageable. 
For the convenience of the reader, the derivatives of the NN part of the PANN model are provided here. Note that for the calibration process, cf.~\cref{sec:PANN_calibration}, the derivatives of the NN are not implemented in an explicit way, but the automatic differentiation provided by TensorFlow is applied.

The first derivative is given by
    \begin{equation}\label{eq:NN_first_deriv}
    \partial_{(\subI)_i}\,W^{\text{NN}}
 = \sum_{a=1}^n w_a^2\,\frac{e^{h_a}}{1+e^{h_a}}\,w_{ai}^1\,,\qquad\text{with }\; \vec{h}:=\vec{w}^1\cdot\vec{\mathcal{I}}+\vec{b}\,,
  \end{equation}
  where we made use of the derivative of the \emph{Softplus} function 
  \begin{equation}
   \partial_x \, \mathcal{SP}(x)=\frac{e^x}{1+e^x}\,.
  \end{equation}
  Actually, $\partial_x \, \mathcal{SP}(x)$ is the widely used \emph{Sigmoid} activation function. This shows another benefit of applying the \emph{Softplus} activation function for the representation of hyperelastic potentials. Since also its derivative is a commonly used activation function, the derivatives of the NN potential benefit from the properties of activation functions \cite{huang2022a}, which are specifically chosen for the application in NNs. Note that in \cref{eq:NN_first_deriv}, the derivatives of the NN potential w.r.t.\ each single invariant also depends on the remaining invariants.
 The second derivative follows as
      \begin{equation}
       \partial^2_{(\subI)_i(\subI)_j}\,W^{\text{NN}} = \sum_{a=1}^n w_a^2\,\frac{e^{h_a}}{(1+e^{h_a})^2}\,w_{ai}^1\,w_{aj}^1\,.
    \end{equation}
    
\section{Continuum formulation of finite deformation elasticity} \label{sec:cont_form}
While in the previous section, material specific constitutive models were discussed, in this section, general balance laws of finite deformation elasticity are introduced.
For this, we consider the potential energy \(\Pi\), which is comprised of an internal part \(\Pi^{\text{int}}\) and an external part \(\Pi^{\text{ext}}\), respectively, such that
\begin{equation}\label{equation:potentials}
  \Pi(\vec{\varphi}) =  \Pi^{\text{int}} + \Pi^{\text{ext}}\,,
\end{equation}
where the external part is given by 
\begin{equation}
  \Pi^{\text{ext}} = -\int_{\B}\bar{\vec{B}}\cdot\vec{\varphi}\,\d[V] - \int_{\partial\B^{\subT}}\bar{\vec{T}}\cdot\vec{\varphi}\,\d[A]\,.
\end{equation}
Therein, \(\bar{\vec{B}}:\B\rightarrow\R^3\) is the prescribed body force and \(\bar{\vec{T}}:\partial\B^{\subT}\rightarrow\R^3\) denotes the prescribed Piola-Kirchoff traction vector on \(\partial\B^{\subT}\subset\partial\B\), where the boundary is comprised of a Dirichlet boundary \(\partial\B^{\subVarphi}\) with prescribed deformations \(\bar{\vec{\varphi}}:\partial\B^{\subVarphi}\rightarrow\R^3\) and a Neumann boundary \(\partial\B^{\subT}\) with prescribed tractions \(\bar{\vec{T}}:\partial\B^{\subT}\) such that 
\(\partial\B = \partial\B^{\subT}\cup\partial\B^{\subVarphi}\) and \(\partial\B^{\subT}\cap\partial\B^{\subVarphi}=\emptyset
\). 
Furthermore, in \cref{equation:potentials}, \(\Pi^{\text{int}}\) basically denotes the elastic energy stored internally by the deformed body. 
This can be a purely displacement-based version
\begin{equation}
\Pi^{\text{int}} = \int_{\B}\WF(\F)\,\d[V]\,,
\end{equation}
which itself depends on the hyperelastic potential introduced in \cref{eq:pot}. 
In the above, the solution function \(\vec{\varphi}\in\mathds{V}_{\subVarphi}\) is subject to the space 
\begin{equation}
  \mathds{V}_{\subVarphi} = \{\vec{\varphi}:\B\rightarrow\R^3\,\rvert\,\varphi_i\in H^1(\B)\land \vec{\varphi}=\bar{\vec{\varphi}}\:\text{on}\: \partial\B^{\subVarphi}\land\det(\FPhi)>0\:\text{in}\:\B\}\,,
\end{equation}
where \(H^1\) denotes the Sobolev space.
The principle of stationary potential energy requires the satisfaction of the stationary condition, such that
\begin{equation}
  \delta\Pi = \delta\Pi^{\text{int}} + \delta\Pi^{\text{ext}} \stackrel{!}{=} 0\,,
\end{equation}
which has to hold for arbitrary \(\delta\vec{\varphi}\in\mathds{V}^0_{\subVarphi}\) with the space
\begin{equation}
  \mathds{V}_{\subVarphi}^0 = \{\delta\vec{\varphi}:\B\rightarrow\R^3\,\rvert\,\delta\varphi_i\in H^1(\B)\land \delta\vec{\varphi}=\vec{0}\:\text{on}\: \partial\B^{\subVarphi}\}\,,
\end{equation}
where the variation of \(\Pi^{\text{ext}}\) leads to \begin{equation}
  \delta\Pi^{\text{ext}} = -\int_{\B}\bar{\vec{B}}\cdot\delta\vec{\varphi}\,\d[V] - \int_{\partial\B^{\subT}}\bar{\vec{T}}\cdot\delta\vec{\varphi}\,\d[A]\,.
\end{equation}
The variation of the internal potential \(\Pi^{\text{int}}\) will be dealt with subsequently after discussing different displacement-based and mixed formulations. 
\subsection{Displacement-based formulations}
In solid mechanics, the primary strain measure is the deformation gradient \(\F\). 
As already pointed out in \cref{sec:hyperel_cond}, the notion of polyconvexity introduced by Ball \cite{ball1976,Ball1977} suggests to not only consider the deformation gradient \(\F\) in advanced numerical methods, but also its cofactor $\tens{H}$ and determinant $J$, cf.~\cref{eq:pc}. 
In particular, hyperelastic potentials are usually formulated in this extended set of arguments of $\F$, i.e., they are formulated in \((\F,\,\tens{H},\,J)\). Considering this suggests to use the tensor cross product firstly introduced by \cite{deBoer1982}, which facilitates both the analytical expressions and the implementation of advanced numerical methods, as it leads to neat derivatives of $(\F,\,\tens{H},\,J)$ w.r.t. $\F$. This has already been realized in purely displacement-based and mixed methods \cite{bonet2015}, as well as energy-momentum schemes \cite{betsch2018}.
Starting with the former, \cite{bonet2015} proposed a new computational framework based on polyconvex hyperelastic potentials of the form
\begin{equation}
 \begin{aligned}\label{equation:FHJStrainEnergy}
 \WFt\colon\T^+_3\times\T^+_3\times\bbR_+\rightarrow\bbR\,,\qquad(\F,\tens{H},J)\mapsto \WFt(\F,\tens{H},J)\,.
 \end{aligned}
 \end{equation}
The above stored energy function inspired the objective version thereof, given by  
\begin{equation}\label{equation:CGcStrainEnergy}
\begin{aligned}
\WCt\colon\S^3_+\times\S^3_+\times\bbR_+\rightarrow\bbR\,,\qquad(\C,\G,C)\mapsto\WCt(\C,\G,C)\,.
\end{aligned}
\end{equation}
Introduced by \cite{betsch2018}, this formulation facilitates the design of an energy-momentum scheme (EMS).

\medskip

Furthermore, as has been outlined in \cref{sec:hyperel_inv}, hyperelastic potentials are usually formulated in terms of strain invariants, which basically are nonlinear functions of $(\F,\,\tens{H},\,J)$. 
Thus, the invariant-based formulation
\begin{equation}\label{equation:invariantStrainEnergy}
\begin{aligned}
\WI\colon\R_+\times\R_+\times\R_+\rightarrow\R\,,\qquad (\I,\II,J)\mapsto\WI(\I,\II,J)\,,
\end{aligned}
\end{equation}
can be seen as a consecutive development of established polyconvexity-inspired numerical methods \cite{bonet2015,betsch2018}, now setting the focus directly on invariants.
Variation of the displacement-based potential energy \({}^{\W}\Pi^{\text{int}}\), which depends on the strain energy $\WI$ as defined in \cref{equation:invariantStrainEnergy}, yields 
      \begin{equation}\label{equation:variationDisplacementInvariantFormulation}
        \begin{aligned}
          \delta({}^{\W}\Pi^{\text{int}}) &= \int_{\B}\delta\W(\I,\,\II,\,J)\,\d V\\
          &= \int_{\B}2\,\left(\partial_{\I}\W\,\tens{I} + \partial_{\II}\W\,\tens{I}\wedge\C + \tfrac{1}{2}\,\partial_{J}\W\,J^{-1}\,\G\right):\frac{1}{2}\,\delta\C\,\d[V]\,,
        \end{aligned}
      \end{equation}
      which has to hold for arbitrary \(\delta\vec{\varphi}\in\mathds{V}^0_{\subVarphi}\). Note that in the above expression, one can identify the second Piola-Kirchhoff stress, cf.~\cref{eq:PK2}. 
      In the numerical examples presented in \cref{sec:num}, the invariant-based formulation of the potential is applied for both the analytical Mooney-Rivlin model, cf.~\cref{equation:MR}, and the neural network constitutive model, cf.~\cref{sec:PANN}.

For analytical material models, the derivatives of the hyperelastic potential w.r.t. the strain invariants mostly are simple, constant expressions, cf.~\cref{equation:MR}. In contrast to that, the derivatives of NN-based potentials w.r.t. the strain invariants become quite extensive, cf.~\cref{sec:derivs}. This holds for both the first and the second derivative required in numerical methods. On the other side, the tensor generators, cf.~\cref{eq:generators}, obviously remain the same for both analytical and NN-based models. From this perspective, the essential difference between analytical and NN-based models lies in the increased complexity of their functional relationship in the invariants. Thus, NN-based constitutive models naturally set the focus on invariant-based formulations such as \cref{equation:invariantStrainEnergy}, to take this increased complextiy into account when designing advanced discretization techniques.
Nevertheless, it should be noted that also the formulations \cref{equation:FHJStrainEnergy} and \cref{equation:CGcStrainEnergy} can be applied for NN-based constitutive models, by simple application of the chain rule.
However, in the subsequent section on time discretization, it becomes obvious that the invariant-based strain energy formulation offers implementation advantages specificially related to the discrete gradients. 
This advantage arises from its scalar dependencies, where further details on handling these structurally simple discrete gradients are discussed in \cref{sec:time_disc}.  
\subsection{Mixed formulations}\label{section:mixedFormulations}
The formulation discussed in the previous section is displacement-based and thus prone to locking behaviour, e.g., for nearly incompressible material models and bending dominated deformations. 
To circumvent this issue, a variety of mixed formulations have been proposed. 
The purpose of the present study is not only to demonstrate the ability of the PANN constitutive model to be employed within a mixed formulation, but also to develop a mixed method which is tailored for the PANN model.  
As in the previous section, the mixed formulations discussed here are inspired by the notion of polyconvexity, in the sense that they consider the arguments of the polyconvexity condition as additional unknowns, cf.~\cref{eq:pc}.

In this regard, the Hu-Washizu like 7-field mixed formulation from \cite{bonet2015} corresponding to the displacement-based version \cref{equation:FHJStrainEnergy} is worth mentioning.
There, the consistency of \((\F,\H,J)\) with $\vec{\varphi}$ is enforced by means of the Lagrange multipliers \((\tens{\Lambda}^{\subF},\tens{\Lambda}^{\subH},\Lambda^J)\), respectively, and the total set of unknowns is \((\vec{\varphi},\F,\H,J,\tens{\Lambda}^{\subF},\tens{\Lambda}^{\subH},\Lambda^J)\). 
A further notable version is proposed by \cite{betsch2018}, which corresponds to the displacement-based version \cref{equation:CGcStrainEnergy}. 
There, the consistency of \((\C,\G,\c)\) is enforced by means of the Lagrange multipliers \((\LambdaC,\LambdaG,\Lambdac)\), with the total set of unknowns \((\vec{\varphi},\C,\G,\c,\LambdaC,\LambdaG,\Lambdac)\) employed. 
The extension to enhanced assumed strain (EAS) variants of the above mentioned 7-field mixed formulation should be mentioned and can also be employed in a straight forward manner, cf.\ \cite{pfefferkorn2020}. 
Here, we restrict our consideration without loss of generality to the 5-field mixed formulation \({}^{\WG}\Pi_5^{\text{int}}\) proposed by \cite{kraus2019}, which the authors in \cite{kraus2019} claim to be robust, efficient and locking-free.
It is given by\footnote{In this subsection, the subscript $\subVarphi$ indicates that a quantity is calculated through the displacement field rather than being an additional unknown.} 
\begin{equation}\label{equation:5fieldFormulation}
  {}^{\WG}\Pi_5^{\text{int}} = \int_{\B}\WG(\CPhi,\G,J) + \LambdaG:(\cof(\CPhi) - \G) + \Lambda^J\,(\det(\FPhi) - J)\,\d V\,,
\end{equation}
where \(\vec{\varphi}\in\mathds{V}_{\subVarphi}\). 
Furthermore, \(\G\in\mathds{V}_{\subA}\) and \(J\in\mathds{V}_{A}\) are enforced by means of Lagrange multipliers \(\LambdaG\in\mathds{V}_{\subA}\) and \(\Lambda^J\in\mathds{V}_{A}\) with generalized spaces 
\begin{equation}
  \mathds{V}_{\subA} = \mathds{V}_{\subA}^0 = \{\vec{A}:\B\rightarrow\S^3_+\,\rvert\,A_{ij}\in\mathds{L}_2(\B)\},\quad \mathds{V}_{A} = \mathds{V}_{A}^0 = \{A:\B\rightarrow\R_+\,\rvert\,A\in\mathds{L}_2(\B)\}\,.
\end{equation}
The corresponding variation yields 
\begin{equation}
\begin{aligned}
  \delta({}^{\WG}\Pi_5^{\text{int}}) = \int_{\B}&2\,(\partial_{\subC_{\subVarphi}}\WG+\LambdaG\wedge\CPhi+\tfrac{1}{2}\,\Lambda^J\,\JPhi^{-1}\,\GPhi):\tfrac{1}{2}\,\delta\CPhi\\
  & + (\partial_{\subG}\WG-\LambdaG):\delta\G\\
  & + (\partial_{J}\WG-\Lambda^J)\,\delta J\\
  & + (\cof(\CPhi) - \G):\delta\LambdaG\\
  & + (\det(\FPhi) - J)\,\delta\Lambdac\,\d V\,,
\end{aligned}
\end{equation}    
which has to hold for arbitrary \(\delta\vec{\varphi}\in\mathds{V}^0_{\subVarphi}\), \(\delta\G,\delta\LambdaG\in\mathds{V}^0_{\subA}\), and \(\delta J,\delta\Lambda^J\in\mathds{V}^0_{A}\). 
Based on the above newly proposed displacement formulation in \cref{equation:invariantStrainEnergy}, we also suggest a new Hu-Washizu like 7-field mixed formulation \({}^{\W}\Pi_7^{\text{int}}\) given by\footnote{The initial idea to use a strain energy depending solely on the invariants of \(\C\), i.e.\ \(W=W(\I,\II,\III)\) resulted in the occurrence of hourglassing phenomena for nearly incompressible materials under shear loading conditions. Consequently, this approach was deemed unsuitable and subsequently discarded.}
\begin{equation}\label{equation:new7fieldFormulation}
  \begin{aligned}
        {}^{\W}\Pi_7^{\text{int}} = \int_{\B}\W(\I,\II,J) &+ \lambda^{\I}\,(\tr\CPhi - \I)\\
        &+ \lambda^{\II}\,(\tr(\cof\CPhi) - \II)\\
        &+ \lambda^{J}\,(\det\FPhi - J)\,\d V\,,
  \end{aligned}
\end{equation}
with \(\vec{\varphi}\in\mathds{V}_{\subVarphi}\) and \(\I,\II,J,\lambda^{\I},\lambda^{\II},\lambda^J\in\mathds{V}_A\). 
The above formulation can also be regarded as tailor made for the employed PANN constitutive model. 
Variation of this Hu-Washizu like 7-field mixed formulation yields
    \begin{equation}
      \begin{aligned}
        \delta({}^{\W}\Pi_7^{\text{int}}) = \int_{\B}&2\,(\lambda^{\I}\,\tens{I}+\lambda^{\II}\,(\tens{I}\wedge\CPhi)+\tfrac{1}{2}\,\lambda^{J}\,\JPhi^{-1}\,\GPhi):\tfrac{1}{2}\,\delta\CPhi\\
        & + (\partial_{\I}\W-\lambda^{\I})\,\delta\I\\
          & + (\partial_{\II}\W-\lambda^{\II})\,\delta\II\\
            & + (\partial_{J}\W-\lambda^{J})\,\delta J\\
              & + (\tr\CPhi - \I)\,\delta\lambda^{\I}\\
              &+ (\tr\GPhi - \II)\,\delta\lambda^{\II}\\
              &+ (\det\FPhi - J)\,\delta\lambda^{J}\,\d V\,,
      \end{aligned}
    \end{equation}
which has to hold for arbitrary \(\delta\vec{\varphi}\in\mathds{V}^0_{\subVarphi}\) and \(\delta\I,\delta\II,\delta J,\delta\lambda^{\I},\delta\lambda^{\II},\delta\lambda^J\in\mathds{V}^0_{A}\). 
\subsection{Dynamic formulation}
For the dynamic formulation, we restrict our consideration to the displacement-based formulation given in \cref{equation:variationDisplacementInvariantFormulation}. 
In so doing, we introduce the time interval of interest with range \(\mathcal{R} = [0,T]\), where \(T\in\R_+\).
The motion of the deformation field \(\vec{\varphi}_t\in\mathds{V}_{\subVarphi}\) denotes the placement of a particle at \(\vec{X}\in\B\) at time \(t\in\mathcal{R}\). 
We introduce the velocity field \(\mathds{V}_{\subv}\ni\vec{v}_t:\B\times\mathcal{R}\rightarrow\R^3\) with space
\begin{equation}
  \mathds{V}_{\subv} = \{\vec{v}:\B\times\mathcal{R}\rightarrow\R^3\,\rvert\,v_i\in \mathds{L}_2(\B)\}\,,
\end{equation}
and the mass density in the reference configuration \(\rho_0:\B\rightarrow\R^{+}\).
With that we are able to introduce the action functional
\begin{equation}
  S(\vec{\varphi}_t,\vec{v}_t) = \int_{t_0}^{t_e}\int_{\B}(\vec{\dot{\varphi}}_t-\tfrac{1}{2}\,\vec{v}_t)\cdot\vec{v}_t\,\rho_0\,\d{V}-({}^{\W}\Pi^{\text{int}}+\Pi^{\text{ext}})\,\d{V}\,\d{t}\,.
\end{equation}
By means of Hamilton's principle, the stationary conditions of the action \(S\) w.r.t.\ the variations yield 
\begin{equation}\label{equation:weak}
  \begin{aligned}
    \int_{\B}\delta\vec{v}_t\cdot(\dot{\vec{\varphi}}_t-\vec{v}_t)\,\rho_0\,\d{V} &= 0\\
    \int_{\B}\delta\vec{\varphi}\cdot\rho_0\,\dot{\vec{v}}_t + \tens{S}_t:\tfrac{1}{2}\,\delta\C\,\d{V} + \Pi^{\text{ext}}(\delta\vec{\varphi}) &= 0\,,
  \end{aligned}
\end{equation}
which have to hold for arbitrary \(\delta\vec{v}\in\mathds{V}_{\subv}^0\) and \(\delta\vec{\varphi}\in\mathds{V}_{\subVarphi}\) with space
\begin{equation}
\mathds{V}_{\subv}^0 = \{\vec{v}:\B\rightarrow\R^3\,\rvert\,v_i\in \mathds{L}_2(\B)\}\,.
\end{equation}
In the above equation, due to \cref{equation:variationDisplacementInvariantFormulation}, the PK2 stress tensor is  defined as
\begin{equation}\label{equation:2ndPKStress}
\tens{S}_t = 2\,\left(\partial_{\I}\W\,\tens{I} + \partial_{\II}\W\,\tens{I}\wedge\CPhit + \tfrac{1}{2}\,\partial_{J}\W\,\JPhit^{-1}\,\GPhit\right)\,.
\end{equation}
The verification of the balance laws for the time-continuous system is detailed in \cref{section:balanceLaws}, which serves as a comprehensive reference for ensuring the validity and consistency of the sytem's fundamental principles. 
\section{Time discretization with consistent energy-momentum scheme}\label{sec:time_disc}
Regarding the time discretization, the use of the midpoint rule in nonlinear continuum mechanics typically introduces a violation of the energy consistency solely due to the discretization of the variation of the (strain) energy within the weak form, while all other terms involved do not give rise to any issue. 
In order to circumvent this issue and to obtain an energy and momentum consistent time integration scheme, a midpoint-type discretization with application of the concept of the discrete gradient, as defined in \cite{gonzalez1996}, is employed. 
Specifically, for the weak form in \cref{equation:weak} the following steps are pursued to ensure energy and momentum consistency:  
\begin{itemize}
\item substitution of time rates \((\dot{\bullet})\) with \(\frac{(\bullet)_{n+1} - (\bullet)_n}{\Delta t} = \frac{\Delta(\bullet)}{\Delta t}\),
\item algorithmic evaluation of the variation of the (strain) energy \(\W\), which leads to an algorithmic  PK2 stress tensor \(\tens{S}_{\text{algo}}\) based on the concept of the discrete gradients in the sense of \cite{gonzalez1996},
\item midpoint evaluation of remaining terms, i.e.\ \(\big(\bullet)_{n+\frac{1}{2}}=\frac{1}{2}\,((\bullet)_n + (\bullet)_{n+1}\big)\).
\end{itemize}
Employing the above steps to the weak form \cref{equation:weak} yields the semi-discrete weak form as 
\begin{equation}\label{equation:semiDiscreteWeak}
  \begin{aligned}
    \int_{\B}\delta\vec{v}\cdot\frac{\Delta\vec{\varphi}}{\Delta t}\,\rho_0\,\d{V} &= \int_{\B}\delta\vec{v}\cdot\vec{v}_{n+\frac{1}{2}}\,\rho_0\,\d{V}\,\\
    \int_{\B}\delta\vec{\varphi}\cdot\rho_0\,\frac{\Delta\vec{v}}{\Delta t}\,\d{V} &= -\int_{\B}\tens{S}_{\text{algo}}:\frac{1}{2}\left(\delta\F{\transp}\,\tens{F}_{{n+\frac{1}{2}}} + \tens{F}_{{n+\frac{1}{2}}}{\transp}\,\delta\F\right)\,\d{V} - \Pi_{n+\frac{1}{2}}^{\text{ext}}(\delta\vec{\varphi})\,.
  \end{aligned}
\end{equation}
Therein, the particular form of the algorithmic version of the PK2 stress tensor \(\tens{S}_{\text{algo}}\) follows from the so-called directionality property, which is a sufficient condition for energy consistency. 
The directionality property itself serves as a time-discrete representation of the continuum stress power \(\dot\W=\tens{S}:\frac{1}{2}\,\dot{\tens{C}}\), acting as a sufficient criterion for energy conservation, cf.~\cite{romero2012}. 
In that regard, the directionality property for the present system yields 
\begin{equation}
\begin{aligned}
\tens{S}_{\text{algo}}:\frac{1}{2}\,\Delta\tens{C} &= \D_{\I}\W\,\Delta\I + \D_{\II}\W\,\Delta\II + \D_{J}\W\,\Delta \JPhi\\
    &=\W(\IPhiNOne,\IIPhiNOne,\JPhiNOne) - \W(\IPhiN,\IIPhiN,\JPhiN)\label{equation:directionalityPropertyShort}\,.
    \end{aligned}
\end{equation}
The satisfaction of this condition motivates for proposing the algorithmic PK2 stress tensor as 
\begin{equation}\label{equation:algorithmic2ndPKStress}
  \tens{S}_{\text{algo}} = 2\,\left(\D_{\I}\W\,\tens{I} + \D_{\II}\W\,\tens{I}\wedge\C_{\text{algo}} + \frac{1}{2}\,\D_{J}\W\,(J_{\text{algo}})^{-1}\G_{\text{algo}}\right)\,,
\end{equation}
which serves as consistent time-discrete variant of \cref{equation:2ndPKStress} and ensures energy consistency while maintaining the second-order accuracy of the midpoint rule. 
In case of midpoint time integration, the only difference compared to the EM scheme is the evaluation of the PK2 stress tensor.
In particular, instead of \cref{equation:algorithmic2ndPKStress} we would evaluate the PK2 stress tensor as follows 
\begin{equation}\label{equation:midpoint2ndPKStress}
\begin{aligned}
  \tens{S}_{{n+\frac{1}{2}}} =\; &2\,(\partial_{\I}\W(\F_{{n+\frac{1}{2}}})\,\tens{I} + \partial_{\II}\W(\F_{{n+\frac{1}{2}}})\,\tens{I}\wedge\C((\F_{{n+\frac{1}{2}}})\\
  &+ \frac{1}{2}\,\partial_{J}\W(\F_{{n+\frac{1}{2}}})\,(J((\F_{{n+\frac{1}{2}}}))^{-1}\G(\F_{{n+\frac{1}{2}}}))\,.
  \end{aligned}
\end{equation}
The algorthmic form of the PK2 stress tensor in \cref{equation:algorithmic2ndPKStress} is the basic feature of the energy-momentum method and contains the subsequently introduced partitioned discrete derivatives \(\D_{\I}\W, \D_{\II}\W,\) \(\D_{J}\W\) in replacement of the partial derivatives of the strain energy from \cref{equation:2ndPKStress}. 
The construction of these particular discrete gradients is provided by \cref{appendix:discreteGradient}, which eventually yields the final Greenspan-like versions 
  \begin{equation}\label{equation:discreteGradientsIIIJ}
  \begin{aligned}
  D_{\I}\W =&\left(\W(\IPhiNOne,\IIPhiN,\JPhiN)-\W(\IPhiN,\IIPhiN,\JPhiN)\right.\\
            &\left.+ \W(\IPhiNOne,\IIPhiNOne,\JPhiNOne)-\W(\IPhiN,\IIPhiNOne,\JPhiNOne)\right)/(2\,\Delta\I)\,,\\
  D_{\II}\W =&\left(\W(\IPhiNOne,\IIPhiNOne,\JPhiN)-\W(\IPhiNOne,\IIPhiN,\JPhiN)\right.\\
            &\left.+ \W(\IPhiN,\IIPhiNOne,\JPhiNOne)-\W(\IPhiN,\IIPhiN,\JPhiNOne)\right)/(2\,\Delta\II)\,,\\
  D_{J}\W =&\left(\W(\IPhiNOne,\IIPhiNOne,\JPhiNOne)-\W(\IPhiNOne,\IIPhiNOne,\JPhiN)\right.\\
            &\left.+ \W(\IPhiN,\IIPhiN,\JPhiNOne)-\W(\IPhiN,\IIPhiN,\JPhiN)\right)/(2\,\Delta J)\,.
    \end{aligned}
  \end{equation}
As shown in \cref{equation:algorithmic2ndPKStress}, algorithmic versions of the kinematic relations \(\C_{\text{algo}}\), \(\G_{\text{algo}}\), \(J_{\text{algo}}\) are employed, which directly follows from the directionality condition \cref{equation:directionalityProperty}. 
Alternatively, the algorithmic PK2 tensor from  \cref{equation:algorithmic2ndPKStress} with its algorithmic kinematic relations can be kind of naturally achieved when deriving it from the corresponding dynamic Hu-Washizu like mixed formulation given based on \cref{equation:CGcStrainEnergy}, see \cite{betsch2018} for more details. 
In any case, the algorithmic kinematic relations \(\C_{\text{algo}}\), \(\G_{\text{algo}}\), \(J_{\text{algo}}\) are chosen as
\begin{equation}\label{equation:algorithmicKinematic}
\begin{aligned}
  \C_{\text{algo}} &= \frac{1}{2}\,(\C_{{n+1}} + \C_{{n}})\\
  \G_{\text{algo}} &= \frac{1}{3}\,(\C_{\text{algo}}\wedge\C_{\text{algo}} + \widetilde{\G}_{\text{algo}}),\quad\text{with }\; \widetilde{\G}_{\text{algo}} = \frac{1}{2}\,(\G_{{n+1}} + \G_{{n}})\,,\\
  J_{\text{algo}} &= \frac{1}{2}\,(J_{{n+1}} + J_{{n}})\,,
  \end{aligned}
\end{equation}
and fulfill the necessary directionality property of \cref{equation:directionalityPropertyShort}. 
It is important to remark that the discrete gradients possess the following properties, see \cite{gonzalez1996} for more information. 
In particular, the discrete gradients 
\begin{itemize}
    \item satisfy the directionality property \cref{equation:directionalityProperty}, which serves as sufficient condition for energy consistency,
    \item are well defined in the limit \(\|\Delta\Pi^i\|\rightarrow 0\) where \(\|\Delta\Pi^i\|=\sqrt{\langle\Delta\Pi^i,\Delta\Pi^i\rangle}\), \(\Delta\Pi^i=\Pi^i_{n+1}-\Pi^i_n\), here \(\tens{\Pi} = (\Pi^1, \Pi^2, \Pi^3) = (\I, \II, J)\), cf.~\cref{appendix:discreteGradient}, and \(\langle\bullet,\bullet\rangle\) denotes the inner product, 
    \item are a second-order approximation of the corresponding midpoint evaluations of the partial derivatives,
    \item and are undefined for handling zero magnitude \(\|\Delta\Pi^i\| = 0\), which regarding the implementation occurs e.g.\ within a very first iteration of Newton's method, and can be avoided by midpoint-type evaluation of the partial derivatives \(\D_{\Pi^i}\W=\partial_{\Pi^i}\W(\Pi^i_{\nmid})\),
    \item results with the aforementioned algorithmic stress evaluation in the generation of a tangent stiffness matrix that exhibits non-symmetry. 
\end{itemize}
\begin{remark}
  As becomes obvious in \cref{appendix:CGcDiscreteGradient}, the choice of the strain energy \(\WCt(\C,\G,C)\) in \cref{equation:CGcStrainEnergy} leads to far more complex versions of the discrete gradients \(\D_{\subC}\WCt\), \(\D_{\subG}\WCt\) and \(\D_{C}\WCt\) when compared to the choice of strain energy \(\W(\I,\II,J)\) in \cref{equation:invariantStrainEnergy} with just Greenspan versions of the partitioned discrete gradients \(\D_{\I}\W\), \(\D_{\II}\W\) and \(\D_{J}\W\) in \cref{equation:discreteGradientsIIIJ}.
  Also the traditional, projection-based discrete gradient \(\D_{\subC}\WC(\C)\) associated with strain energy \(\WC(\C)\) from \cref{equation:CGcStrainEnergy} provided by \cref{appendix:CDiscreteGradient} can be employed. 
  However, the consistent linearization of the tensor-valued discrete gradient \(\D_{\subC}\WC(\C)\) and the application of the chain rule become more complex when employing the PANN constitutive model with its inputs \(\vec{\mathcal{I}}\) in comparison to the proposed version outlined in \cref{equation:discreteGradientsIIIJ}. 
\end{remark}
Standard integrators like the 2nd order accurate midpoint rule usually violate the conserving properties, particularly the balance of angular momentum and total energy. 
However, the proposed EM scheme ensures that these balance laws are preserved in the time-discrete system. 
The verification and detailed analysis of the balance laws for the time-discrete system can be found in \cref{section:semiDiscreteBalanceLaws}. 
\section{Spatial discretization with the finite element method}\label{sec:spat_disc}
In the following, we focus on the semi-discrete displacement-based system from \cref{sec:time_disc}, which  makes a variety of spatial discretizations possible.
Here, we employ the finite element (FE) method for the spatial discretization of \cref{equation:semiDiscreteWeak}.
To this end, the body \(\B\) is subdivided into \(n_e\in\mathbb{N}\) finite elements
\begin{equation}
  \B\approx\B^{\h}=\bigcup_{e=1}^{n_e}\B^e\,,
\end{equation}
where \(e=1,...,n_e\) denotes the respective element.
In particular, we use the isoparametric concept, cf.\ \cite{hughes2012}, by using the same shape functions for the approximation of the geometry with
\begin{equation}
  \vec{X}^{\h}\big|_{\B^e} = \sum_{a=1}^{n_{\text{node}}}N^a\,\vec{X}^a\,,
\end{equation}
and the deformation field \(\vec{\varphi}_t^h\in\mathds{V}_{\subVarphi}^{\h}\subset\mathds{V}_{\subVarphi}\) and the velocity field \(\vec{v}_t^h\in\mathds{V}_{\subv}^{\h}\subset\mathds{V}_{\subv}\) with spaces 
\begin{equation}
  \spaceV_{\subVarphi}^{\h}=\left\{\vec{\varphi}_t\in\spaceV_{\subVarphi}\,:\,\vec{\varphi}^{\h}_t\big|_{\B^e}=\sum_{a=1}^{n_\textrm{node}}N^a\vec{\varphi}^a_t\right\},\quad\spaceV_{\subv}^{\h}= \left\{\vec{v}_t\in\spaceV_{\subv}\,:\,\vec{v}^{\h}_t\big|_{\B^e}=\sum_{a=1}^{n_\textrm{node}}N^a\vec{v}^a_t\right\}\,,
\end{equation}
where \(N^a:\B\rightarrow\R\) denote nodal shape functions with corresponding nodal values \(\vec{X}^a\in\R^3\), \(\vec{\varphi}_a:\mathcal{R}\rightarrow\R^3\), and \(\vec{v}_a:\mathcal{R}\rightarrow\R^3\). 
According to a Bubnov-Galerkin FE ansatz, we employ the same Ansatz functions for solution and variations \(\delta\vec{\varphi}\in\mathds{V}_{\subVarphi}^{0,\h}\subset\mathds{V}_{\subVarphi}^0\), and \(\delta\vec{v}\in\mathds{V}_{\subv}^{0,\h}\subset\mathds{V}_{\subv}^0\) with spaces 
\begin{equation}
  \spaceV_{\subVarphi}^{0,\h}=\left\{\delta\vec{\varphi}\in\spaceV_{\subVarphi}^0\,:\,\delta\vec{\varphi}^{\h}\big|_{\B^e}=\sum_{a=1}^{n_\textrm{node}}N^a\,\delta\vec{\varphi}^a\right\},\quad\spaceV_{\subv}^{0,\h}= \left\{\delta\vec{v}\in\spaceV_{\subv}^0\,:\,\delta\vec{v}^{\h}\big|_{\B^e}=\sum_{a=1}^{n_\textrm{node}}N^a\delta\vec{v}^a\right\}\,.
\end{equation}
We insert the above discretized solution and variations into the semi-discrete weak form \cref{equation:semiDiscreteWeak} and obtain
\begin{equation}
  \begin{aligned}
    &\assem_{e=1}^{n_e}{\sum_{a=1}^{n_\textrm{node}}\delta\vec{v}^a\cdot\vec{R}_{\subv}^{a,e}=0}\,,\\
    &\assem_{e=1}^{n_e}{\sum_{a=1}^{n_\textrm{node}}\delta\vec{\varphi}^a\cdot\vec{R}_{\subVarphi}^{a,e}=0}\,,
  \end{aligned}
\end{equation}
where $\assemtext_{e=1}^{n_e}$ denotes the assembly operator. 
In the above, the nodal residuals \(\vec{R}_{\subv}^{a,e}\in\R^3\) and \(\vec{R}_{\subVarphi}^{a,e}\in\R^3\) are defined as
\begin{equation}\label{equation:fullDiscreteWeakForm}
\begin{aligned}
\vec{R}_{\subv}^{a,e}& =\vec{M}^{ab,e}\,\left(\frac{1}{\Delta t}\Delta\vec{\varphi}^{b}-\vec{v}^{b}_{\nmid}\right)\,,\\
\vec{R}_{\subVarphi}^{a,e}&=\vec{M}^{ab,e}\,\frac{1}{\Delta t}\Delta\vec{v}^{b}+\int_{\B^e}\tens{B}{a\transp}(\vec{\varphi}^h_{\nmid})\vec{S}^{\h,\voigt}_{\text{algo}}\d{V}-\int_{\B^e}N^a\bar{\vec{B}}^{\h}(\vec{\varphi}^h_{\nmid})\d{V}\,,
\end{aligned}
\end{equation}
where \(\vec{B}^a\in\R^{6\times 3}\) denotes the standard nodal operator matrix, \((\bullet)^{\text{V}}\) symbolizes Voigt's vector notation and \(\vec{M}^{ab,e}\) denotes the elemental mass matrix defined as
\begin{equation}
\vec{M}^{ab,e} = \int_{\B^e}\rho_0\,N^a\,N^b\,\d{V}\,\vec{I}\,.
\end{equation}
In order to pursue an efficient implementation, cf.~\cite{betsch2018},  \cref{equation:fullDiscreteWeakForm}\(_1\) can be rewritten as
\begin{equation}
  \vec{v}^{b}_{n+1} = \frac{2}{\Delta t}\Delta\vec{\varphi}^{b}_{n+1}-\vec{v}^{b}_{n}\,,
\end{equation}
and inserting it into \cref{equation:fullDiscreteWeakForm}\(_2\) yields the nodal residual vector \(\widehat{\vec{R}}_{\subVarphi}^{a,e}\in\R^3\) given by
\begin{equation}
  \widehat{\vec{R}}_{\subVarphi}^{a,e}=\frac{2}{\Delta t}\,\left(\frac{1}{\Delta t}\,\vec{M}^{ab,e}\,\Delta\vec{\varphi}^{b}-\vec{M}^{ab,e}\,\vec{v}^{b}_{n}\right)+\tens{B}^{a\transp}_{\nmid}\vec{S}^{\h,\voigt}_{\text{algo}}\d{V}-\int_{\B^e}N^a\bar{\vec{B}}^{\h}(\vec{\varphi}^h_{\nmid})\d{V}\,.
\end{equation}
Accordingly, the velocity field is condensed out and we have a pure displacement-based formulation. 
Note that, for the midpoint rule, the algorithmic PK2 stress tensor \(\vec{S}^{\h}_{\text{algo}}\) is simply replaced with the midpoint evaluated PK2 stress tensor from \cref{equation:midpoint2ndPKStress}. 

It is important to remark that the discretization of the mixed approaches in \cref{section:mixedFormulations} is straight forward, where of course the additional fields have to be discretized with suitable spaces considering the inf-sup conditions for the particular choice of approximation formulas, see \cite{kraus2019} for more information.  

\section{Numerical examples}\label{sec:num}
In this section, stability and robustness of the proposed framework are examined in representative finite element simulations. In \cref{section:preparationNN}, the PANN constitutive model is calibrated to data generated with an analytical potential.
Afterwards in \cref{section:cooksMembrane} and in \cref{section:LShapedBody}, the PANN constitutive model is applied in both static and dynamic numerical examples. 
\subsection{Preparation of the neural network-based constitutive model}\label{section:preparationNN}
The PANN constitutive model is now calibrated to data generated with an analytical potential. Details on the data generation are presented in \cref{section:dataGeneration}, while details on the model calibration are presented in \cref{section:modelCalibration}.
\subsubsection{Data generation}\label{section:dataGeneration}
For the following investigations, the PANN model introduced in \cref{sec:PANN} is calibrated to data generated with the analytical Mooney-Rivlin model, cf.~\cref{equation:MR}. For this, two sets of material parameters are considered, which lead to a compressible and a nearly incompressible model, respectively, cf.~\cref{tab:params}. If not stated otherwise, the parameter set for a compressible material behavior is applied.
In the following, the Mooney-Rivlin model is referred to as the ground truth (GT) model, to which the results obtained with the calibrated PANN model are compared. 

Following \cite{Fernandez2020}, the calibration dataset consists of (i) a uniaxial tension stress state, (ii) an equibiaxial tension stress state, and (iii) a simple shear deformation state, which are applied on the analytical Mooney-Rivlin model. The uniaxial tension is applied in $x$-direction with $F_{11}\in[0.75,\,1.75]$, the equibiaxial tension is applied in $x-y$-direction with $F_{11}=F_{22}\in[0.75,\,1.755]$, and simple shear is applied with $F_{12}\in[-0.25,\,0.75]$.
In addition, for the test dataset, a mixed shear-tension deformation is applied (\enquote{test 3} in \cite{Fernandez2020}), which represents a fairly general deformation mode, cf.~\cref{figure:test}. Each load case consists of 100 datapoints, which makes 300 strain-stress tuples in the calibration dataset. 
It should be mentioned that besides these experimentally motivated load cases, it is also possible to sample the deformation states either by analytical considerations on physically admissible deformation states \cite{kunc2019} or by investigating the deformation states which occur in the engineering component under investigation \cite{kalina2023}. 
\begin{table}[h]
 \centering
 \begin{tabular}{ p{7cm}|p{1cm}p{1.5cm}p{1.2cm}  }
compressible case& $a$ & $831.25$ & Pa\\
 & $b$ & $166.25$ & Pa\\
 & $c$ & $10000$ & Pa\\
 & $d$ & $2327.5$ & Pa\\
 mass density (static/transient examples) & $\rho_0$ & $0/100$ & $\mathrm{kg\,m}^{-3}$\\
 \\
 \hline
 \\
nearly incompressible case& $a$ & $126$ & Pa\\
 & $b$ & $252$ & Pa\\
 & $c$ & $81512$ & Pa\\
 & $d$ & $1260$ & Pa\\
 mass density & $\rho_0$ & $0$ & $\mathrm{kg\,m}^{-3}$
 \end{tabular}
 \caption{Mooney-Rivlin material parameters employed for the numerical examples.}
 \label{tab:params}
\end{table}
\subsubsection{Model calibration}\label{section:modelCalibration}
For the PANN model calibration, six different network architectures with $n\in\{4,\,8,\,16,\,32,\,64,\,128\}$ nodes in the hidden layer are considered for the compressible material behavior, which corresponds to $|\vec{\mathcal{P}}|=24,\,48,\,96,\,192,\,384,\,768$ NN model parameters.
In this way, the computational performance of the algorithms can be examined, and furthermore, it is demonstrated that the choice of nodes $n$ in the hidden layer is generic for the approach introduced in this work. In general, this also holds for the number of hidden layers, of which the proposed framework is also independendent, with the caveat that the explicit derivatives of the NN, cf.~\cref{sec:derivs}, must be adapted when more than one hidden layer is considered. 
For the nearly incompressible material behavior, $n=8$ nodes are considered.
At this point, it should be noted that even the smallest NN architecture with 4 nodes is flexible enough to perfectly represent the Mooney-Rivlin model, and, furthermore, calibrating a PANN model to such a simple analytical potential should be considered as an academic example. In the following examples, PANN$n$ denotes a PANN model with $n$ nodes in the hidden layer.

For the calibration, the models are implemented in TensorFlow 2.5.0 using Python 3.9.9. 
The optimization is carried out using the ADAM algorithm with default settings. 
Therefore, the full batch of calibration data is used for $5000$ calibration epochs. 
Each model architecture is calibrated one time. 
For the reproducibility of the results, the calibrated model parameters are provided in the GitHub repository \url{https://github.com/CPShub/sim-data}.

\begin{figure}[h]
\centering
  \begin{tabular}{cc}
     \setlength{\figH}{0.2\textheight}
   \setlength{\figW}{0.3\textwidth}
%
%

\pgfplotstableset{
create on use/C_1ax/.style={create col/copy column from table={figures/calibration/C_test3.txt}{0}}
}

\definecolor{mycolor1}{rgb}{0.00000,0.44700,0.74100}%
\definecolor{mycolor2}{rgb}{0.85000,0.32500,0.09800}%
\begin{tikzpicture}

\begin{axis}[%
width=0.951\figW,
height=\figH,
at={(0\figW,0\figH)},
scale only axis,
scaled ticks=false,
ytick = {-3,-2,-1,0,1},
yticklabels={-30,-20,-10,0,10},
xlabel style={font=\color{white!15!black}},
xlabel={$C_{11}$},
ymin=-3,
ymax=1.5,
grid=major,
ylabel style={font=\color{white!15!black}},
ylabel={$S_{ij}$},
axis background/.style={fill=white},
legend style={legend cell align=left,align=left, draw=white!15!black},
legend pos = south east
]

\addplot[mark repeat = 10, only marks, mark=*, color=CPSdarkblue,line width=1.5pt] table [x=C_1ax, y index=0] {figures/calibration/S_test3.txt}; 
\addlegendentry{$S_{11}$}

\addplot[mark repeat = 10, only marks, mark=square*, color=CPSred,  line width=1.5pt] table [x=C_1ax, y index=1] {figures/calibration/S_test3.txt}; 
\addlegendentry{$S_{12}$}

\addplot[mark repeat = 10, only marks, mark=triangle*, color=CPSorange,  line width=1.5pt] table [x=C_1ax, y index=8] {figures/calibration/S_test3.txt}; 
\addlegendentry{$S_{33}$}

\addplot[mark =none, color=CPSdarkblue,line width=1.5pt] table [x=C_1ax, y index=0] {figures/calibration/S_test3_m.txt};

\addplot[mark  = none, color=CPSred,  line width=1.5pt] table [x=C_1ax, y index=1] {figures/calibration/S_test3_m.txt};

\addplot[mark  = none, color=CPSorange,  line width=1.5pt] table [x=C_1ax, y index=8] {figures/calibration/S_test3_m.txt};

\end{axis}
\end{tikzpicture}%
   &
        \setlength{\figH}{0.2\textheight}
   \setlength{\figW}{0.3\textwidth}
%
%

\pgfplotstableset{
create on use/C_1ax/.style={create col/copy column from table={figures/calibration/C_test3_inc.txt}{0}}
}

\definecolor{mycolor1}{rgb}{0.00000,0.44700,0.74100}%
\definecolor{mycolor2}{rgb}{0.85000,0.32500,0.09800}%
\begin{tikzpicture}

\begin{axis}[%
width=0.951\figW,
height=\figH,
at={(0\figW,0\figH)},
scale only axis,
scaled ticks=false,
ytick = {-1.5,-1,-0.5,0,0.5},
yticklabels={-15,-10,-5,0,5},
xlabel style={font=\color{white!15!black}},
xlabel={$C_{11}$},
ymin=-1.5,
ymax=0.75,
grid=major,
ylabel style={font=\color{white!15!black}},
ylabel={$S_{ij}$},
axis background/.style={fill=white},
legend style={legend cell align=left,align=left, draw=white!15!black},
legend pos = south east
]

\addplot[mark repeat = 10, only marks, mark=*, color=CPSdarkblue,line width=1.5pt] table [x=C_1ax, y index=0] {figures/calibration/S_test3_inc.txt}; 
\addlegendentry{$S_{11}$}

\addplot[mark repeat = 10, only marks, mark=square*, color=CPSred,  line width=1.5pt] table [x=C_1ax, y index=1] {figures/calibration/S_test3_inc.txt}; 
\addlegendentry{$S_{12}$}

\addplot[mark repeat = 10, only marks, mark=triangle*, color=CPSorange,  line width=1.5pt] table [x=C_1ax, y index=8] {figures/calibration/S_test3_inc.txt}; 
\addlegendentry{$S_{33}$}

\addplot[mark =none, color=CPSdarkblue,line width=1.5pt] table [x=C_1ax, y index=0] {figures/calibration/S_test3_m_inc.txt};

\addplot[mark  = none, color=CPSred,  line width=1.5pt] table [x=C_1ax, y index=1] {figures/calibration/S_test3_m_inc.txt};

\addplot[mark  = none, color=CPSorange,  line width=1.5pt] table [x=C_1ax, y index=8] {figures/calibration/S_test3_m_inc.txt};

\end{axis}
\end{tikzpicture}%
   \end{tabular}
\caption{Mixed shear-tension test evaluated for the GT models (marks) and the calibrated PANN8 models (continous lines) for both the compressible (left) and the nearly incompressible (right) case. Stress in MPa.}
  \label{figure:test}
\end{figure}
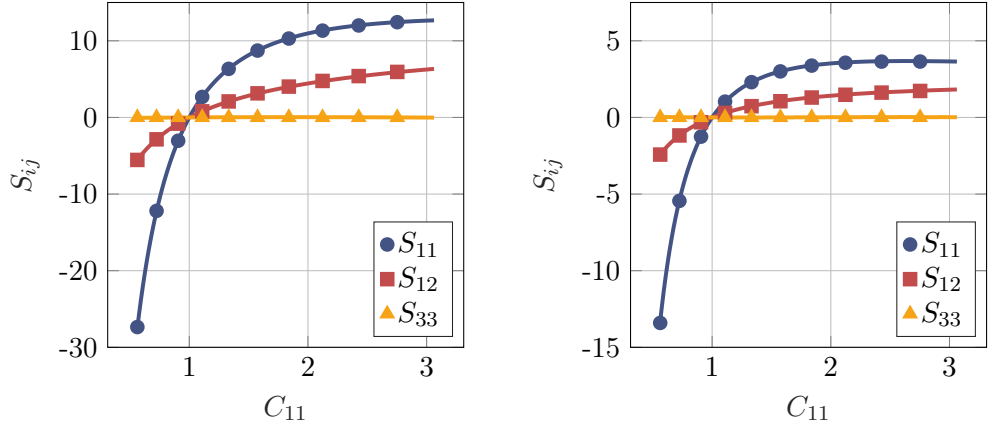

All model architectures perform excellent, both on the calibration and the test dataset, and both for the compressible and the nearly incompressible material behavior, cf.~\cref{tab:loss}, where the losses of the calibrated models are provided. While there is a general trend for a decreasing loss for larger model architectures, it should be noted that even the small architectures are able to nearly perfectly represent the GT models, which becomes evident in \cref{figure:test}, where the test case is evaluated for the PANN8 models.

\begin{table}[h]
 \centering
 \begin{tabular}
 { l r |  r r | r r }
& & \multicolumn{2}{l|}{compressible model} & \multicolumn{2}{l}{nearly incompressible model} \\
&& calibration & test & \hspace{3em}calibration & test \\
\hline
PANN& 4 & 1.70 & 0.89 &  &  \\
&8 & 0.84 & 0.61 & 1.05 & 0.23 \\
&16 & 0.93 & 0.61 &  &  \\
&32& 0.25 & 0.15 &  &  \\
&64 & -0.23 & 0.17 &  &  \\
&128 & -0.11 & 0.10 &  &  \\
 \end{tabular}
 \caption{log$_{10}(\mathcal{MSE})$ according to \cref{eq:MSE} for the calibrated PANN models.}
 \label{tab:loss}
\end{table}
\subsection{Static deformation of Cook's membrane}\label{section:cooksMembrane}
Here, a static example is investigated. In \cref{section:cooksCompressible}, compressible material behavior is assumed, and the computation time of the PANN constitutive model compared to the GT model is examined. 
In \cref{section:cooksNearlyIncompressible}, nearly incompressible material behavior is assumed, and the performance of the mixed formulations is demonstrated.
\subsubsection{Compressible case}\label{section:cooksCompressible}
As a first example, a three-dimensional version of the well-known Cook's membrane with compressible material behavior is considered, see e.g. \cite{betsch2018}. 
With this example, the spatial performance of the PANN constitutive model compared to the GT model in a purely static, large deformation solid mechanical scenario is examined. 
Furthermore, a detailed investigation of the computation times for both PANN and GT models is carried out.

\begin{figure}[t]
\centering
  \begin{tabular}{ccc}
    \psfrag{16}[][]{\(16\)}
    \psfrag{44}[][]{\(44\)}
    \psfrag{48}[][]{\(48\)}
    \psfrag{m}[][]{\([m]\)}
    \psfrag{t}[][]{\(t=4\)}
    \psfrag{ex}[][]{\(\vec{e}_1\)}
    \psfrag{ey}[][]{\(\vec{e}_2\)}
    \psfrag{ez}[][]{\(\vec{e}_3\)}
    \includegraphics[width=0.3\textwidth]{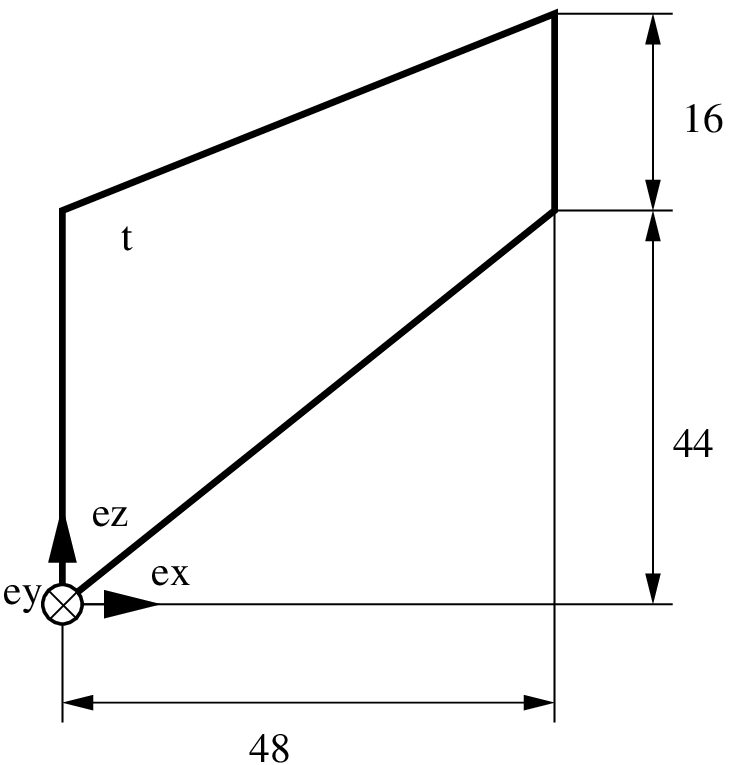}
    &
    \hspace*{1cm}
    &
    \newlength{\svgwidth}
    \setlength{\svgwidth}{0.3\textwidth}
\begingroup%
  \makeatletter%
  \providecommand\color[2][]{%
    \errmessage{(Inkscape) Color is used for the text in Inkscape, but the package 'color.sty' is not loaded}%
    \renewcommand\color[2][]{}%
  }%
  \providecommand\transparent[1]{%
    \errmessage{(Inkscape) Transparency is used (non-zero) for the text in Inkscape, but the package 'transparent.sty' is not loaded}%
    \renewcommand\transparent[1]{}%
  }%
  \providecommand\rotatebox[2]{#2}%
  \newcommand*\fsize{\dimexpr\f@size pt\relax}%
  \newcommand*\lineheight[1]{\fontsize{\fsize}{#1\fsize}\selectfont}%
  \ifx\svgwidth\undefined%
    \setlength{\unitlength}{314.70533315bp}%
    \ifx\svgscale\undefined%
      \relax%
    \else%
      \setlength{\unitlength}{\unitlength * \real{\svgscale}}%
    \fi%
  \else%
    \setlength{\unitlength}{\svgwidth}%
  \fi%
  \global\let\svgwidth\undefined%
  \global\let\svgscale\undefined%
  \makeatother%
  \begin{picture}(1,1.23680815)%
    \lineheight{1}%
    \setlength\tabcolsep{0pt}%
    \put(0,0){\includegraphics[width=\unitlength]{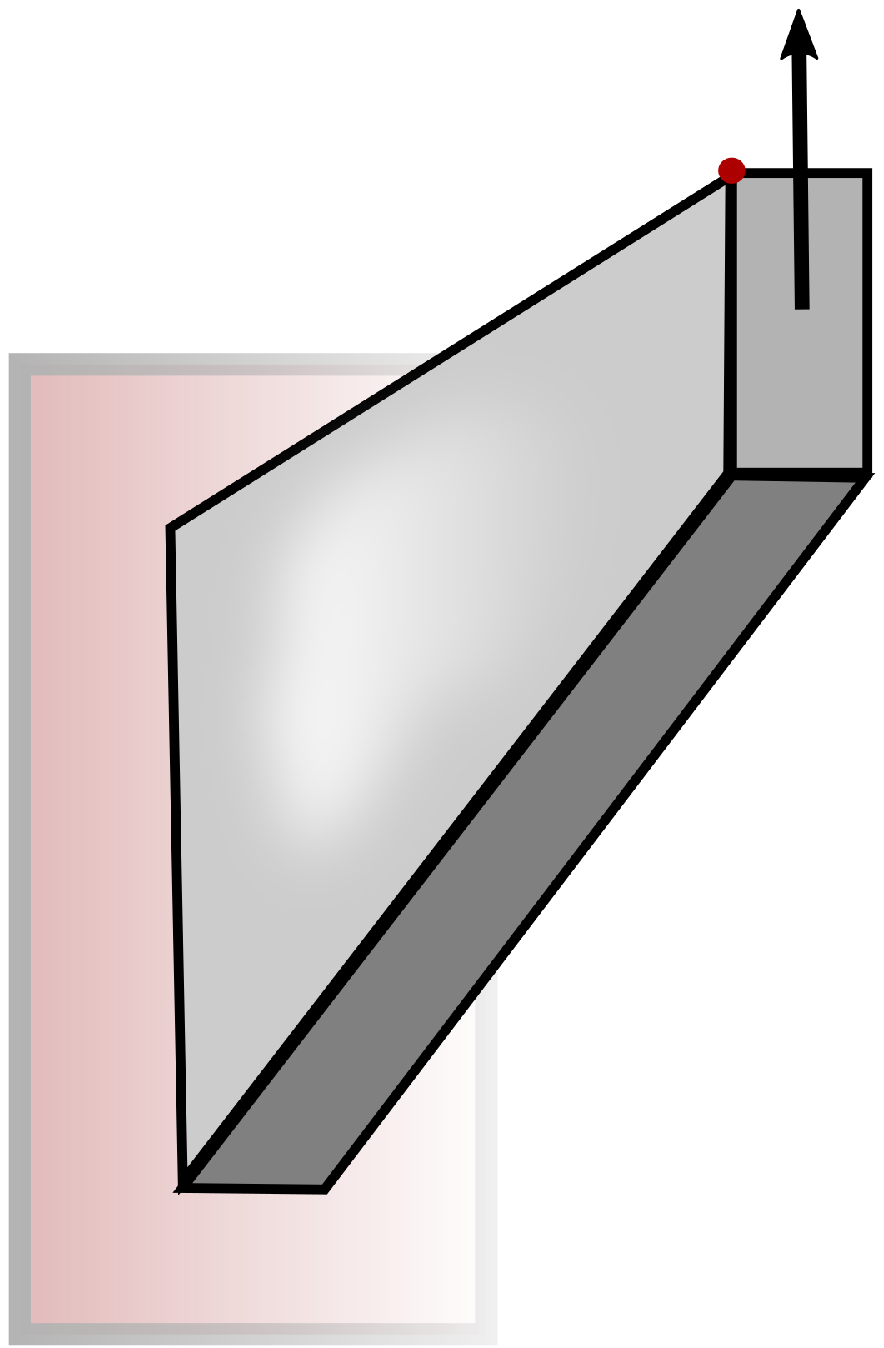}}%
    \put(0.9466169,1.18475962){\color[rgb]{0,0,0}\makebox(0,0)[lt]{\lineheight{1.25}\smash{\begin{tabular}[t]{l}\(p\)\end{tabular}}}}%
    \put(0.69772589,1.09840025){\color[rgb]{0,0,0}\makebox(0,0)[lt]{\lineheight{1.25}\smash{\begin{tabular}[t]{l}\(A\)\end{tabular}}}}%
    \put(-0.00753083,0.93225502){\color[rgb]{0,0,0}\makebox(0,0)[lt]{\lineheight{1.25}\smash{\begin{tabular}[t]{l}\(\vec{u}(X_1=0)=\vec{0}\)\end{tabular}}}}%
  \end{picture}%
\endgroup%

  \end{tabular}
  \caption{2D projection of the geometry (left) and 3D illustration of the  boundary conditions (right) of the Cook's membrane.}
  \label{figure:CookBoundaryConditions}
\end{figure}

The geometry of the membrane is provided by \cref{figure:CookBoundaryConditions} (left). 
The membrane is subject to a clamping on its left side while a pressure load is applied to shear the membrane on the right side, as illustrated in \cref{figure:CookBoundaryConditions} (right). 
The resultant force of the pressure load has the magnitude of \(p=200\) Pa and shows into the positive \(\vec{e}_3\)-direction. 
For the discretization, displacement-based trilinear hexahedron elements based on to \cref{equation:invariantStrainEnergy} are employed. 

\begin{figure}[t]
\centering
  \begin{tabular}{cccl}
\includegraphics[width=0.15\textwidth]{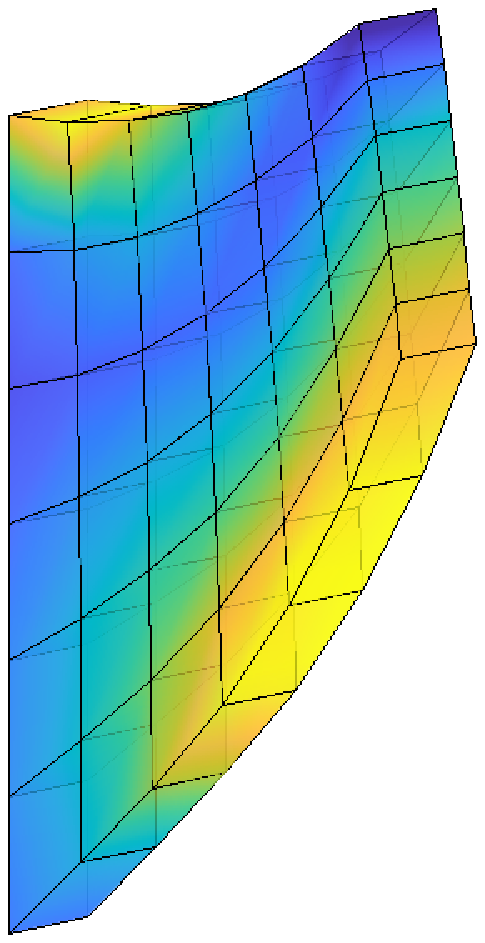}
&
\hspace*{3cm}
&
\includegraphics[width=0.15\textwidth]{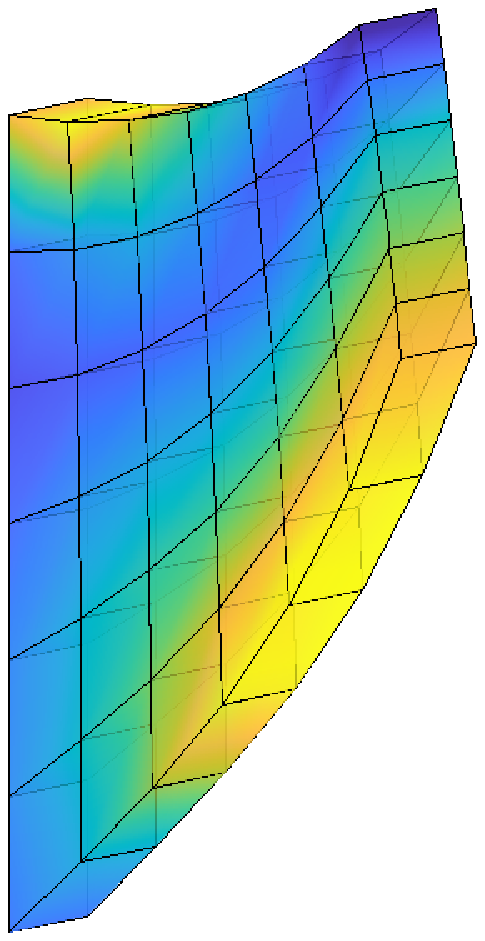}
&
   \setlength{\figH}{0.2\textheight}
   \setlength{\figW}{0.2\textwidth}
%
%
\begin{tikzpicture}

\begin{axis}[%
width=0.891\figW,
height=\figH,
at={(0\figW,0\figH)},
scale only axis,
point meta min=50,
point meta max=550,
xmin=0,
xmax=1,
ymin=0,
ymax=1,
axis line style={draw=none},
ticks=none,
axis x line*=bottom,
axis y line*=left,
colormap={mymap}{[1pt] rgb(0pt)=(0.2422,0.1504,0.6603); rgb(1pt)=(0.2444,0.1534,0.6728); rgb(2pt)=(0.2464,0.1569,0.6847); rgb(3pt)=(0.2484,0.1607,0.6961); rgb(4pt)=(0.2503,0.1648,0.7071); rgb(5pt)=(0.2522,0.1689,0.7179); rgb(6pt)=(0.254,0.1732,0.7286); rgb(7pt)=(0.2558,0.1773,0.7393); rgb(8pt)=(0.2576,0.1814,0.7501); rgb(9pt)=(0.2594,0.1854,0.761); rgb(11pt)=(0.2628,0.1932,0.7828); rgb(12pt)=(0.2645,0.1972,0.7937); rgb(13pt)=(0.2661,0.2011,0.8043); rgb(14pt)=(0.2676,0.2052,0.8148); rgb(15pt)=(0.2691,0.2094,0.8249); rgb(16pt)=(0.2704,0.2138,0.8346); rgb(17pt)=(0.2717,0.2184,0.8439); rgb(18pt)=(0.2729,0.2231,0.8528); rgb(19pt)=(0.274,0.228,0.8612); rgb(20pt)=(0.2749,0.233,0.8692); rgb(21pt)=(0.2758,0.2382,0.8767); rgb(22pt)=(0.2766,0.2435,0.884); rgb(23pt)=(0.2774,0.2489,0.8908); rgb(24pt)=(0.2781,0.2543,0.8973); rgb(25pt)=(0.2788,0.2598,0.9035); rgb(26pt)=(0.2794,0.2653,0.9094); rgb(27pt)=(0.2798,0.2708,0.915); rgb(28pt)=(0.2802,0.2764,0.9204); rgb(29pt)=(0.2806,0.2819,0.9255); rgb(30pt)=(0.2809,0.2875,0.9305); rgb(31pt)=(0.2811,0.293,0.9352); rgb(32pt)=(0.2813,0.2985,0.9397); rgb(33pt)=(0.2814,0.304,0.9441); rgb(34pt)=(0.2814,0.3095,0.9483); rgb(35pt)=(0.2813,0.315,0.9524); rgb(36pt)=(0.2811,0.3204,0.9563); rgb(37pt)=(0.2809,0.3259,0.96); rgb(38pt)=(0.2807,0.3313,0.9636); rgb(39pt)=(0.2803,0.3367,0.967); rgb(40pt)=(0.2798,0.3421,0.9702); rgb(41pt)=(0.2791,0.3475,0.9733); rgb(42pt)=(0.2784,0.3529,0.9763); rgb(43pt)=(0.2776,0.3583,0.9791); rgb(44pt)=(0.2766,0.3638,0.9817); rgb(45pt)=(0.2754,0.3693,0.984); rgb(46pt)=(0.2741,0.3748,0.9862); rgb(47pt)=(0.2726,0.3804,0.9881); rgb(48pt)=(0.271,0.386,0.9898); rgb(49pt)=(0.2691,0.3916,0.9912); rgb(50pt)=(0.267,0.3973,0.9924); rgb(51pt)=(0.2647,0.403,0.9935); rgb(52pt)=(0.2621,0.4088,0.9946); rgb(53pt)=(0.2591,0.4145,0.9955); rgb(54pt)=(0.2556,0.4203,0.9965); rgb(55pt)=(0.2517,0.4261,0.9974); rgb(56pt)=(0.2473,0.4319,0.9983); rgb(57pt)=(0.2424,0.4378,0.9991); rgb(58pt)=(0.2369,0.4437,0.9996); rgb(59pt)=(0.2311,0.4497,0.9995); rgb(60pt)=(0.225,0.4559,0.9985); rgb(61pt)=(0.2189,0.462,0.9968); rgb(62pt)=(0.2128,0.4682,0.9948); rgb(63pt)=(0.2066,0.4743,0.9926); rgb(64pt)=(0.2006,0.4803,0.9906); rgb(65pt)=(0.195,0.4861,0.9887); rgb(66pt)=(0.1903,0.4919,0.9867); rgb(67pt)=(0.1869,0.4975,0.9844); rgb(68pt)=(0.1847,0.503,0.9819); rgb(69pt)=(0.1831,0.5084,0.9793); rgb(70pt)=(0.1818,0.5138,0.9766); rgb(71pt)=(0.1806,0.5191,0.9738); rgb(72pt)=(0.1795,0.5244,0.9709); rgb(73pt)=(0.1785,0.5296,0.9677); rgb(74pt)=(0.1778,0.5349,0.9641); rgb(75pt)=(0.1773,0.5401,0.9602); rgb(76pt)=(0.1768,0.5452,0.956); rgb(77pt)=(0.1764,0.5504,0.9516); rgb(78pt)=(0.1755,0.5554,0.9473); rgb(79pt)=(0.174,0.5605,0.9432); rgb(80pt)=(0.1716,0.5655,0.9393); rgb(81pt)=(0.1686,0.5705,0.9357); rgb(82pt)=(0.1649,0.5755,0.9323); rgb(83pt)=(0.161,0.5805,0.9289); rgb(84pt)=(0.1573,0.5854,0.9254); rgb(85pt)=(0.154,0.5902,0.9218); rgb(86pt)=(0.1513,0.595,0.9182); rgb(87pt)=(0.1492,0.5997,0.9147); rgb(88pt)=(0.1475,0.6043,0.9113); rgb(89pt)=(0.1461,0.6089,0.908); rgb(90pt)=(0.1446,0.6135,0.905); rgb(91pt)=(0.1429,0.618,0.9022); rgb(92pt)=(0.1408,0.6226,0.8998); rgb(93pt)=(0.1383,0.6272,0.8975); rgb(94pt)=(0.1354,0.6317,0.8953); rgb(95pt)=(0.1321,0.6363,0.8932); rgb(96pt)=(0.1288,0.6408,0.891); rgb(97pt)=(0.1253,0.6453,0.8887); rgb(98pt)=(0.1219,0.6497,0.8862); rgb(99pt)=(0.1185,0.6541,0.8834); rgb(100pt)=(0.1152,0.6584,0.8804); rgb(101pt)=(0.1119,0.6627,0.877); rgb(102pt)=(0.1085,0.6669,0.8734); rgb(103pt)=(0.1048,0.671,0.8695); rgb(104pt)=(0.1009,0.675,0.8653); rgb(105pt)=(0.0964,0.6789,0.8609); rgb(106pt)=(0.0914,0.6828,0.8562); rgb(107pt)=(0.0855,0.6865,0.8513); rgb(108pt)=(0.0789,0.6902,0.8462); rgb(109pt)=(0.0713,0.6938,0.8409); rgb(110pt)=(0.0628,0.6972,0.8355); rgb(111pt)=(0.0535,0.7006,0.8299); rgb(112pt)=(0.0433,0.7039,0.8242); rgb(113pt)=(0.0328,0.7071,0.8183); rgb(114pt)=(0.0234,0.7103,0.8124); rgb(115pt)=(0.0155,0.7133,0.8064); rgb(116pt)=(0.0091,0.7163,0.8003); rgb(117pt)=(0.0046,0.7192,0.7941); rgb(118pt)=(0.0019,0.722,0.7878); rgb(119pt)=(0.0009,0.7248,0.7815); rgb(120pt)=(0.0018,0.7275,0.7752); rgb(121pt)=(0.0046,0.7301,0.7688); rgb(122pt)=(0.0094,0.7327,0.7623); rgb(123pt)=(0.0162,0.7352,0.7558); rgb(124pt)=(0.0253,0.7376,0.7492); rgb(125pt)=(0.0369,0.74,0.7426); rgb(126pt)=(0.0504,0.7423,0.7359); rgb(127pt)=(0.0638,0.7446,0.7292); rgb(128pt)=(0.077,0.7468,0.7224); rgb(129pt)=(0.0899,0.7489,0.7156); rgb(130pt)=(0.1023,0.751,0.7088); rgb(131pt)=(0.1141,0.7531,0.7019); rgb(132pt)=(0.1252,0.7552,0.695); rgb(133pt)=(0.1354,0.7572,0.6881); rgb(134pt)=(0.1448,0.7593,0.6812); rgb(135pt)=(0.1532,0.7614,0.6741); rgb(136pt)=(0.1609,0.7635,0.6671); rgb(137pt)=(0.1678,0.7656,0.6599); rgb(138pt)=(0.1741,0.7678,0.6527); rgb(139pt)=(0.1799,0.7699,0.6454); rgb(140pt)=(0.1853,0.7721,0.6379); rgb(141pt)=(0.1905,0.7743,0.6303); rgb(142pt)=(0.1954,0.7765,0.6225); rgb(143pt)=(0.2003,0.7787,0.6146); rgb(144pt)=(0.2061,0.7808,0.6065); rgb(145pt)=(0.2118,0.7828,0.5983); rgb(146pt)=(0.2178,0.7849,0.5899); rgb(147pt)=(0.2244,0.7869,0.5813); rgb(148pt)=(0.2318,0.7887,0.5725); rgb(149pt)=(0.2401,0.7905,0.5636); rgb(150pt)=(0.2491,0.7922,0.5546); rgb(151pt)=(0.2589,0.7937,0.5454); rgb(152pt)=(0.2695,0.7951,0.536); rgb(153pt)=(0.2809,0.7964,0.5266); rgb(154pt)=(0.2929,0.7975,0.517); rgb(155pt)=(0.3052,0.7985,0.5074); rgb(156pt)=(0.3176,0.7994,0.4975); rgb(157pt)=(0.3301,0.8002,0.4876); rgb(158pt)=(0.3424,0.8009,0.4774); rgb(159pt)=(0.3548,0.8016,0.4669); rgb(160pt)=(0.3671,0.8021,0.4563); rgb(161pt)=(0.3795,0.8026,0.4454); rgb(162pt)=(0.3921,0.8029,0.4344); rgb(163pt)=(0.405,0.8031,0.4233); rgb(164pt)=(0.4184,0.803,0.4122); rgb(165pt)=(0.4322,0.8028,0.4013); rgb(166pt)=(0.4463,0.8024,0.3904); rgb(167pt)=(0.4608,0.8018,0.3797); rgb(168pt)=(0.4753,0.8011,0.3691); rgb(169pt)=(0.4899,0.8002,0.3586); rgb(170pt)=(0.5044,0.7993,0.348); rgb(171pt)=(0.5187,0.7982,0.3374); rgb(172pt)=(0.5329,0.797,0.3267); rgb(173pt)=(0.547,0.7957,0.3159); rgb(175pt)=(0.5748,0.7929,0.2941); rgb(176pt)=(0.5886,0.7913,0.2833); rgb(177pt)=(0.6024,0.7896,0.2726); rgb(178pt)=(0.6161,0.7878,0.2622); rgb(179pt)=(0.6297,0.7859,0.2521); rgb(180pt)=(0.6433,0.7839,0.2423); rgb(181pt)=(0.6567,0.7818,0.2329); rgb(182pt)=(0.6701,0.7796,0.2239); rgb(183pt)=(0.6833,0.7773,0.2155); rgb(184pt)=(0.6963,0.775,0.2075); rgb(185pt)=(0.7091,0.7727,0.1998); rgb(186pt)=(0.7218,0.7703,0.1924); rgb(187pt)=(0.7344,0.7679,0.1852); rgb(188pt)=(0.7468,0.7654,0.1782); rgb(189pt)=(0.759,0.7629,0.1717); rgb(190pt)=(0.771,0.7604,0.1658); rgb(191pt)=(0.7829,0.7579,0.1608); rgb(192pt)=(0.7945,0.7554,0.157); rgb(193pt)=(0.806,0.7529,0.1546); rgb(194pt)=(0.8172,0.7505,0.1535); rgb(195pt)=(0.8281,0.7481,0.1536); rgb(196pt)=(0.8389,0.7457,0.1546); rgb(197pt)=(0.8495,0.7435,0.1564); rgb(198pt)=(0.86,0.7413,0.1587); rgb(199pt)=(0.8703,0.7392,0.1615); rgb(200pt)=(0.8804,0.7372,0.165); rgb(201pt)=(0.8903,0.7353,0.1695); rgb(202pt)=(0.9,0.7336,0.1749); rgb(203pt)=(0.9093,0.7321,0.1815); rgb(204pt)=(0.9184,0.7308,0.189); rgb(205pt)=(0.9272,0.7298,0.1973); rgb(206pt)=(0.9357,0.729,0.2061); rgb(207pt)=(0.944,0.7285,0.2151); rgb(208pt)=(0.9523,0.7284,0.2237); rgb(209pt)=(0.9606,0.7285,0.2312); rgb(210pt)=(0.9689,0.7292,0.2373); rgb(211pt)=(0.977,0.7304,0.2418); rgb(212pt)=(0.9842,0.733,0.2446); rgb(213pt)=(0.99,0.7365,0.2429); rgb(214pt)=(0.9946,0.7407,0.2394); rgb(215pt)=(0.9966,0.7458,0.2351); rgb(216pt)=(0.9971,0.7513,0.2309); rgb(217pt)=(0.9972,0.7569,0.2267); rgb(218pt)=(0.9971,0.7626,0.2224); rgb(219pt)=(0.9969,0.7683,0.2181); rgb(220pt)=(0.9966,0.774,0.2138); rgb(221pt)=(0.9962,0.7798,0.2095); rgb(222pt)=(0.9957,0.7856,0.2053); rgb(223pt)=(0.9949,0.7915,0.2012); rgb(224pt)=(0.9938,0.7974,0.1974); rgb(225pt)=(0.9923,0.8034,0.1939); rgb(226pt)=(0.9906,0.8095,0.1906); rgb(227pt)=(0.9885,0.8156,0.1875); rgb(228pt)=(0.9861,0.8218,0.1846); rgb(229pt)=(0.9835,0.828,0.1817); rgb(230pt)=(0.9807,0.8342,0.1787); rgb(231pt)=(0.9778,0.8404,0.1757); rgb(232pt)=(0.9748,0.8467,0.1726); rgb(233pt)=(0.972,0.8529,0.1695); rgb(234pt)=(0.9694,0.8591,0.1665); rgb(235pt)=(0.9671,0.8654,0.1636); rgb(236pt)=(0.9651,0.8716,0.1608); rgb(237pt)=(0.9634,0.8778,0.1582); rgb(238pt)=(0.9619,0.884,0.1557); rgb(239pt)=(0.9608,0.8902,0.1532); rgb(240pt)=(0.9601,0.8963,0.1507); rgb(241pt)=(0.9596,0.9023,0.148); rgb(242pt)=(0.9595,0.9084,0.145); rgb(243pt)=(0.9597,0.9143,0.1418); rgb(244pt)=(0.9601,0.9203,0.1382); rgb(245pt)=(0.9608,0.9262,0.1344); rgb(246pt)=(0.9618,0.932,0.1304); rgb(247pt)=(0.9629,0.9379,0.1261); rgb(248pt)=(0.9642,0.9437,0.1216); rgb(249pt)=(0.9657,0.9494,0.1168); rgb(250pt)=(0.9674,0.9552,0.1116); rgb(251pt)=(0.9692,0.9609,0.1061); rgb(252pt)=(0.9711,0.9667,0.1001); rgb(253pt)=(0.973,0.9724,0.0938); rgb(254pt)=(0.9749,0.9782,0.0872); rgb(255pt)=(0.9769,0.9839,0.0805)},
colorbar,
    colorbar style={
        height=4.5cm}
]
\end{axis}
\end{tikzpicture}%
\end{tabular}
  \caption{Deformed configuration with von Mises stress distribution \(\sigma_{vM}\) [Pa] with a typical mesh and load step for compressible GT model (left) and the corresponding PANN8 model (mid).}
  \label{figure:CookStress}
\end{figure}

\begin{figure}[t]
  \begin{tabular}{cc}
   \setlength{\figH}{0.2\textheight}
   \setlength{\figW}{0.3\textwidth}
   \input{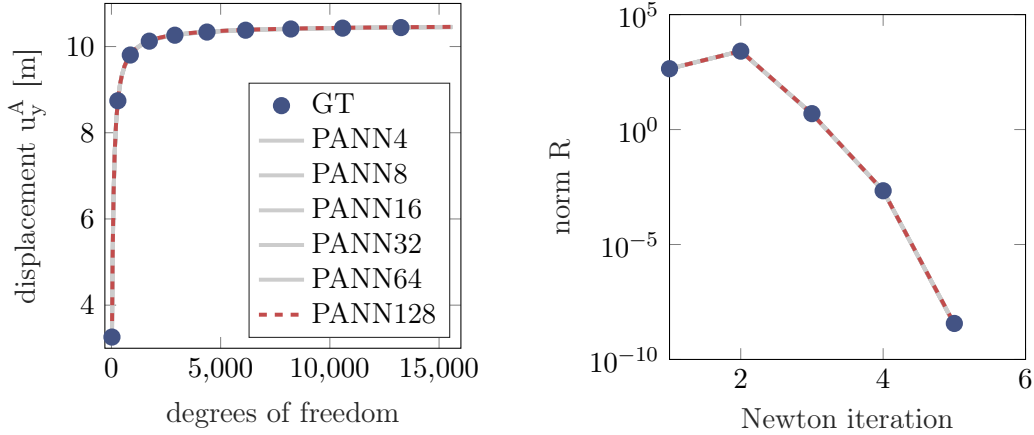}
   &
   \setlength{\figH}{0.2\textheight}
   \setlength{\figW}{0.3\textwidth}
%
%
\begin{tikzpicture}

\begin{axis}[%
width=0.951\figW,
height=\figH,
at={(0\figW,0\figH)},
scale only axis,
xmin=1,
xmax=6,
xlabel style={font=\color{white!15!black}},
xlabel={Newton iteration},
ymode=log,
ymin=1e-10,
ymax=100000,
yminorticks=true,
ylabel style={font=\color{white!15!black}},
ylabel={norm R},
axis background/.style={fill=white},
legend style={at={(0.03,0.03)}, anchor=south west, legend cell align=left, align=left, draw=white!15!black}
]
\addplot [color=CPSdarkblue, mark=*, mark size = 2.5pt, line width=1.5pt, only marks]
  table[row sep=crcr]{%
1	4.42217e+02\\
2	2.57276e+03\\
3	4.91968e+00\\
4	2.17400e-03\\
5	3.67960e-09\\
};
\addlegendentry{GT}

\addplot [color=CPSgrey, line width=1.5pt]
  table[row sep=crcr]{%
1	4.42217e+02\\
2	2.58704e+03\\
3	4.78327e+00\\
4	2.19841e-03\\
5	3.21035e-09\\
};
\addlegendentry{ANN4}

\addplot [color=CPSgrey, line width=1.5pt]
  table[row sep=crcr]{%
1	4.42217e+02\\
2	2.59066e+03\\
3	4.75067e+00\\
4	2.19462e-03\\
5	3.28250e-09\\
};
\addlegendentry{ANN8}

\addplot [color=CPSgrey, line width=1.5pt]
  table[row sep=crcr]{%
1	4.42217e+02\\
2	2.59724e+03\\
3	4.72073e+00\\
4	2.21488e-03\\
5	3.28541e-09\\
};
\addlegendentry{ANN16}

\addplot [color=CPSgrey, line width=1.5pt]
  table[row sep=crcr]{%
1	4.42217e+02\\
2	2.58176e+03\\
3	4.90759e+00\\
4	2.19730e-03\\
5	3.44974e-09\\
};
\addlegendentry{ANN32}

\addplot [color=CPSgrey, line width=1.5pt]
  table[row sep=crcr]{%
1	4.42217e+02\\
2	2.58038e+03\\
3	4.89247e+00\\
4	2.19556e-03\\
5	3.58395e-09\\
};
\addlegendentry{ANN64}

\addplot [color=CPSred, dashed, line width=1.5pt]
  table[row sep=crcr]{%
1	4.42217e+02\\
2	2.58043e+03\\
3	4.88397e+00\\
4	2.19762e-03\\
5	3.50953e-09\\
};
\addlegendentry{ANN128}

\legend{}
\end{axis}
\end{tikzpicture}%
\end{tabular}
\caption{Convergence plot with vertical displacement of point A (see \cref{figure:CookBoundaryConditions})  vs.\ mesh refinement (left) and Newton iterations for a typical mesh and load step (right). 
}
  \label{figure:CookConvergenceIterations}
\end{figure}

In \cref{figure:CookStress}, typical deformed configurations for both the GT and the PANN8 model are provided. The mesh convergence study in \cref{figure:CookConvergenceIterations} (left) shows a practically identical behavior for the GT model and the PANN model for all choices of nodes. Unsurprisingly, the PANN model is able to perfectly represent the behavior of the GT model, even for small numbers of nodes. 
This is also reflected in the Newton iterations required to solve the system of nonlinear equations, which are visualized in \cref{figure:CookConvergenceIterations} (right) for a typical mesh and load step. For all choices of nodes in the PANN model, the required Newton iterations and the corresponding norm of the residual are equal to the GT model. This shows that the PANN model is not only able to excellently predict the stress behavior of the GT model, which is closely related to the first derivative of the hyperelastic potential, but also excellently predicts the second derivative, which is essential for the solution of the nonlinear FEA, although the second derivative was not included in the calibration process, cf.~\cref{sec:PANN_calibration}.

\begin{figure}[t]
\center
   \setlength{\figH}{0.2\textheight}
   \setlength{\figW}{0.3\textwidth}
%
%

\colorlet{mycolor1}{CPSdarkblue}
\colorlet{mycolor2}{CPSorange}
\begin{tikzpicture}

\begin{axis}[%
width=0.979\figW,
height=\figH,
at={(0\figW,0\figH)},
scale only axis,
bar shift auto,
xmin=0.514285714285714,
xmax=7.48571428571429,
xtick={1,2,3,4,5,6,7},
xticklabels={{GT},{4},{8},{16},{32},{64},{128}},
xlabel style={font=\color{white!15!black}},
xlabel={GT or number of neurons},
ymin=0,
ymax=1000,
ylabel style={font=\color{white!15!black}},
ylabel={computation time [s]},
axis background/.style={fill=white},
legend style={at={(0.5,0.97)}, anchor=north, legend cell align=left, align=left, draw=white!15!black}
]
\addplot[ybar, bar width=0.229, fill = CPSlightblue!60, draw=black, area legend] table[row sep=crcr] {%
1	375.9467\\
2	674.2047\\
3	700.8182\\
4	683.769\\
5	663.5408\\
6	700.3094\\
7	893.0973\\
};
\addplot[forget plot, color=white!15!black] table[row sep=crcr] {%
0.514285714285714	0\\
7.48571428571429	0\\
};
\addlegendentry{1 core}

\addplot[ybar, bar width=0.229, fill=CPSred!10,
postaction={
        pattern=north east lines, pattern color = CPSred
    }, 
    draw=black, area legend] table[row sep=crcr] {%
1	111.0852\\
2	118.7766\\
3	121.4616\\
4	116.2959\\
5	116.143\\
6	116.8476\\
7	119.6818\\
};
\addplot[forget plot, color=white!15!black] table[row sep=crcr] {%
0.514285714285714	0\\
7.48571428571429	0\\
};
\addlegendentry{24 cores}

\end{axis}
\end{tikzpicture}%
\caption{Computation times with 1 and 24 core setup for GT and different PANN models, evaluated for the simulation of Cook's membran with $36015$ DOFs.}
  \label{figure:CookComputationTime}
\end{figure}
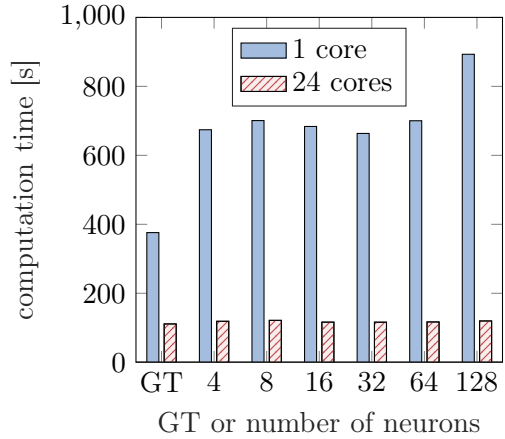

In \cref{figure:CookPieChartComputationTimeGT} and \cref{figure:CookPieChartComputationTime}, the percentage computation times for the simulation of Cook's membrane with $14406$ degree of freedom (DOFs) on one single core are presented for the GT model and the PANN128 model, respectively. On the element level of the GT model, the evaluation of the derivatives of the analytical Mooney-Rivlin potential (\enquote{GT derivatives}) only makes up 5\% of the computation time and takes even less time than operations such as the evaluation of the tensor cross product. For the PANN128 model, however, the evaluation of the NN potential is significantly more expensive and takes over 50\% of the computation time on element level. Thus, the overall simulation time is increased. 
In \cref{figure:CookComputationTime}, the computation time for the GT model and the PANN model with different numbers of nodes on one single core are visualized. Depending on the number of nodes in the NN, the computation time of the PANN model is more than twice as long compared to the GT model, when the simulation is carried out on a single core.
However, when the FEA is parallelised on 24 cores, the computation time of the PANN model is equal to the computation time of the analytical GT model, and, within the examined bounds, even independent of the number of nodes in the hidden layer. 
The parallelization is employed within the research code \textsc{moofeKIT} \cite{moofeKIT} written in \textsc{Matlab} using the \textsc{Parallel Computing Toolbox}.
In particular, the element routine is parallelized within a parallel for loop environment \texttt{parfor}. 
Obviously, the computation time is highly influenced by many factors, such as computer hardware and programming language. Still, our findings suggest that with a suitable implementation of the NN-based constitutive model, computation times comparable to analytical constitutive models can be achieved. 

 \begin{figure}[t]
     \centering
     \resizebox{0.95\textwidth}{!}{
     \begin{tikzpicture}
\pie[
    color = {
    CPSlightblue!40,
    CPSorange!40,
    CPSgreen!40,
    CPSgrey!70},
    rotate=-360*0.159,
       explode=0.1,
        radius=3,
]
{   65.9/element evaluation, 
    10.8/direct solver,
    23.3/other} 
\pie[
pos={10,0},
    color = {
    CPSlightblue!40,
    CPSorange!40,
    CPSgreen!40,
    CPSred!40,
    CPSgrey!40,    
    CPSgrey!70
    },
        explode=0.1,
        radius=3,
        rotate=-0.55*360,
] 
{   5/GT derivatives,
    18.4/\(J\) and \(\partial_{\subX}N_I\),
    9.5/tensor cross product,
    34.3/B matrix,
    5/\(\mathcal{D}(\tens{A})\),    
    27.8/other}
\end{tikzpicture}
     }
\caption{Percentage computation time of the GT model on global level (left) with 1.6\% assembly of $\tens{K}$ included in the slice \enquote{other}, and percentage computation time on element level (right), where the operator matrix \(\mathcal{D}(\tens{A})\) is employed (see \cite[Appx.\ B]{betsch2018}).
Evaluated for the simulation of Cook's membrane with $14406$ DOFs on one core.}
  \label{figure:CookPieChartComputationTimeGT}
 \end{figure}
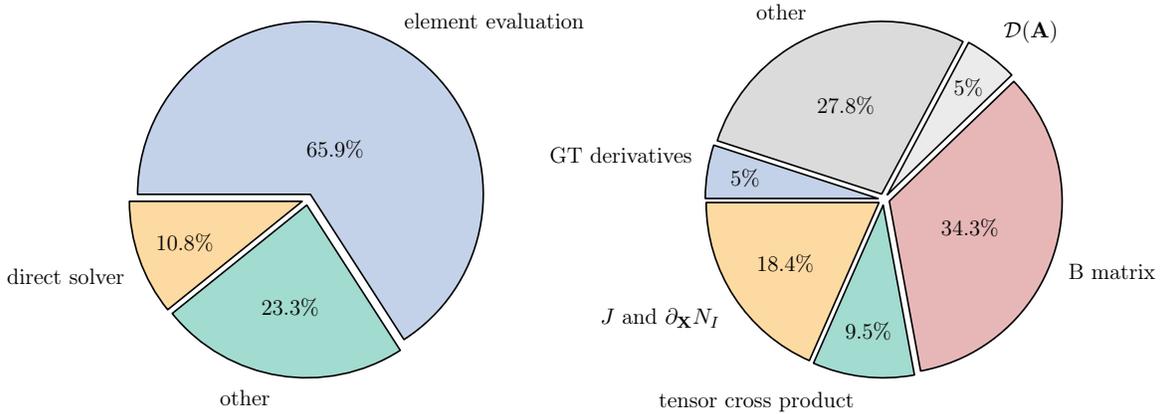
 
 \begin{figure}[t]
     \centering
     \resizebox{0.95\textwidth}{!}{
     \begin{tikzpicture}
\pie[
    color = {
    CPSlightblue!40,
    CPSorange!40,
    CPSgreen!40,
    CPSgrey!70},
    rotate=-360*0.376,
       explode=0.1,
        radius=3,
] 
{   87.6/element evaluation, 
    2.9/direct solver,
    9.5/other} 
\pie[
pos={10,0},
    color = {
    CPSlightblue!40,
    CPSorange!40,
    CPSgreen!40,
    CPSred!40,
    CPSgrey!70},
        explode=0.1,
        radius=3,
        rotate=-0.042*360,
]
{   54.2/PANN128 derivatives,     
    5.6/\(J\) and \(\partial_{\subX}N_I\),
    8.7/tensor cross product,
    10.6/B matrix,
    20.9/other 
    }
\end{tikzpicture}
     }
     \caption{Percentage computation time of the PANN128 model on global level (left) with 0.5\% assembly of $\tens{K}$ included in the slice \enquote{other}, and percentage computation time on element level (right).
Evaluated for the simulation of Cook's membrane with $14406$ DOFs on one core.}
  \label{figure:CookPieChartComputationTime}
 \end{figure}
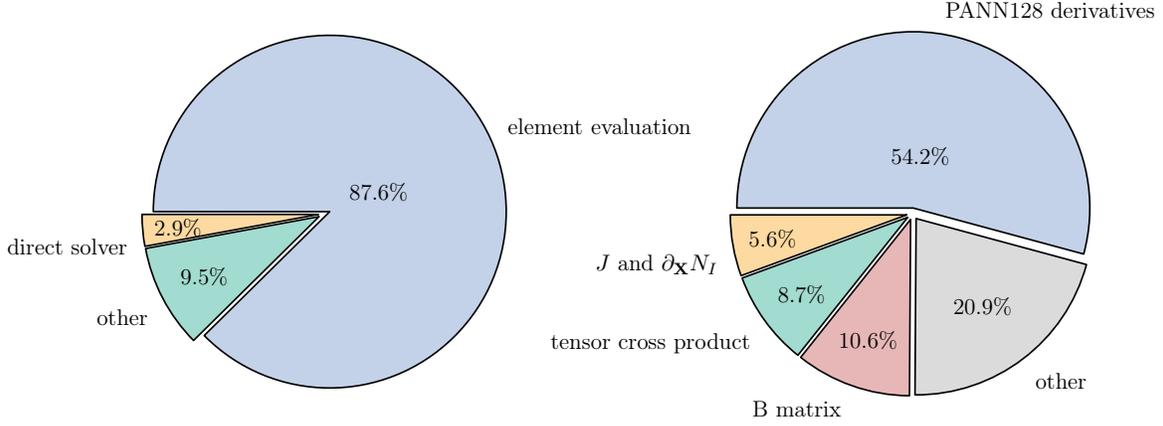
 
\subsubsection{Nearly incompressible case}\label{section:cooksNearlyIncompressible}
In the next step, Cook's membrane is simulated with a nearly incompressible material behavior of both the GT and the PANN model. Thereby, the performance of the PANN constitutive model when applied in mixed formulations, which avoid the locking effect, is investigated.

\begin{figure}[H]
\centering
  \begin{tabular}{cccl}
\includegraphics[width=0.15\textwidth]{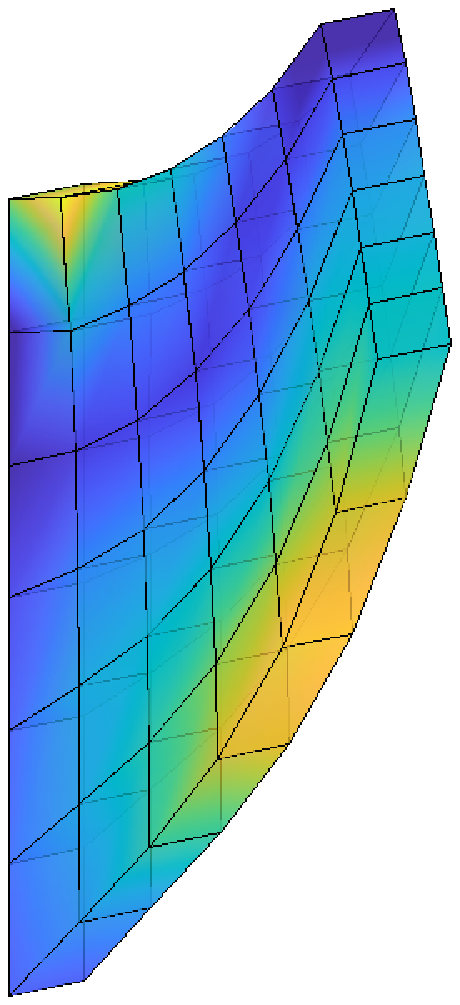}
&
\hspace*{3cm}
&
\includegraphics[width=0.15\textwidth]{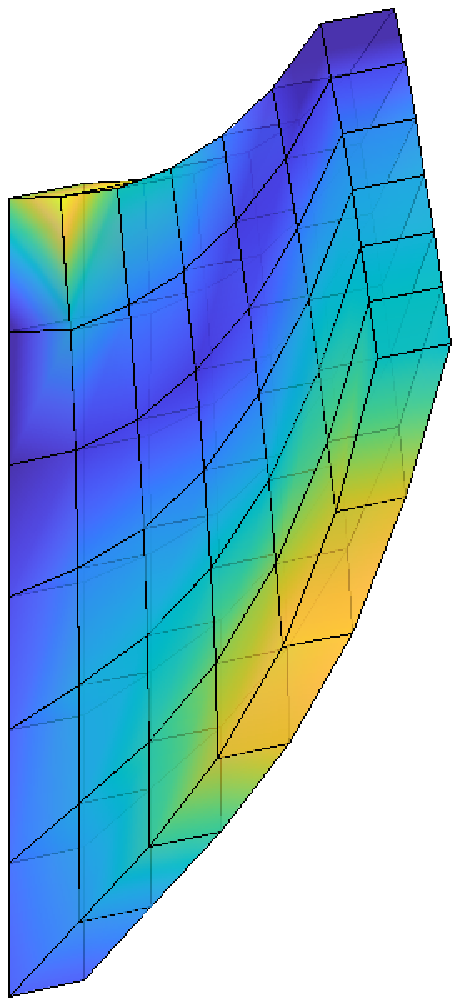}
&
   \setlength{\figH}{0.2\textheight}
   \setlength{\figW}{0.2\textwidth}
%
%
\begin{tikzpicture}

\begin{axis}[%
width=0.891\figW,
height=\figH,
at={(0\figW,0\figH)},
scale only axis,
point meta min=50,
point meta max=350,
xmin=0,
xmax=1,
ymin=0,
ymax=1,
axis line style={draw=none},
ticks=none,
axis x line*=bottom,
axis y line*=left,
colormap={mymap}{[1pt] rgb(0pt)=(0.2422,0.1504,0.6603); rgb(1pt)=(0.2444,0.1534,0.6728); rgb(2pt)=(0.2464,0.1569,0.6847); rgb(3pt)=(0.2484,0.1607,0.6961); rgb(4pt)=(0.2503,0.1648,0.7071); rgb(5pt)=(0.2522,0.1689,0.7179); rgb(6pt)=(0.254,0.1732,0.7286); rgb(7pt)=(0.2558,0.1773,0.7393); rgb(8pt)=(0.2576,0.1814,0.7501); rgb(9pt)=(0.2594,0.1854,0.761); rgb(11pt)=(0.2628,0.1932,0.7828); rgb(12pt)=(0.2645,0.1972,0.7937); rgb(13pt)=(0.2661,0.2011,0.8043); rgb(14pt)=(0.2676,0.2052,0.8148); rgb(15pt)=(0.2691,0.2094,0.8249); rgb(16pt)=(0.2704,0.2138,0.8346); rgb(17pt)=(0.2717,0.2184,0.8439); rgb(18pt)=(0.2729,0.2231,0.8528); rgb(19pt)=(0.274,0.228,0.8612); rgb(20pt)=(0.2749,0.233,0.8692); rgb(21pt)=(0.2758,0.2382,0.8767); rgb(22pt)=(0.2766,0.2435,0.884); rgb(23pt)=(0.2774,0.2489,0.8908); rgb(24pt)=(0.2781,0.2543,0.8973); rgb(25pt)=(0.2788,0.2598,0.9035); rgb(26pt)=(0.2794,0.2653,0.9094); rgb(27pt)=(0.2798,0.2708,0.915); rgb(28pt)=(0.2802,0.2764,0.9204); rgb(29pt)=(0.2806,0.2819,0.9255); rgb(30pt)=(0.2809,0.2875,0.9305); rgb(31pt)=(0.2811,0.293,0.9352); rgb(32pt)=(0.2813,0.2985,0.9397); rgb(33pt)=(0.2814,0.304,0.9441); rgb(34pt)=(0.2814,0.3095,0.9483); rgb(35pt)=(0.2813,0.315,0.9524); rgb(36pt)=(0.2811,0.3204,0.9563); rgb(37pt)=(0.2809,0.3259,0.96); rgb(38pt)=(0.2807,0.3313,0.9636); rgb(39pt)=(0.2803,0.3367,0.967); rgb(40pt)=(0.2798,0.3421,0.9702); rgb(41pt)=(0.2791,0.3475,0.9733); rgb(42pt)=(0.2784,0.3529,0.9763); rgb(43pt)=(0.2776,0.3583,0.9791); rgb(44pt)=(0.2766,0.3638,0.9817); rgb(45pt)=(0.2754,0.3693,0.984); rgb(46pt)=(0.2741,0.3748,0.9862); rgb(47pt)=(0.2726,0.3804,0.9881); rgb(48pt)=(0.271,0.386,0.9898); rgb(49pt)=(0.2691,0.3916,0.9912); rgb(50pt)=(0.267,0.3973,0.9924); rgb(51pt)=(0.2647,0.403,0.9935); rgb(52pt)=(0.2621,0.4088,0.9946); rgb(53pt)=(0.2591,0.4145,0.9955); rgb(54pt)=(0.2556,0.4203,0.9965); rgb(55pt)=(0.2517,0.4261,0.9974); rgb(56pt)=(0.2473,0.4319,0.9983); rgb(57pt)=(0.2424,0.4378,0.9991); rgb(58pt)=(0.2369,0.4437,0.9996); rgb(59pt)=(0.2311,0.4497,0.9995); rgb(60pt)=(0.225,0.4559,0.9985); rgb(61pt)=(0.2189,0.462,0.9968); rgb(62pt)=(0.2128,0.4682,0.9948); rgb(63pt)=(0.2066,0.4743,0.9926); rgb(64pt)=(0.2006,0.4803,0.9906); rgb(65pt)=(0.195,0.4861,0.9887); rgb(66pt)=(0.1903,0.4919,0.9867); rgb(67pt)=(0.1869,0.4975,0.9844); rgb(68pt)=(0.1847,0.503,0.9819); rgb(69pt)=(0.1831,0.5084,0.9793); rgb(70pt)=(0.1818,0.5138,0.9766); rgb(71pt)=(0.1806,0.5191,0.9738); rgb(72pt)=(0.1795,0.5244,0.9709); rgb(73pt)=(0.1785,0.5296,0.9677); rgb(74pt)=(0.1778,0.5349,0.9641); rgb(75pt)=(0.1773,0.5401,0.9602); rgb(76pt)=(0.1768,0.5452,0.956); rgb(77pt)=(0.1764,0.5504,0.9516); rgb(78pt)=(0.1755,0.5554,0.9473); rgb(79pt)=(0.174,0.5605,0.9432); rgb(80pt)=(0.1716,0.5655,0.9393); rgb(81pt)=(0.1686,0.5705,0.9357); rgb(82pt)=(0.1649,0.5755,0.9323); rgb(83pt)=(0.161,0.5805,0.9289); rgb(84pt)=(0.1573,0.5854,0.9254); rgb(85pt)=(0.154,0.5902,0.9218); rgb(86pt)=(0.1513,0.595,0.9182); rgb(87pt)=(0.1492,0.5997,0.9147); rgb(88pt)=(0.1475,0.6043,0.9113); rgb(89pt)=(0.1461,0.6089,0.908); rgb(90pt)=(0.1446,0.6135,0.905); rgb(91pt)=(0.1429,0.618,0.9022); rgb(92pt)=(0.1408,0.6226,0.8998); rgb(93pt)=(0.1383,0.6272,0.8975); rgb(94pt)=(0.1354,0.6317,0.8953); rgb(95pt)=(0.1321,0.6363,0.8932); rgb(96pt)=(0.1288,0.6408,0.891); rgb(97pt)=(0.1253,0.6453,0.8887); rgb(98pt)=(0.1219,0.6497,0.8862); rgb(99pt)=(0.1185,0.6541,0.8834); rgb(100pt)=(0.1152,0.6584,0.8804); rgb(101pt)=(0.1119,0.6627,0.877); rgb(102pt)=(0.1085,0.6669,0.8734); rgb(103pt)=(0.1048,0.671,0.8695); rgb(104pt)=(0.1009,0.675,0.8653); rgb(105pt)=(0.0964,0.6789,0.8609); rgb(106pt)=(0.0914,0.6828,0.8562); rgb(107pt)=(0.0855,0.6865,0.8513); rgb(108pt)=(0.0789,0.6902,0.8462); rgb(109pt)=(0.0713,0.6938,0.8409); rgb(110pt)=(0.0628,0.6972,0.8355); rgb(111pt)=(0.0535,0.7006,0.8299); rgb(112pt)=(0.0433,0.7039,0.8242); rgb(113pt)=(0.0328,0.7071,0.8183); rgb(114pt)=(0.0234,0.7103,0.8124); rgb(115pt)=(0.0155,0.7133,0.8064); rgb(116pt)=(0.0091,0.7163,0.8003); rgb(117pt)=(0.0046,0.7192,0.7941); rgb(118pt)=(0.0019,0.722,0.7878); rgb(119pt)=(0.0009,0.7248,0.7815); rgb(120pt)=(0.0018,0.7275,0.7752); rgb(121pt)=(0.0046,0.7301,0.7688); rgb(122pt)=(0.0094,0.7327,0.7623); rgb(123pt)=(0.0162,0.7352,0.7558); rgb(124pt)=(0.0253,0.7376,0.7492); rgb(125pt)=(0.0369,0.74,0.7426); rgb(126pt)=(0.0504,0.7423,0.7359); rgb(127pt)=(0.0638,0.7446,0.7292); rgb(128pt)=(0.077,0.7468,0.7224); rgb(129pt)=(0.0899,0.7489,0.7156); rgb(130pt)=(0.1023,0.751,0.7088); rgb(131pt)=(0.1141,0.7531,0.7019); rgb(132pt)=(0.1252,0.7552,0.695); rgb(133pt)=(0.1354,0.7572,0.6881); rgb(134pt)=(0.1448,0.7593,0.6812); rgb(135pt)=(0.1532,0.7614,0.6741); rgb(136pt)=(0.1609,0.7635,0.6671); rgb(137pt)=(0.1678,0.7656,0.6599); rgb(138pt)=(0.1741,0.7678,0.6527); rgb(139pt)=(0.1799,0.7699,0.6454); rgb(140pt)=(0.1853,0.7721,0.6379); rgb(141pt)=(0.1905,0.7743,0.6303); rgb(142pt)=(0.1954,0.7765,0.6225); rgb(143pt)=(0.2003,0.7787,0.6146); rgb(144pt)=(0.2061,0.7808,0.6065); rgb(145pt)=(0.2118,0.7828,0.5983); rgb(146pt)=(0.2178,0.7849,0.5899); rgb(147pt)=(0.2244,0.7869,0.5813); rgb(148pt)=(0.2318,0.7887,0.5725); rgb(149pt)=(0.2401,0.7905,0.5636); rgb(150pt)=(0.2491,0.7922,0.5546); rgb(151pt)=(0.2589,0.7937,0.5454); rgb(152pt)=(0.2695,0.7951,0.536); rgb(153pt)=(0.2809,0.7964,0.5266); rgb(154pt)=(0.2929,0.7975,0.517); rgb(155pt)=(0.3052,0.7985,0.5074); rgb(156pt)=(0.3176,0.7994,0.4975); rgb(157pt)=(0.3301,0.8002,0.4876); rgb(158pt)=(0.3424,0.8009,0.4774); rgb(159pt)=(0.3548,0.8016,0.4669); rgb(160pt)=(0.3671,0.8021,0.4563); rgb(161pt)=(0.3795,0.8026,0.4454); rgb(162pt)=(0.3921,0.8029,0.4344); rgb(163pt)=(0.405,0.8031,0.4233); rgb(164pt)=(0.4184,0.803,0.4122); rgb(165pt)=(0.4322,0.8028,0.4013); rgb(166pt)=(0.4463,0.8024,0.3904); rgb(167pt)=(0.4608,0.8018,0.3797); rgb(168pt)=(0.4753,0.8011,0.3691); rgb(169pt)=(0.4899,0.8002,0.3586); rgb(170pt)=(0.5044,0.7993,0.348); rgb(171pt)=(0.5187,0.7982,0.3374); rgb(172pt)=(0.5329,0.797,0.3267); rgb(173pt)=(0.547,0.7957,0.3159); rgb(175pt)=(0.5748,0.7929,0.2941); rgb(176pt)=(0.5886,0.7913,0.2833); rgb(177pt)=(0.6024,0.7896,0.2726); rgb(178pt)=(0.6161,0.7878,0.2622); rgb(179pt)=(0.6297,0.7859,0.2521); rgb(180pt)=(0.6433,0.7839,0.2423); rgb(181pt)=(0.6567,0.7818,0.2329); rgb(182pt)=(0.6701,0.7796,0.2239); rgb(183pt)=(0.6833,0.7773,0.2155); rgb(184pt)=(0.6963,0.775,0.2075); rgb(185pt)=(0.7091,0.7727,0.1998); rgb(186pt)=(0.7218,0.7703,0.1924); rgb(187pt)=(0.7344,0.7679,0.1852); rgb(188pt)=(0.7468,0.7654,0.1782); rgb(189pt)=(0.759,0.7629,0.1717); rgb(190pt)=(0.771,0.7604,0.1658); rgb(191pt)=(0.7829,0.7579,0.1608); rgb(192pt)=(0.7945,0.7554,0.157); rgb(193pt)=(0.806,0.7529,0.1546); rgb(194pt)=(0.8172,0.7505,0.1535); rgb(195pt)=(0.8281,0.7481,0.1536); rgb(196pt)=(0.8389,0.7457,0.1546); rgb(197pt)=(0.8495,0.7435,0.1564); rgb(198pt)=(0.86,0.7413,0.1587); rgb(199pt)=(0.8703,0.7392,0.1615); rgb(200pt)=(0.8804,0.7372,0.165); rgb(201pt)=(0.8903,0.7353,0.1695); rgb(202pt)=(0.9,0.7336,0.1749); rgb(203pt)=(0.9093,0.7321,0.1815); rgb(204pt)=(0.9184,0.7308,0.189); rgb(205pt)=(0.9272,0.7298,0.1973); rgb(206pt)=(0.9357,0.729,0.2061); rgb(207pt)=(0.944,0.7285,0.2151); rgb(208pt)=(0.9523,0.7284,0.2237); rgb(209pt)=(0.9606,0.7285,0.2312); rgb(210pt)=(0.9689,0.7292,0.2373); rgb(211pt)=(0.977,0.7304,0.2418); rgb(212pt)=(0.9842,0.733,0.2446); rgb(213pt)=(0.99,0.7365,0.2429); rgb(214pt)=(0.9946,0.7407,0.2394); rgb(215pt)=(0.9966,0.7458,0.2351); rgb(216pt)=(0.9971,0.7513,0.2309); rgb(217pt)=(0.9972,0.7569,0.2267); rgb(218pt)=(0.9971,0.7626,0.2224); rgb(219pt)=(0.9969,0.7683,0.2181); rgb(220pt)=(0.9966,0.774,0.2138); rgb(221pt)=(0.9962,0.7798,0.2095); rgb(222pt)=(0.9957,0.7856,0.2053); rgb(223pt)=(0.9949,0.7915,0.2012); rgb(224pt)=(0.9938,0.7974,0.1974); rgb(225pt)=(0.9923,0.8034,0.1939); rgb(226pt)=(0.9906,0.8095,0.1906); rgb(227pt)=(0.9885,0.8156,0.1875); rgb(228pt)=(0.9861,0.8218,0.1846); rgb(229pt)=(0.9835,0.828,0.1817); rgb(230pt)=(0.9807,0.8342,0.1787); rgb(231pt)=(0.9778,0.8404,0.1757); rgb(232pt)=(0.9748,0.8467,0.1726); rgb(233pt)=(0.972,0.8529,0.1695); rgb(234pt)=(0.9694,0.8591,0.1665); rgb(235pt)=(0.9671,0.8654,0.1636); rgb(236pt)=(0.9651,0.8716,0.1608); rgb(237pt)=(0.9634,0.8778,0.1582); rgb(238pt)=(0.9619,0.884,0.1557); rgb(239pt)=(0.9608,0.8902,0.1532); rgb(240pt)=(0.9601,0.8963,0.1507); rgb(241pt)=(0.9596,0.9023,0.148); rgb(242pt)=(0.9595,0.9084,0.145); rgb(243pt)=(0.9597,0.9143,0.1418); rgb(244pt)=(0.9601,0.9203,0.1382); rgb(245pt)=(0.9608,0.9262,0.1344); rgb(246pt)=(0.9618,0.932,0.1304); rgb(247pt)=(0.9629,0.9379,0.1261); rgb(248pt)=(0.9642,0.9437,0.1216); rgb(249pt)=(0.9657,0.9494,0.1168); rgb(250pt)=(0.9674,0.9552,0.1116); rgb(251pt)=(0.9692,0.9609,0.1061); rgb(252pt)=(0.9711,0.9667,0.1001); rgb(253pt)=(0.973,0.9724,0.0938); rgb(254pt)=(0.9749,0.9782,0.0872); rgb(255pt)=(0.9769,0.9839,0.0805)},
colorbar,
    colorbar style={
        height=4.5cm}
]
\end{axis}
\end{tikzpicture}%
\end{tabular}
  \caption{Deformed configuration with von Mises stress distribution \(\sigma_{vM}\) [Pa] with a typical mesh and load step for the nearly incompressible GT model (left) and the corresponding PANN8 model (mid).}
  \label{figure:CookStressNearlyIncompressible}
\end{figure}

\begin{figure}[H]
 \centering
  \setlength{\figH}{0.2\textheight}
   \setlength{\figW}{0.3\textwidth}
    \input{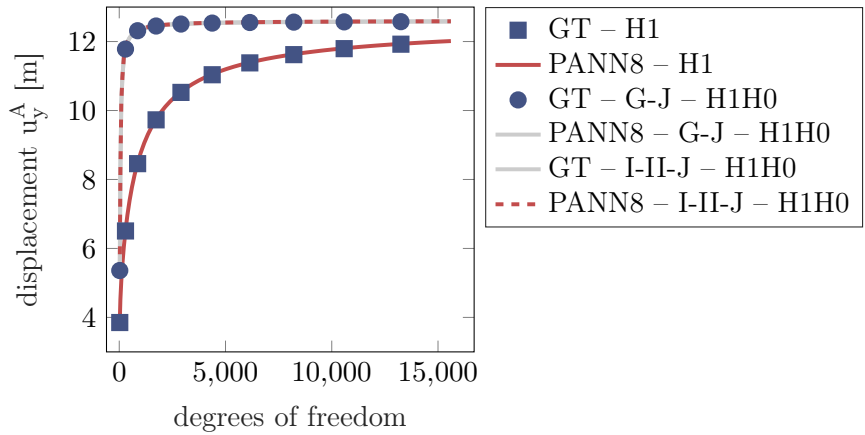}
     \caption{Convergence plot with displacement vs.\ mesh refinement for the nearly incompressible Cook's membrane. Different material models (GT and PANN8) and elements (H1: displacement-based trilinear hexahedron, G-J -- H1H0: 5-field mixed formulation according to \cref{equation:5fieldFormulation} with trilinear / constant hexahedron elements, and I-II-J -- H1H0: new Hu-Washizu like 7-field mixed formulation according to \cref{equation:new7fieldFormulation} with trilinear / constant hexahedron elements) are employed. 
     Note that the two alternative constitutive model formulations under investigation essentially lead to the same results for each finite element formulation.
    }
  \label{figure:CookConvergenceIncompressible}
\end{figure}

The resultant force of the pressure load has the magnitude of \(p=100\) Pa and shows into the positive \(\vec{e}_3\)-direction, see \cref{figure:CookBoundaryConditions}. 
Typical deformed configurations with von Mises stress are provided in \cref{figure:CookStressNearlyIncompressible}. 
A convergence study is provided in \cref{figure:CookConvergenceIncompressible} (left), where different material models and element types are examined.
The solutions for both the GT and the PANN8 model are almost indistinguishable for all elements employed. As to be expected in this nearly incompressible, shear-loaded problem, the locking effect becomes obvious for both displacement-based models with trilinear hexahedron elements (GT -- H1 and PANN8 -- H1). 
In contrast to that, the 5-field mixed formulation presented in \cref{equation:5fieldFormulation} with trilinear / constant hexahedron elements (G-J) and the new Hu-Washizu like 7-field mixed formulation in \cref{equation:new7fieldFormulation} with trilinear / constant hexahedron elements (I-II-J) exhibit fast, locking-free convergence. 
In particular, although the I-II-J element only considers the \emph{scalar} quantities \(\I\), \(\II\), and \(J\) as additional kinematic unknowns, it shows the same excellent spatial performance as the G-J element, which, in contrast, includes an additional \emph{tensor-valued} unknown.

\subsection{Dynamics of L-shaped body}\label{section:LShapedBody}
The objective of this investigation is to demonstrate the discrete conservation properties of the proposed EM integrator, particularly applied in the context of the PANN model.
In addition to that, the robustness and long-term stability of the midpoint (MP) and the EM integrator are compared for the PANN model. 

\begin{figure}[t]
\centering
\begin{tabular}{cccc}
    \phantom{XXXXXX}
    &
    \psfrag{7}[][]{\(7\)}
    \psfrag{3}[][]{\(3\)}
    \psfrag{6}[][]{\(6\)}
    \psfrag{m}[][]{\([m]\)}
    \psfrag{t}[][]{\(t=3\)}
    \includegraphics[width=0.3\textwidth]{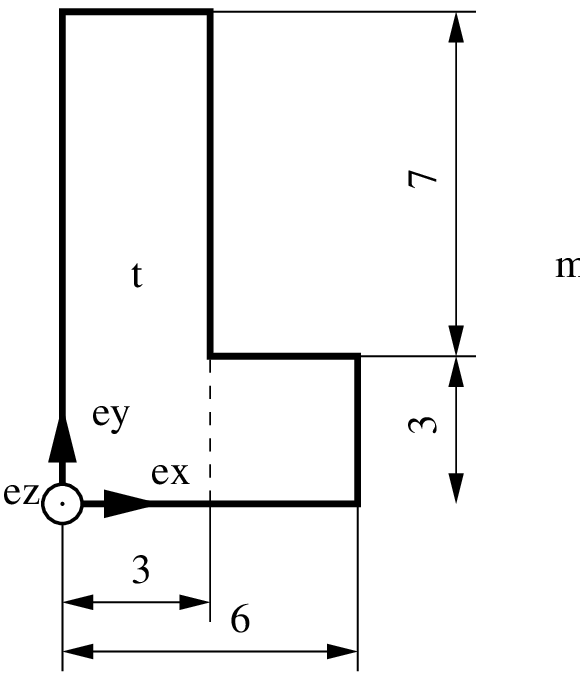}
    &
    \phantom{XXXXXX}
    &
    \newlength{\svgwidth}
    \setlength{\svgwidth}{0.6\textwidth}
\begingroup%
  \makeatletter%
  \providecommand\color[2][]{%
    \errmessage{(Inkscape) Color is used for the text in Inkscape, but the package 'color.sty' is not loaded}%
    \renewcommand\color[2][]{}%
  }%
  \providecommand\transparent[1]{%
    \errmessage{(Inkscape) Transparency is used (non-zero) for the text in Inkscape, but the package 'transparent.sty' is not loaded}%
    \renewcommand\transparent[1]{}%
  }%
  \providecommand\rotatebox[2]{#2}%
  \newcommand*\fsize{\dimexpr\f@size pt\relax}%
  \newcommand*\lineheight[1]{\fontsize{\fsize}{#1\fsize}\selectfont}%
  \ifx\svgwidth\undefined%
    \setlength{\unitlength}{1252.79752185bp}%
    \ifx\svgscale\undefined%
      \relax%
    \else%
      \setlength{\unitlength}{\unitlength * \real{\svgscale}}%
    \fi%
  \else%
    \setlength{\unitlength}{\svgwidth}%
  \fi%
  \global\let\svgwidth\undefined%
  \global\let\svgscale\undefined%
  \makeatother%
  \begin{picture}(1,0.59803116)%
    \lineheight{1}%
    \setlength\tabcolsep{0pt}%
    \put(0,0){\includegraphics[width=\unitlength]{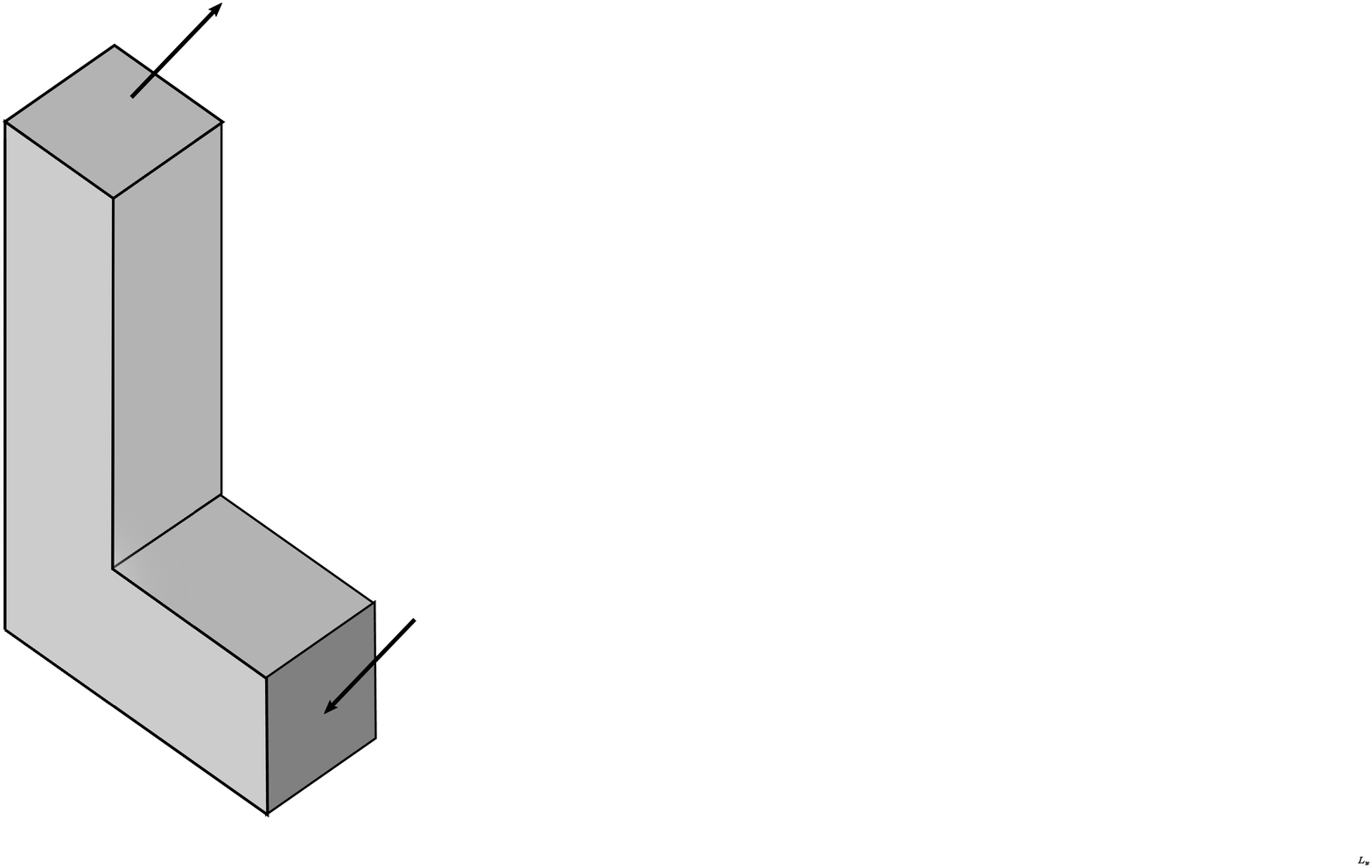}}%
    \put(0.33062186,0.11788326){\color[rgb]{0,0,0}\makebox(0,0)[lt]{\lineheight{1.25}\smash{\begin{tabular}[t]{l}\(\vec{P}_2\)\end{tabular}}}}%
    \put(0.21603792,0.56119821){\color[rgb]{0,0,0}\makebox(0,0)[lt]{\lineheight{1.25}\smash{\begin{tabular}[t]{l}\(\vec{P}_1\)\end{tabular}}}}%
  \end{picture}%
\endgroup%

    \end{tabular}
  \caption{2D projection of the geometry (left) and 3D illustration of the boundary conditions (right) of the L-shaped body with \(t=3m\).}
  \label{figure:LshapeBoundaryConditions}
\end{figure}

The initial boundary value problem of an L-shaped body is considered, which is inspired by \cite{simo1992,betsch2018} (in case of pure mechanics) and \cite{franke2019,franke2022} (in case of thermo-mechanics and electro-thermo-mechanics, respectively). 
The geometry of the L-shaped body and the initial setting with boundary conditions are depicted in \cref{figure:LshapeBoundaryConditions}. 
Therein the applied dead load pressures \(\vec{P}_1:\mathcal{T}\rightarrow\mathds{R}^3\) and \(\vec{P}_2:\mathcal{T}\rightarrow\mathds{R}^3\) follow the functions
\begin{equation}
  \begin{aligned}
    \vec{P}_1(t) =f(t)
    \begin{pmatrix}
      256/9\\
      512/9\\
      768/9\\
    \end{pmatrix}
    \frac{\text{N}}{\text{m}^2}= -\vec{P}_2(t),\quad \text{with} \quad f(t)
    = \begin{cases}
      t\quad   & \text{for}\quad t\le 2.5 \text{ s}       \\
      5-t\quad & \text{for}\quad 2.5\text{ s}\leq t \leq5\text{ s} \\
      0\quad & \text{for}\quad t> 5\text{ s}
    \end{cases}\,.
  \end{aligned}
\end{equation}

\begin{figure}[tp]
\begin{minipage}{\textwidth}
\begin{minipage}{0.75\textwidth}
  \begin{tabular}{ccc}
   \includegraphics[width=0.33\textwidth]{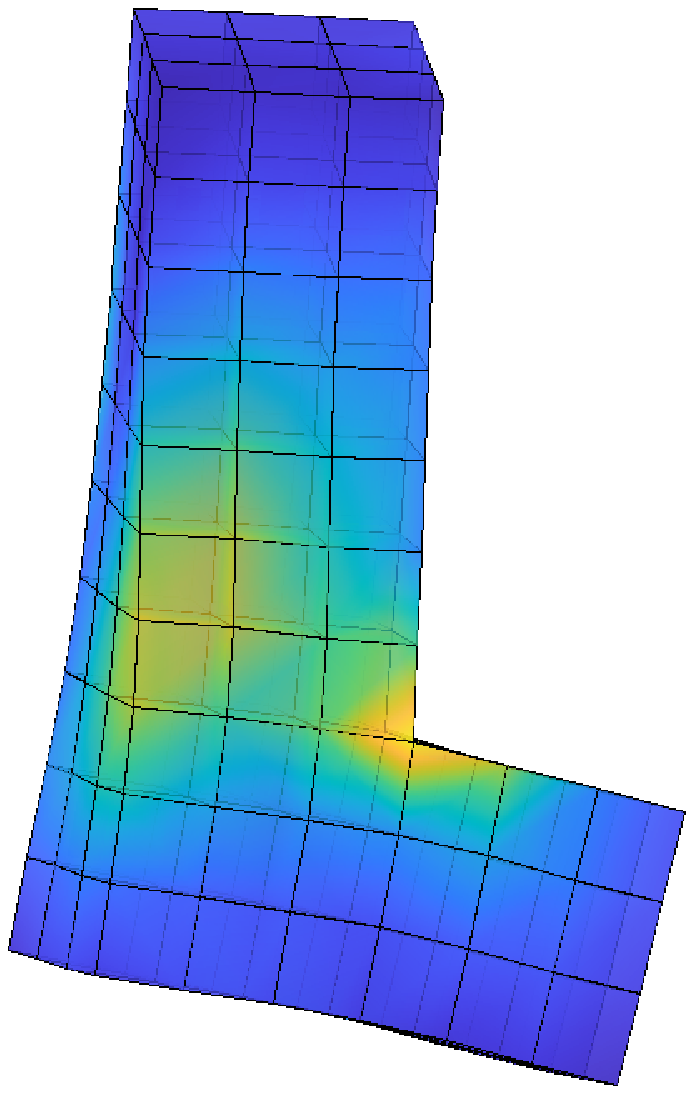}
    &
    \includegraphics[width=0.33\textwidth]{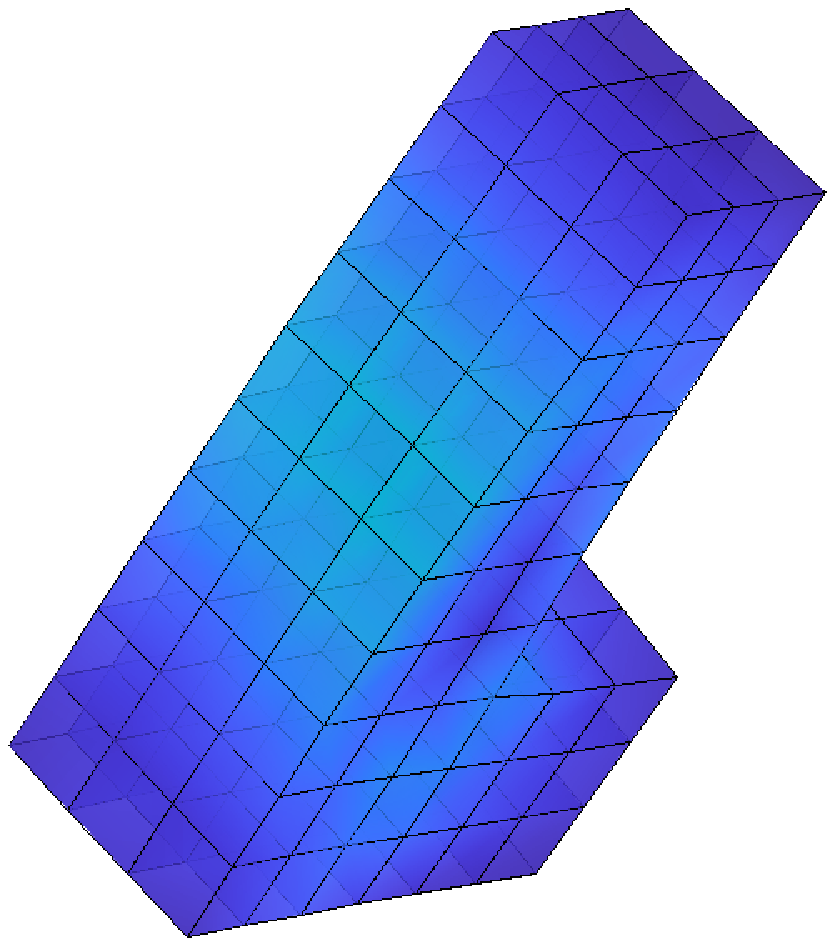}
    &
    \includegraphics[width=0.33\textwidth]{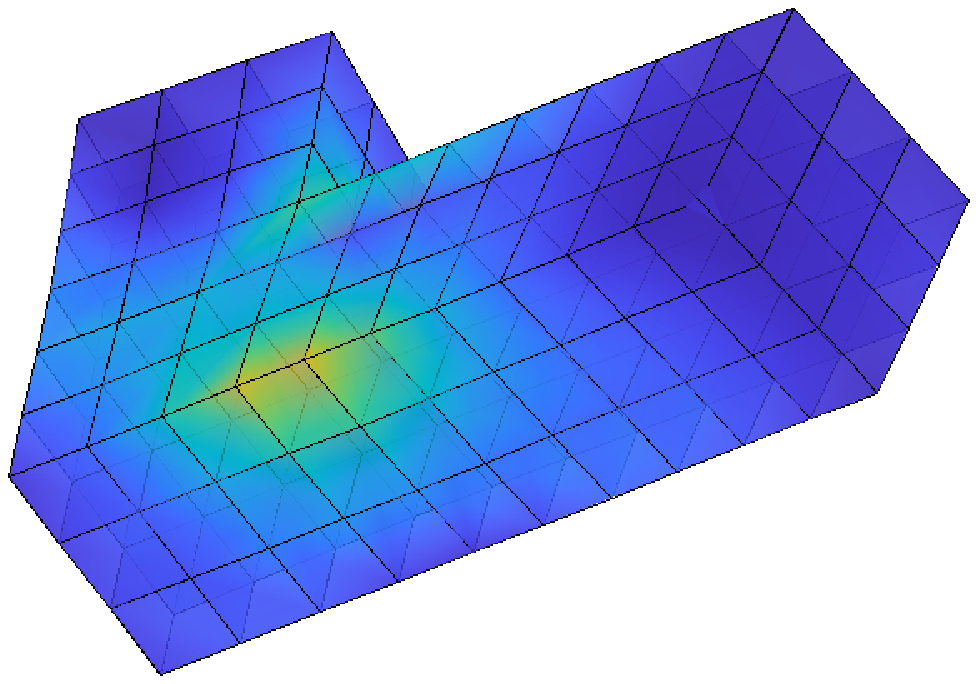}\\
    \includegraphics[width=0.33\textwidth]{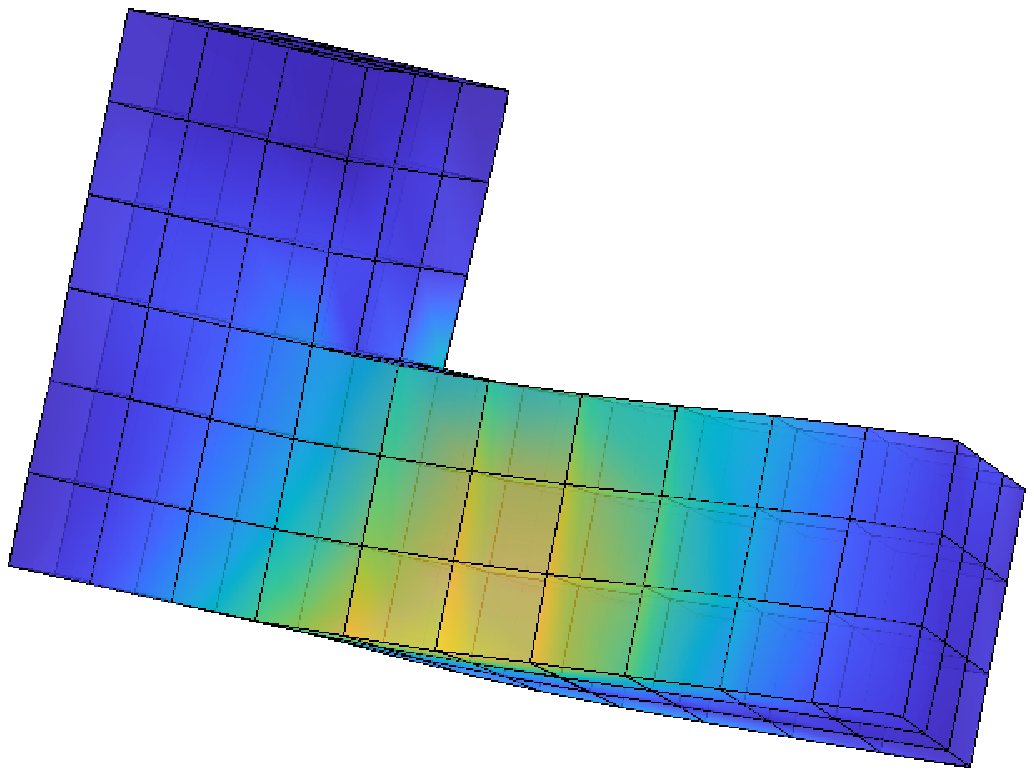}
    &
    \includegraphics[width=0.33\textwidth]{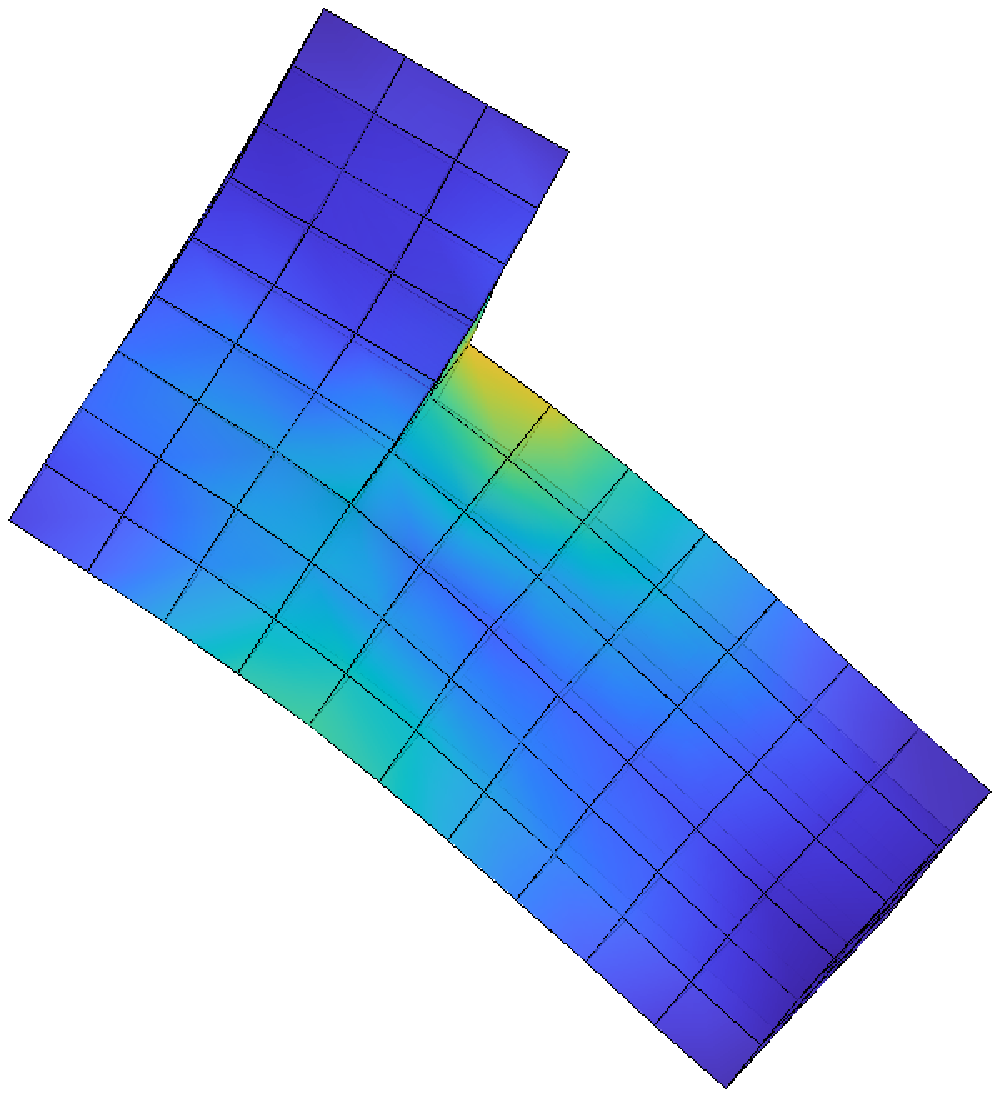}
    &
    \includegraphics[width=0.33\textwidth]{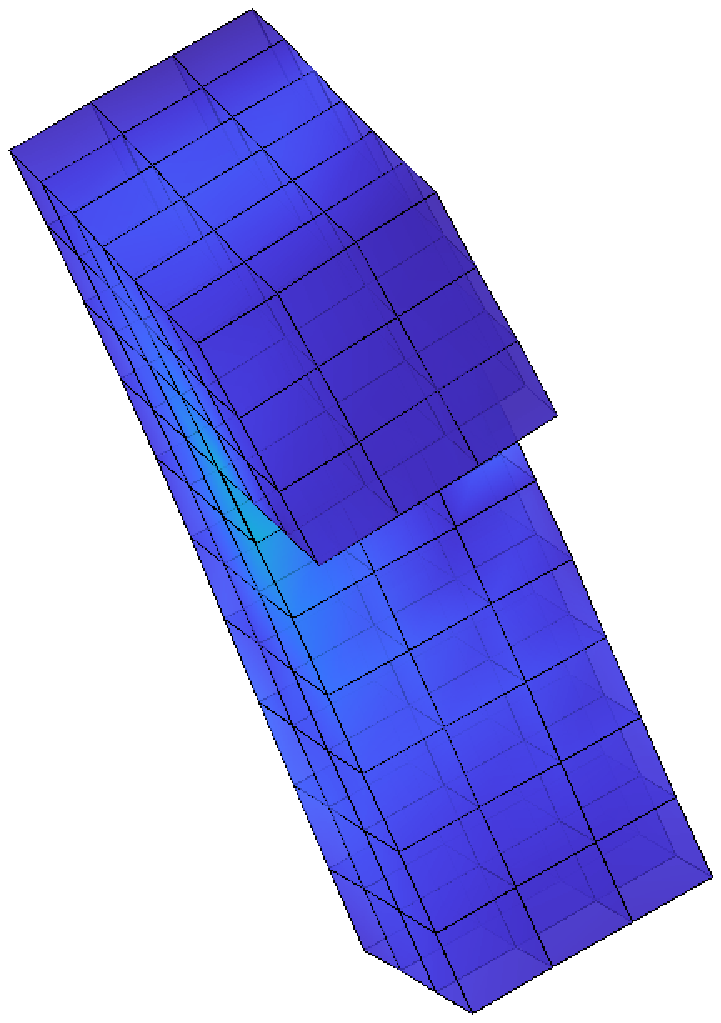}
    \end{tabular}
\end{minipage}
\begin{minipage}{0.225\textwidth}
    \setlength{\figH}{0.1\textheight}
   \setlength{\figW}{0.1\textwidth}
%
%
\begin{tikzpicture}

\begin{axis}[%
width=0.581\figW,
height=\figH,
at={(0\figW,0\figH)},
scale only axis,
point meta min=0,
point meta max=600,
xmin=-4,
xmax=10,
ymin=-4,
ymax=14,
axis line style={draw=none},
ticks=none,
axis x line*=bottom,
axis y line*=left,
colormap={mymap}{[1pt] rgb(0pt)=(0.2422,0.1504,0.6603); rgb(1pt)=(0.2444,0.1534,0.6728); rgb(2pt)=(0.2464,0.1569,0.6847); rgb(3pt)=(0.2484,0.1607,0.6961); rgb(4pt)=(0.2503,0.1648,0.7071); rgb(5pt)=(0.2522,0.1689,0.7179); rgb(6pt)=(0.254,0.1732,0.7286); rgb(7pt)=(0.2558,0.1773,0.7393); rgb(8pt)=(0.2576,0.1814,0.7501); rgb(9pt)=(0.2594,0.1854,0.761); rgb(11pt)=(0.2628,0.1932,0.7828); rgb(12pt)=(0.2645,0.1972,0.7937); rgb(13pt)=(0.2661,0.2011,0.8043); rgb(14pt)=(0.2676,0.2052,0.8148); rgb(15pt)=(0.2691,0.2094,0.8249); rgb(16pt)=(0.2704,0.2138,0.8346); rgb(17pt)=(0.2717,0.2184,0.8439); rgb(18pt)=(0.2729,0.2231,0.8528); rgb(19pt)=(0.274,0.228,0.8612); rgb(20pt)=(0.2749,0.233,0.8692); rgb(21pt)=(0.2758,0.2382,0.8767); rgb(22pt)=(0.2766,0.2435,0.884); rgb(23pt)=(0.2774,0.2489,0.8908); rgb(24pt)=(0.2781,0.2543,0.8973); rgb(25pt)=(0.2788,0.2598,0.9035); rgb(26pt)=(0.2794,0.2653,0.9094); rgb(27pt)=(0.2798,0.2708,0.915); rgb(28pt)=(0.2802,0.2764,0.9204); rgb(29pt)=(0.2806,0.2819,0.9255); rgb(30pt)=(0.2809,0.2875,0.9305); rgb(31pt)=(0.2811,0.293,0.9352); rgb(32pt)=(0.2813,0.2985,0.9397); rgb(33pt)=(0.2814,0.304,0.9441); rgb(34pt)=(0.2814,0.3095,0.9483); rgb(35pt)=(0.2813,0.315,0.9524); rgb(36pt)=(0.2811,0.3204,0.9563); rgb(37pt)=(0.2809,0.3259,0.96); rgb(38pt)=(0.2807,0.3313,0.9636); rgb(39pt)=(0.2803,0.3367,0.967); rgb(40pt)=(0.2798,0.3421,0.9702); rgb(41pt)=(0.2791,0.3475,0.9733); rgb(42pt)=(0.2784,0.3529,0.9763); rgb(43pt)=(0.2776,0.3583,0.9791); rgb(44pt)=(0.2766,0.3638,0.9817); rgb(45pt)=(0.2754,0.3693,0.984); rgb(46pt)=(0.2741,0.3748,0.9862); rgb(47pt)=(0.2726,0.3804,0.9881); rgb(48pt)=(0.271,0.386,0.9898); rgb(49pt)=(0.2691,0.3916,0.9912); rgb(50pt)=(0.267,0.3973,0.9924); rgb(51pt)=(0.2647,0.403,0.9935); rgb(52pt)=(0.2621,0.4088,0.9946); rgb(53pt)=(0.2591,0.4145,0.9955); rgb(54pt)=(0.2556,0.4203,0.9965); rgb(55pt)=(0.2517,0.4261,0.9974); rgb(56pt)=(0.2473,0.4319,0.9983); rgb(57pt)=(0.2424,0.4378,0.9991); rgb(58pt)=(0.2369,0.4437,0.9996); rgb(59pt)=(0.2311,0.4497,0.9995); rgb(60pt)=(0.225,0.4559,0.9985); rgb(61pt)=(0.2189,0.462,0.9968); rgb(62pt)=(0.2128,0.4682,0.9948); rgb(63pt)=(0.2066,0.4743,0.9926); rgb(64pt)=(0.2006,0.4803,0.9906); rgb(65pt)=(0.195,0.4861,0.9887); rgb(66pt)=(0.1903,0.4919,0.9867); rgb(67pt)=(0.1869,0.4975,0.9844); rgb(68pt)=(0.1847,0.503,0.9819); rgb(69pt)=(0.1831,0.5084,0.9793); rgb(70pt)=(0.1818,0.5138,0.9766); rgb(71pt)=(0.1806,0.5191,0.9738); rgb(72pt)=(0.1795,0.5244,0.9709); rgb(73pt)=(0.1785,0.5296,0.9677); rgb(74pt)=(0.1778,0.5349,0.9641); rgb(75pt)=(0.1773,0.5401,0.9602); rgb(76pt)=(0.1768,0.5452,0.956); rgb(77pt)=(0.1764,0.5504,0.9516); rgb(78pt)=(0.1755,0.5554,0.9473); rgb(79pt)=(0.174,0.5605,0.9432); rgb(80pt)=(0.1716,0.5655,0.9393); rgb(81pt)=(0.1686,0.5705,0.9357); rgb(82pt)=(0.1649,0.5755,0.9323); rgb(83pt)=(0.161,0.5805,0.9289); rgb(84pt)=(0.1573,0.5854,0.9254); rgb(85pt)=(0.154,0.5902,0.9218); rgb(86pt)=(0.1513,0.595,0.9182); rgb(87pt)=(0.1492,0.5997,0.9147); rgb(88pt)=(0.1475,0.6043,0.9113); rgb(89pt)=(0.1461,0.6089,0.908); rgb(90pt)=(0.1446,0.6135,0.905); rgb(91pt)=(0.1429,0.618,0.9022); rgb(92pt)=(0.1408,0.6226,0.8998); rgb(93pt)=(0.1383,0.6272,0.8975); rgb(94pt)=(0.1354,0.6317,0.8953); rgb(95pt)=(0.1321,0.6363,0.8932); rgb(96pt)=(0.1288,0.6408,0.891); rgb(97pt)=(0.1253,0.6453,0.8887); rgb(98pt)=(0.1219,0.6497,0.8862); rgb(99pt)=(0.1185,0.6541,0.8834); rgb(100pt)=(0.1152,0.6584,0.8804); rgb(101pt)=(0.1119,0.6627,0.877); rgb(102pt)=(0.1085,0.6669,0.8734); rgb(103pt)=(0.1048,0.671,0.8695); rgb(104pt)=(0.1009,0.675,0.8653); rgb(105pt)=(0.0964,0.6789,0.8609); rgb(106pt)=(0.0914,0.6828,0.8562); rgb(107pt)=(0.0855,0.6865,0.8513); rgb(108pt)=(0.0789,0.6902,0.8462); rgb(109pt)=(0.0713,0.6938,0.8409); rgb(110pt)=(0.0628,0.6972,0.8355); rgb(111pt)=(0.0535,0.7006,0.8299); rgb(112pt)=(0.0433,0.7039,0.8242); rgb(113pt)=(0.0328,0.7071,0.8183); rgb(114pt)=(0.0234,0.7103,0.8124); rgb(115pt)=(0.0155,0.7133,0.8064); rgb(116pt)=(0.0091,0.7163,0.8003); rgb(117pt)=(0.0046,0.7192,0.7941); rgb(118pt)=(0.0019,0.722,0.7878); rgb(119pt)=(0.0009,0.7248,0.7815); rgb(120pt)=(0.0018,0.7275,0.7752); rgb(121pt)=(0.0046,0.7301,0.7688); rgb(122pt)=(0.0094,0.7327,0.7623); rgb(123pt)=(0.0162,0.7352,0.7558); rgb(124pt)=(0.0253,0.7376,0.7492); rgb(125pt)=(0.0369,0.74,0.7426); rgb(126pt)=(0.0504,0.7423,0.7359); rgb(127pt)=(0.0638,0.7446,0.7292); rgb(128pt)=(0.077,0.7468,0.7224); rgb(129pt)=(0.0899,0.7489,0.7156); rgb(130pt)=(0.1023,0.751,0.7088); rgb(131pt)=(0.1141,0.7531,0.7019); rgb(132pt)=(0.1252,0.7552,0.695); rgb(133pt)=(0.1354,0.7572,0.6881); rgb(134pt)=(0.1448,0.7593,0.6812); rgb(135pt)=(0.1532,0.7614,0.6741); rgb(136pt)=(0.1609,0.7635,0.6671); rgb(137pt)=(0.1678,0.7656,0.6599); rgb(138pt)=(0.1741,0.7678,0.6527); rgb(139pt)=(0.1799,0.7699,0.6454); rgb(140pt)=(0.1853,0.7721,0.6379); rgb(141pt)=(0.1905,0.7743,0.6303); rgb(142pt)=(0.1954,0.7765,0.6225); rgb(143pt)=(0.2003,0.7787,0.6146); rgb(144pt)=(0.2061,0.7808,0.6065); rgb(145pt)=(0.2118,0.7828,0.5983); rgb(146pt)=(0.2178,0.7849,0.5899); rgb(147pt)=(0.2244,0.7869,0.5813); rgb(148pt)=(0.2318,0.7887,0.5725); rgb(149pt)=(0.2401,0.7905,0.5636); rgb(150pt)=(0.2491,0.7922,0.5546); rgb(151pt)=(0.2589,0.7937,0.5454); rgb(152pt)=(0.2695,0.7951,0.536); rgb(153pt)=(0.2809,0.7964,0.5266); rgb(154pt)=(0.2929,0.7975,0.517); rgb(155pt)=(0.3052,0.7985,0.5074); rgb(156pt)=(0.3176,0.7994,0.4975); rgb(157pt)=(0.3301,0.8002,0.4876); rgb(158pt)=(0.3424,0.8009,0.4774); rgb(159pt)=(0.3548,0.8016,0.4669); rgb(160pt)=(0.3671,0.8021,0.4563); rgb(161pt)=(0.3795,0.8026,0.4454); rgb(162pt)=(0.3921,0.8029,0.4344); rgb(163pt)=(0.405,0.8031,0.4233); rgb(164pt)=(0.4184,0.803,0.4122); rgb(165pt)=(0.4322,0.8028,0.4013); rgb(166pt)=(0.4463,0.8024,0.3904); rgb(167pt)=(0.4608,0.8018,0.3797); rgb(168pt)=(0.4753,0.8011,0.3691); rgb(169pt)=(0.4899,0.8002,0.3586); rgb(170pt)=(0.5044,0.7993,0.348); rgb(171pt)=(0.5187,0.7982,0.3374); rgb(172pt)=(0.5329,0.797,0.3267); rgb(173pt)=(0.547,0.7957,0.3159); rgb(175pt)=(0.5748,0.7929,0.2941); rgb(176pt)=(0.5886,0.7913,0.2833); rgb(177pt)=(0.6024,0.7896,0.2726); rgb(178pt)=(0.6161,0.7878,0.2622); rgb(179pt)=(0.6297,0.7859,0.2521); rgb(180pt)=(0.6433,0.7839,0.2423); rgb(181pt)=(0.6567,0.7818,0.2329); rgb(182pt)=(0.6701,0.7796,0.2239); rgb(183pt)=(0.6833,0.7773,0.2155); rgb(184pt)=(0.6963,0.775,0.2075); rgb(185pt)=(0.7091,0.7727,0.1998); rgb(186pt)=(0.7218,0.7703,0.1924); rgb(187pt)=(0.7344,0.7679,0.1852); rgb(188pt)=(0.7468,0.7654,0.1782); rgb(189pt)=(0.759,0.7629,0.1717); rgb(190pt)=(0.771,0.7604,0.1658); rgb(191pt)=(0.7829,0.7579,0.1608); rgb(192pt)=(0.7945,0.7554,0.157); rgb(193pt)=(0.806,0.7529,0.1546); rgb(194pt)=(0.8172,0.7505,0.1535); rgb(195pt)=(0.8281,0.7481,0.1536); rgb(196pt)=(0.8389,0.7457,0.1546); rgb(197pt)=(0.8495,0.7435,0.1564); rgb(198pt)=(0.86,0.7413,0.1587); rgb(199pt)=(0.8703,0.7392,0.1615); rgb(200pt)=(0.8804,0.7372,0.165); rgb(201pt)=(0.8903,0.7353,0.1695); rgb(202pt)=(0.9,0.7336,0.1749); rgb(203pt)=(0.9093,0.7321,0.1815); rgb(204pt)=(0.9184,0.7308,0.189); rgb(205pt)=(0.9272,0.7298,0.1973); rgb(206pt)=(0.9357,0.729,0.2061); rgb(207pt)=(0.944,0.7285,0.2151); rgb(208pt)=(0.9523,0.7284,0.2237); rgb(209pt)=(0.9606,0.7285,0.2312); rgb(210pt)=(0.9689,0.7292,0.2373); rgb(211pt)=(0.977,0.7304,0.2418); rgb(212pt)=(0.9842,0.733,0.2446); rgb(213pt)=(0.99,0.7365,0.2429); rgb(214pt)=(0.9946,0.7407,0.2394); rgb(215pt)=(0.9966,0.7458,0.2351); rgb(216pt)=(0.9971,0.7513,0.2309); rgb(217pt)=(0.9972,0.7569,0.2267); rgb(218pt)=(0.9971,0.7626,0.2224); rgb(219pt)=(0.9969,0.7683,0.2181); rgb(220pt)=(0.9966,0.774,0.2138); rgb(221pt)=(0.9962,0.7798,0.2095); rgb(222pt)=(0.9957,0.7856,0.2053); rgb(223pt)=(0.9949,0.7915,0.2012); rgb(224pt)=(0.9938,0.7974,0.1974); rgb(225pt)=(0.9923,0.8034,0.1939); rgb(226pt)=(0.9906,0.8095,0.1906); rgb(227pt)=(0.9885,0.8156,0.1875); rgb(228pt)=(0.9861,0.8218,0.1846); rgb(229pt)=(0.9835,0.828,0.1817); rgb(230pt)=(0.9807,0.8342,0.1787); rgb(231pt)=(0.9778,0.8404,0.1757); rgb(232pt)=(0.9748,0.8467,0.1726); rgb(233pt)=(0.972,0.8529,0.1695); rgb(234pt)=(0.9694,0.8591,0.1665); rgb(235pt)=(0.9671,0.8654,0.1636); rgb(236pt)=(0.9651,0.8716,0.1608); rgb(237pt)=(0.9634,0.8778,0.1582); rgb(238pt)=(0.9619,0.884,0.1557); rgb(239pt)=(0.9608,0.8902,0.1532); rgb(240pt)=(0.9601,0.8963,0.1507); rgb(241pt)=(0.9596,0.9023,0.148); rgb(242pt)=(0.9595,0.9084,0.145); rgb(243pt)=(0.9597,0.9143,0.1418); rgb(244pt)=(0.9601,0.9203,0.1382); rgb(245pt)=(0.9608,0.9262,0.1344); rgb(246pt)=(0.9618,0.932,0.1304); rgb(247pt)=(0.9629,0.9379,0.1261); rgb(248pt)=(0.9642,0.9437,0.1216); rgb(249pt)=(0.9657,0.9494,0.1168); rgb(250pt)=(0.9674,0.9552,0.1116); rgb(251pt)=(0.9692,0.9609,0.1061); rgb(252pt)=(0.9711,0.9667,0.1001); rgb(253pt)=(0.973,0.9724,0.0938); rgb(254pt)=(0.9749,0.9782,0.0872); rgb(255pt)=(0.9769,0.9839,0.0805)},
colorbar,
    colorbar style={
        height=4.5cm}
]
\end{axis}

\begin{axis}[%
width=1.227\figW,
height=1.227\figH,
at={(-0.305\figW,-0.135\figH)},
scale only axis,
point meta min=0,
point meta max=1,
xmin=0,
xmax=1,
ymin=0,
ymax=1,
axis line style={draw=none},
ticks=none,
axis x line*=bottom,
axis y line*=left
]
\end{axis}
\end{tikzpicture}%
\end{minipage}
\end{minipage}
  \caption{Snapshots with von Mises stress distribution \(\sigma_{vM}\) [Pa] of the GT model at times \(t\in\{4.0, 8.0, 12.0, 16.0, 20.0, 24.0\}s\).}
  \label{figure:LShapedSnapshotsGT}
\end{figure}

\begin{figure}[tp]
\begin{minipage}{\textwidth}
\begin{minipage}{0.75\textwidth}
  \begin{tabular}{ccc}
   \includegraphics[width=0.33\textwidth]{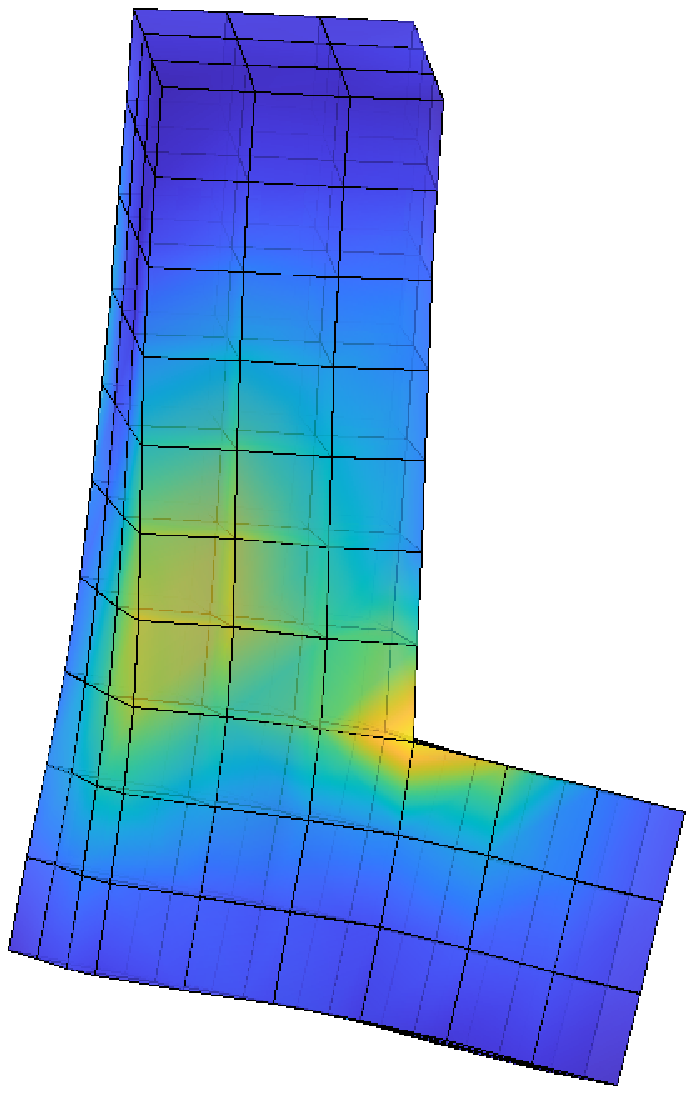}
    &
    \includegraphics[width=0.33\textwidth]{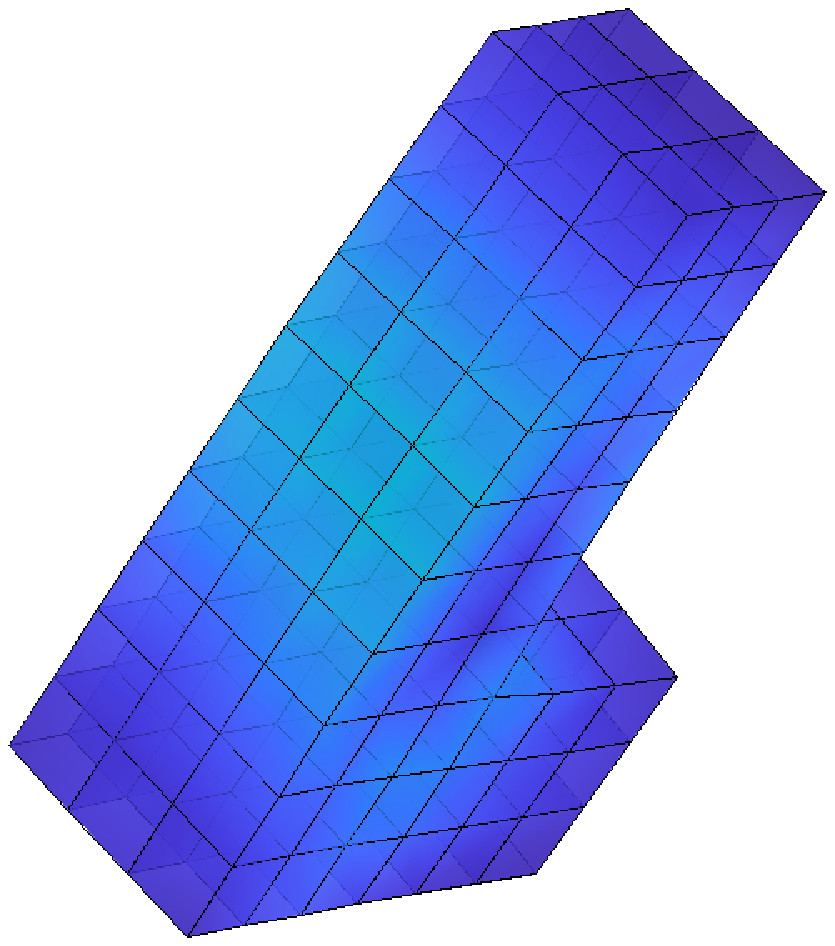}
    &
    \includegraphics[width=0.33\textwidth]{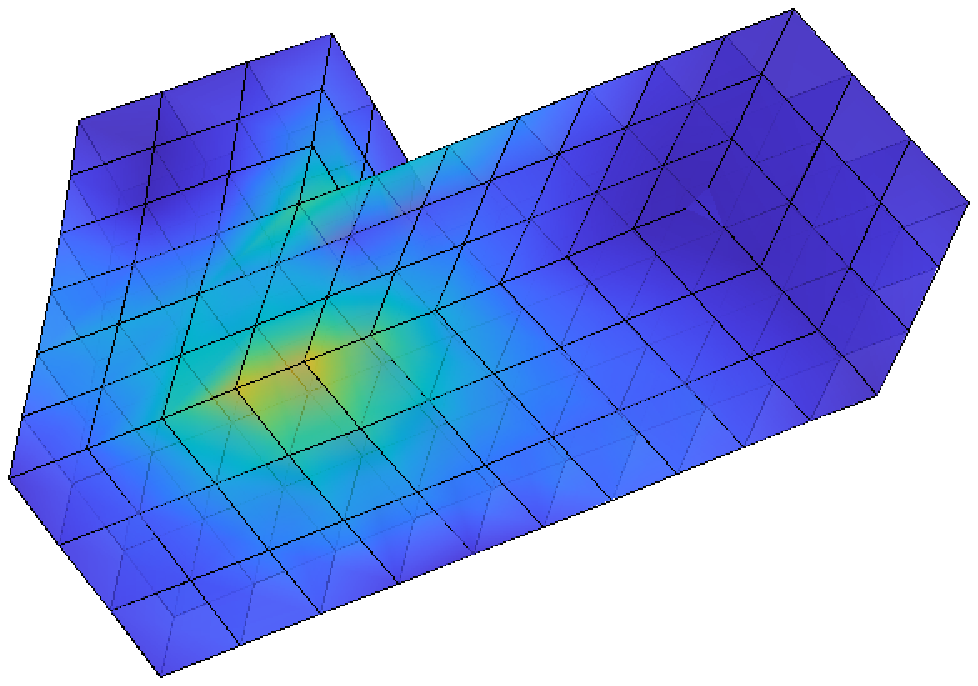}
    \\
    \includegraphics[width=0.33\textwidth]{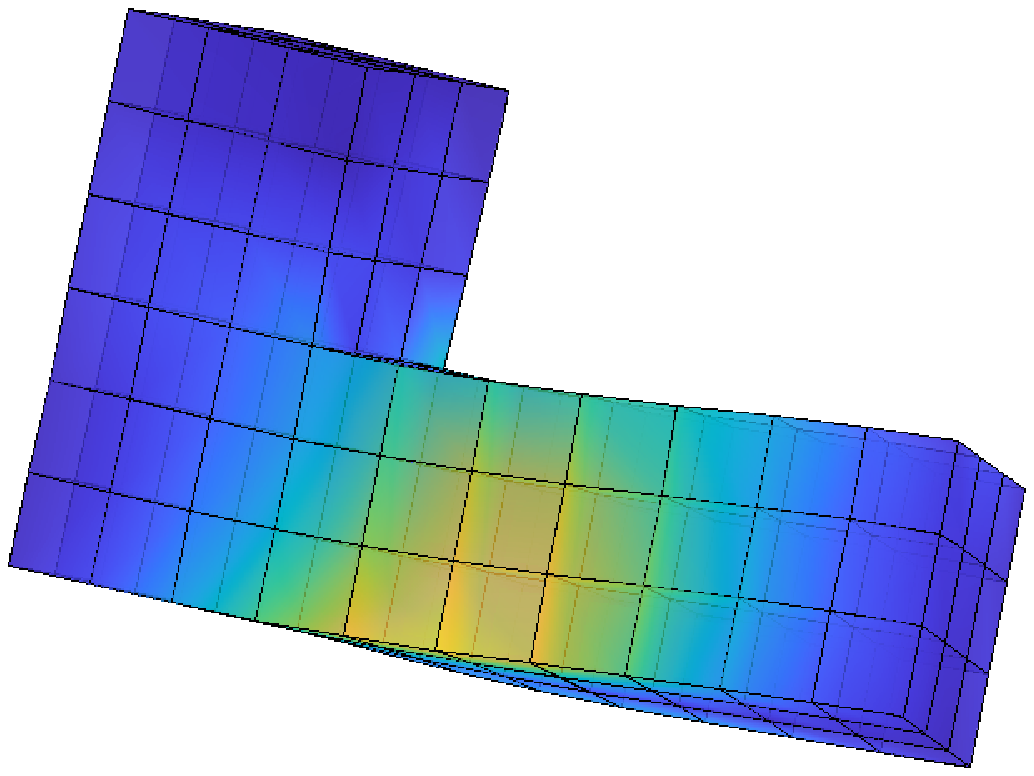}
    &
    \includegraphics[width=0.33\textwidth]{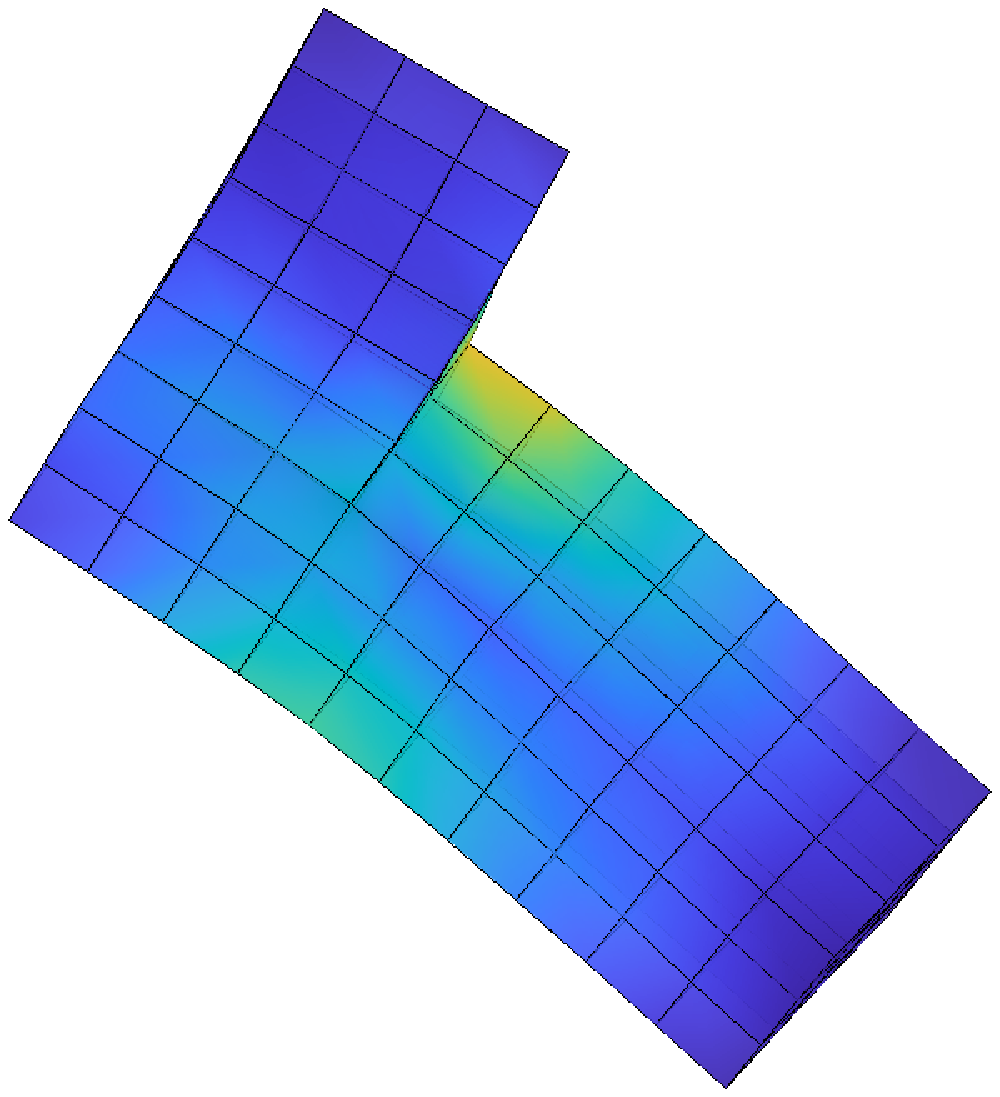}
    &
    \includegraphics[width=0.33\textwidth]{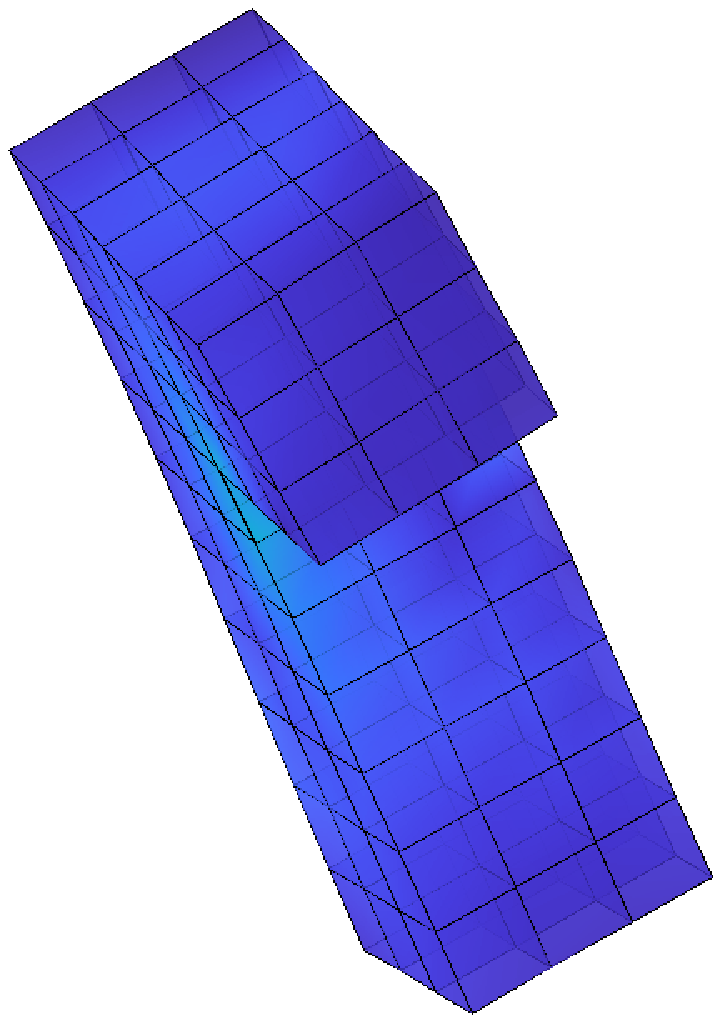}
    \end{tabular}
\end{minipage}
\begin{minipage}{0.225\textwidth}
    \setlength{\figH}{0.1\textheight}
   \setlength{\figW}{0.1\textwidth}
%
%
\begin{tikzpicture}

\begin{axis}[%
width=0.581\figW,
height=\figH,
at={(0\figW,0\figH)},
scale only axis,
point meta min=0,
point meta max=600,
xmin=-4,
xmax=10,
ymin=-4,
ymax=14,
axis line style={draw=none},
ticks=none,
axis x line*=bottom,
axis y line*=left,
colormap={mymap}{[1pt] rgb(0pt)=(0.2422,0.1504,0.6603); rgb(1pt)=(0.2444,0.1534,0.6728); rgb(2pt)=(0.2464,0.1569,0.6847); rgb(3pt)=(0.2484,0.1607,0.6961); rgb(4pt)=(0.2503,0.1648,0.7071); rgb(5pt)=(0.2522,0.1689,0.7179); rgb(6pt)=(0.254,0.1732,0.7286); rgb(7pt)=(0.2558,0.1773,0.7393); rgb(8pt)=(0.2576,0.1814,0.7501); rgb(9pt)=(0.2594,0.1854,0.761); rgb(11pt)=(0.2628,0.1932,0.7828); rgb(12pt)=(0.2645,0.1972,0.7937); rgb(13pt)=(0.2661,0.2011,0.8043); rgb(14pt)=(0.2676,0.2052,0.8148); rgb(15pt)=(0.2691,0.2094,0.8249); rgb(16pt)=(0.2704,0.2138,0.8346); rgb(17pt)=(0.2717,0.2184,0.8439); rgb(18pt)=(0.2729,0.2231,0.8528); rgb(19pt)=(0.274,0.228,0.8612); rgb(20pt)=(0.2749,0.233,0.8692); rgb(21pt)=(0.2758,0.2382,0.8767); rgb(22pt)=(0.2766,0.2435,0.884); rgb(23pt)=(0.2774,0.2489,0.8908); rgb(24pt)=(0.2781,0.2543,0.8973); rgb(25pt)=(0.2788,0.2598,0.9035); rgb(26pt)=(0.2794,0.2653,0.9094); rgb(27pt)=(0.2798,0.2708,0.915); rgb(28pt)=(0.2802,0.2764,0.9204); rgb(29pt)=(0.2806,0.2819,0.9255); rgb(30pt)=(0.2809,0.2875,0.9305); rgb(31pt)=(0.2811,0.293,0.9352); rgb(32pt)=(0.2813,0.2985,0.9397); rgb(33pt)=(0.2814,0.304,0.9441); rgb(34pt)=(0.2814,0.3095,0.9483); rgb(35pt)=(0.2813,0.315,0.9524); rgb(36pt)=(0.2811,0.3204,0.9563); rgb(37pt)=(0.2809,0.3259,0.96); rgb(38pt)=(0.2807,0.3313,0.9636); rgb(39pt)=(0.2803,0.3367,0.967); rgb(40pt)=(0.2798,0.3421,0.9702); rgb(41pt)=(0.2791,0.3475,0.9733); rgb(42pt)=(0.2784,0.3529,0.9763); rgb(43pt)=(0.2776,0.3583,0.9791); rgb(44pt)=(0.2766,0.3638,0.9817); rgb(45pt)=(0.2754,0.3693,0.984); rgb(46pt)=(0.2741,0.3748,0.9862); rgb(47pt)=(0.2726,0.3804,0.9881); rgb(48pt)=(0.271,0.386,0.9898); rgb(49pt)=(0.2691,0.3916,0.9912); rgb(50pt)=(0.267,0.3973,0.9924); rgb(51pt)=(0.2647,0.403,0.9935); rgb(52pt)=(0.2621,0.4088,0.9946); rgb(53pt)=(0.2591,0.4145,0.9955); rgb(54pt)=(0.2556,0.4203,0.9965); rgb(55pt)=(0.2517,0.4261,0.9974); rgb(56pt)=(0.2473,0.4319,0.9983); rgb(57pt)=(0.2424,0.4378,0.9991); rgb(58pt)=(0.2369,0.4437,0.9996); rgb(59pt)=(0.2311,0.4497,0.9995); rgb(60pt)=(0.225,0.4559,0.9985); rgb(61pt)=(0.2189,0.462,0.9968); rgb(62pt)=(0.2128,0.4682,0.9948); rgb(63pt)=(0.2066,0.4743,0.9926); rgb(64pt)=(0.2006,0.4803,0.9906); rgb(65pt)=(0.195,0.4861,0.9887); rgb(66pt)=(0.1903,0.4919,0.9867); rgb(67pt)=(0.1869,0.4975,0.9844); rgb(68pt)=(0.1847,0.503,0.9819); rgb(69pt)=(0.1831,0.5084,0.9793); rgb(70pt)=(0.1818,0.5138,0.9766); rgb(71pt)=(0.1806,0.5191,0.9738); rgb(72pt)=(0.1795,0.5244,0.9709); rgb(73pt)=(0.1785,0.5296,0.9677); rgb(74pt)=(0.1778,0.5349,0.9641); rgb(75pt)=(0.1773,0.5401,0.9602); rgb(76pt)=(0.1768,0.5452,0.956); rgb(77pt)=(0.1764,0.5504,0.9516); rgb(78pt)=(0.1755,0.5554,0.9473); rgb(79pt)=(0.174,0.5605,0.9432); rgb(80pt)=(0.1716,0.5655,0.9393); rgb(81pt)=(0.1686,0.5705,0.9357); rgb(82pt)=(0.1649,0.5755,0.9323); rgb(83pt)=(0.161,0.5805,0.9289); rgb(84pt)=(0.1573,0.5854,0.9254); rgb(85pt)=(0.154,0.5902,0.9218); rgb(86pt)=(0.1513,0.595,0.9182); rgb(87pt)=(0.1492,0.5997,0.9147); rgb(88pt)=(0.1475,0.6043,0.9113); rgb(89pt)=(0.1461,0.6089,0.908); rgb(90pt)=(0.1446,0.6135,0.905); rgb(91pt)=(0.1429,0.618,0.9022); rgb(92pt)=(0.1408,0.6226,0.8998); rgb(93pt)=(0.1383,0.6272,0.8975); rgb(94pt)=(0.1354,0.6317,0.8953); rgb(95pt)=(0.1321,0.6363,0.8932); rgb(96pt)=(0.1288,0.6408,0.891); rgb(97pt)=(0.1253,0.6453,0.8887); rgb(98pt)=(0.1219,0.6497,0.8862); rgb(99pt)=(0.1185,0.6541,0.8834); rgb(100pt)=(0.1152,0.6584,0.8804); rgb(101pt)=(0.1119,0.6627,0.877); rgb(102pt)=(0.1085,0.6669,0.8734); rgb(103pt)=(0.1048,0.671,0.8695); rgb(104pt)=(0.1009,0.675,0.8653); rgb(105pt)=(0.0964,0.6789,0.8609); rgb(106pt)=(0.0914,0.6828,0.8562); rgb(107pt)=(0.0855,0.6865,0.8513); rgb(108pt)=(0.0789,0.6902,0.8462); rgb(109pt)=(0.0713,0.6938,0.8409); rgb(110pt)=(0.0628,0.6972,0.8355); rgb(111pt)=(0.0535,0.7006,0.8299); rgb(112pt)=(0.0433,0.7039,0.8242); rgb(113pt)=(0.0328,0.7071,0.8183); rgb(114pt)=(0.0234,0.7103,0.8124); rgb(115pt)=(0.0155,0.7133,0.8064); rgb(116pt)=(0.0091,0.7163,0.8003); rgb(117pt)=(0.0046,0.7192,0.7941); rgb(118pt)=(0.0019,0.722,0.7878); rgb(119pt)=(0.0009,0.7248,0.7815); rgb(120pt)=(0.0018,0.7275,0.7752); rgb(121pt)=(0.0046,0.7301,0.7688); rgb(122pt)=(0.0094,0.7327,0.7623); rgb(123pt)=(0.0162,0.7352,0.7558); rgb(124pt)=(0.0253,0.7376,0.7492); rgb(125pt)=(0.0369,0.74,0.7426); rgb(126pt)=(0.0504,0.7423,0.7359); rgb(127pt)=(0.0638,0.7446,0.7292); rgb(128pt)=(0.077,0.7468,0.7224); rgb(129pt)=(0.0899,0.7489,0.7156); rgb(130pt)=(0.1023,0.751,0.7088); rgb(131pt)=(0.1141,0.7531,0.7019); rgb(132pt)=(0.1252,0.7552,0.695); rgb(133pt)=(0.1354,0.7572,0.6881); rgb(134pt)=(0.1448,0.7593,0.6812); rgb(135pt)=(0.1532,0.7614,0.6741); rgb(136pt)=(0.1609,0.7635,0.6671); rgb(137pt)=(0.1678,0.7656,0.6599); rgb(138pt)=(0.1741,0.7678,0.6527); rgb(139pt)=(0.1799,0.7699,0.6454); rgb(140pt)=(0.1853,0.7721,0.6379); rgb(141pt)=(0.1905,0.7743,0.6303); rgb(142pt)=(0.1954,0.7765,0.6225); rgb(143pt)=(0.2003,0.7787,0.6146); rgb(144pt)=(0.2061,0.7808,0.6065); rgb(145pt)=(0.2118,0.7828,0.5983); rgb(146pt)=(0.2178,0.7849,0.5899); rgb(147pt)=(0.2244,0.7869,0.5813); rgb(148pt)=(0.2318,0.7887,0.5725); rgb(149pt)=(0.2401,0.7905,0.5636); rgb(150pt)=(0.2491,0.7922,0.5546); rgb(151pt)=(0.2589,0.7937,0.5454); rgb(152pt)=(0.2695,0.7951,0.536); rgb(153pt)=(0.2809,0.7964,0.5266); rgb(154pt)=(0.2929,0.7975,0.517); rgb(155pt)=(0.3052,0.7985,0.5074); rgb(156pt)=(0.3176,0.7994,0.4975); rgb(157pt)=(0.3301,0.8002,0.4876); rgb(158pt)=(0.3424,0.8009,0.4774); rgb(159pt)=(0.3548,0.8016,0.4669); rgb(160pt)=(0.3671,0.8021,0.4563); rgb(161pt)=(0.3795,0.8026,0.4454); rgb(162pt)=(0.3921,0.8029,0.4344); rgb(163pt)=(0.405,0.8031,0.4233); rgb(164pt)=(0.4184,0.803,0.4122); rgb(165pt)=(0.4322,0.8028,0.4013); rgb(166pt)=(0.4463,0.8024,0.3904); rgb(167pt)=(0.4608,0.8018,0.3797); rgb(168pt)=(0.4753,0.8011,0.3691); rgb(169pt)=(0.4899,0.8002,0.3586); rgb(170pt)=(0.5044,0.7993,0.348); rgb(171pt)=(0.5187,0.7982,0.3374); rgb(172pt)=(0.5329,0.797,0.3267); rgb(173pt)=(0.547,0.7957,0.3159); rgb(175pt)=(0.5748,0.7929,0.2941); rgb(176pt)=(0.5886,0.7913,0.2833); rgb(177pt)=(0.6024,0.7896,0.2726); rgb(178pt)=(0.6161,0.7878,0.2622); rgb(179pt)=(0.6297,0.7859,0.2521); rgb(180pt)=(0.6433,0.7839,0.2423); rgb(181pt)=(0.6567,0.7818,0.2329); rgb(182pt)=(0.6701,0.7796,0.2239); rgb(183pt)=(0.6833,0.7773,0.2155); rgb(184pt)=(0.6963,0.775,0.2075); rgb(185pt)=(0.7091,0.7727,0.1998); rgb(186pt)=(0.7218,0.7703,0.1924); rgb(187pt)=(0.7344,0.7679,0.1852); rgb(188pt)=(0.7468,0.7654,0.1782); rgb(189pt)=(0.759,0.7629,0.1717); rgb(190pt)=(0.771,0.7604,0.1658); rgb(191pt)=(0.7829,0.7579,0.1608); rgb(192pt)=(0.7945,0.7554,0.157); rgb(193pt)=(0.806,0.7529,0.1546); rgb(194pt)=(0.8172,0.7505,0.1535); rgb(195pt)=(0.8281,0.7481,0.1536); rgb(196pt)=(0.8389,0.7457,0.1546); rgb(197pt)=(0.8495,0.7435,0.1564); rgb(198pt)=(0.86,0.7413,0.1587); rgb(199pt)=(0.8703,0.7392,0.1615); rgb(200pt)=(0.8804,0.7372,0.165); rgb(201pt)=(0.8903,0.7353,0.1695); rgb(202pt)=(0.9,0.7336,0.1749); rgb(203pt)=(0.9093,0.7321,0.1815); rgb(204pt)=(0.9184,0.7308,0.189); rgb(205pt)=(0.9272,0.7298,0.1973); rgb(206pt)=(0.9357,0.729,0.2061); rgb(207pt)=(0.944,0.7285,0.2151); rgb(208pt)=(0.9523,0.7284,0.2237); rgb(209pt)=(0.9606,0.7285,0.2312); rgb(210pt)=(0.9689,0.7292,0.2373); rgb(211pt)=(0.977,0.7304,0.2418); rgb(212pt)=(0.9842,0.733,0.2446); rgb(213pt)=(0.99,0.7365,0.2429); rgb(214pt)=(0.9946,0.7407,0.2394); rgb(215pt)=(0.9966,0.7458,0.2351); rgb(216pt)=(0.9971,0.7513,0.2309); rgb(217pt)=(0.9972,0.7569,0.2267); rgb(218pt)=(0.9971,0.7626,0.2224); rgb(219pt)=(0.9969,0.7683,0.2181); rgb(220pt)=(0.9966,0.774,0.2138); rgb(221pt)=(0.9962,0.7798,0.2095); rgb(222pt)=(0.9957,0.7856,0.2053); rgb(223pt)=(0.9949,0.7915,0.2012); rgb(224pt)=(0.9938,0.7974,0.1974); rgb(225pt)=(0.9923,0.8034,0.1939); rgb(226pt)=(0.9906,0.8095,0.1906); rgb(227pt)=(0.9885,0.8156,0.1875); rgb(228pt)=(0.9861,0.8218,0.1846); rgb(229pt)=(0.9835,0.828,0.1817); rgb(230pt)=(0.9807,0.8342,0.1787); rgb(231pt)=(0.9778,0.8404,0.1757); rgb(232pt)=(0.9748,0.8467,0.1726); rgb(233pt)=(0.972,0.8529,0.1695); rgb(234pt)=(0.9694,0.8591,0.1665); rgb(235pt)=(0.9671,0.8654,0.1636); rgb(236pt)=(0.9651,0.8716,0.1608); rgb(237pt)=(0.9634,0.8778,0.1582); rgb(238pt)=(0.9619,0.884,0.1557); rgb(239pt)=(0.9608,0.8902,0.1532); rgb(240pt)=(0.9601,0.8963,0.1507); rgb(241pt)=(0.9596,0.9023,0.148); rgb(242pt)=(0.9595,0.9084,0.145); rgb(243pt)=(0.9597,0.9143,0.1418); rgb(244pt)=(0.9601,0.9203,0.1382); rgb(245pt)=(0.9608,0.9262,0.1344); rgb(246pt)=(0.9618,0.932,0.1304); rgb(247pt)=(0.9629,0.9379,0.1261); rgb(248pt)=(0.9642,0.9437,0.1216); rgb(249pt)=(0.9657,0.9494,0.1168); rgb(250pt)=(0.9674,0.9552,0.1116); rgb(251pt)=(0.9692,0.9609,0.1061); rgb(252pt)=(0.9711,0.9667,0.1001); rgb(253pt)=(0.973,0.9724,0.0938); rgb(254pt)=(0.9749,0.9782,0.0872); rgb(255pt)=(0.9769,0.9839,0.0805)},
colorbar,
    colorbar style={
        height=4.5cm}
]
\end{axis}

\begin{axis}[%
width=1.227\figW,
height=1.227\figH,
at={(-0.305\figW,-0.135\figH)},
scale only axis,
point meta min=0,
point meta max=1,
xmin=0,
xmax=1,
ymin=0,
ymax=1,
axis line style={draw=none},
ticks=none,
axis x line*=bottom,
axis y line*=left
]
\end{axis}
\end{tikzpicture}%
\end{minipage}    
\end{minipage}    
  \caption{Snapshots with von Mises stress distribution \(\sigma_{vM}\) [Pa] of the PANN8 model at times \(t\in\{4.0, 8.0, 12.0, 16.0, 20.0, 24.0\}s\).}
  \label{figure:LShapedSnapshotsANN}
\end{figure}

\begin{figure}[tp]
\center
    \setlength{\figH}{0.2\textheight}
    \setlength{\figW}{0.3\textwidth}
    \input{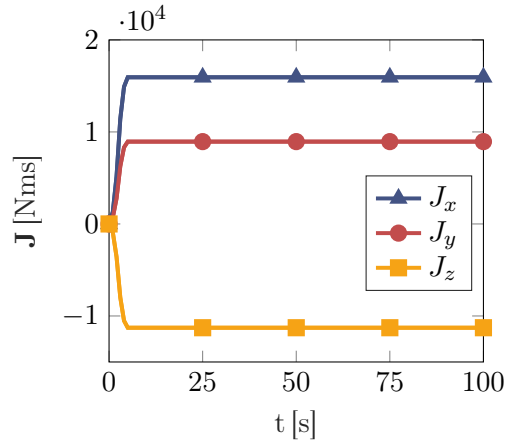}
  \caption{Components of angular momentum of the L-shaped body for the GT model (marks) and for the PANN8 model (continuous lines), both in case of the energy momentum scheme.}
  \label{figure:LShapedBodyAngularMomentum}
\end{figure}

In \cref{figure:LShapedSnapshotsGT} and \cref{figure:LShapedSnapshotsANN} snapshots of the motion of the L-shaped body with von Mises stress distribution for the compressible GT and PANN8 models, respectively, are shown. As can be observed, both stress plots show a remarkable similarity. 
The total energy and the energy difference of both the GT model and corresponding PANN8 model are shown in \cref{figure:LShapedBodyEnergyGT} and \cref{figure:LShapedBodyEnergyANN}, respectively. 
Apparently, the midpoint rule aborts due to energy blow ups for both GT and PANN8 model, even when the time step size is reduced. 
In contrast, the EMS is stable for both the GT and PANN8 model during this long-term simulation and preserves the total energy after the loading phase (\(t>5\) s). 
Furthermore, it is important to note that the energy results obtained for the GT and the PANN8 model are practically identical. 
Furthermore, in \cref{figure:LShapedBodyAngularMomentum} the components of the angular momentum of the L-shaped body for both the GT and the PANN8 model in case of the EMS are shown. 
As evident from the observation, the components of the angular momentum are preserved after the loading phase (\(t>5\) s). 
Similar to the energy results, the components of the angular momentum exhibit practically identical results for both models. 


\begin{figure}[tp]
  \begin{tabular}{cc}
    \setlength{\figH}{0.2\textheight}
    \setlength{\figW}{0.3\textwidth}
    \input{figures/LShape/energyMooneyRivlin.tikz}
   &
    \setlength{\figH}{0.2\textheight}
    \setlength{\figW}{0.3\textwidth}
%
%
\definecolor{mycolor1}{rgb}{0.00000,0.44700,0.74100}%
\begin{tikzpicture}

\begin{axis}[%
width=\figW,
height=\figH,
at={(0\figW,0\figH)},
scale only axis,
xmin=0,
xmax=200,
xlabel style={font=\color{white!15!black}},
xlabel={t\,[s]},
ymin=-1e-06,
ymax=1e-06,
ylabel style={font=\color{white!15!black}},
ylabel={\(E_{n+1}-E_{n}\)\,[J]},
axis background/.style={fill=white}
]
\addplot [color=CPSdarkblue, forget plot, line width=1.5pt]
  table[row sep=crcr]{%
6	-2.9967850423418e-10\\
7	-4.09272615797818e-11\\
8	-9.09494701772928e-13\\
9	0\\
10	-5.45696821063757e-12\\
11	-2.95585778076202e-11\\
12	-7.27595761418343e-12\\
13	-2.72848410531878e-12\\
14	-1.59161572810262e-11\\
15	1.45519152283669e-11\\
16	-6.82121026329696e-12\\
17	-2.04636307898909e-11\\
18	-1.31276465253904e-08\\
19	-4.54747350886464e-13\\
20	-2.15095496969298e-09\\
21	-1.2801137927454e-09\\
22	-8.18545231595635e-12\\
23	2.04636307898909e-11\\
24	-2.89219315163791e-09\\
25	-4.50199877377599e-11\\
26	-8.3718987298198e-10\\
27	-9.36506694415584e-09\\
28	-4.54747350886464e-13\\
29	-7.27595761418343e-12\\
30	7.49878381611779e-10\\
31	-3.81487552658655e-09\\
32	1.31649358081631e-09\\
33	-6.13181327935308e-08\\
34	5.00222085975111e-12\\
35	-3.18323145620525e-12\\
36	-9.09494701772928e-13\\
37	-8.91486706677824e-09\\
38	-4.41559677710757e-10\\
39	-4.83487383462489e-09\\
40	-9.09494701772928e-13\\
41	-1.36424205265939e-12\\
42	2.27373675443232e-12\\
43	-2.27373675443232e-12\\
44	-2.85590431303717e-08\\
45	-9.63609636528417e-10\\
46	-2.5061126507353e-09\\
47	-4.54747350886464e-13\\
48	-5.91171556152403e-12\\
49	1.36424205265939e-12\\
50	-5.27234078617766e-09\\
51	-6.01175997871906e-09\\
52	-5.78458639211021e-07\\
53	0\\
54	-3.63797880709171e-12\\
55	-3.36513039655983e-11\\
56	4.54747350886464e-13\\
57	-3.8835423765704e-09\\
58	-7.85712472861633e-09\\
59	-9.5357700047316e-07\\
60	1.2732925824821e-11\\
61	-1.04591890703887e-11\\
62	-9.09494701772928e-13\\
63	1.81898940354586e-12\\
64	5.91171556152403e-12\\
65	-2.04596290132031e-07\\
66	-3.36513039655983e-11\\
67	-9.64064383879304e-11\\
68	8.45830072648823e-11\\
69	-4.54747350886464e-12\\
70	0\\
71	-7.59882823331282e-10\\
72	-5.31281330040656e-09\\
73	-6.92989488015883e-09\\
74	-9.09494701772928e-13\\
75	-3.63797880709171e-12\\
76	-4.54747350886464e-13\\
77	2.27373675443232e-12\\
78	-3.03771230392158e-10\\
79	-8.64019966684282e-12\\
80	-5.27788870385848e-08\\
81	3.91082721762359e-11\\
82	-4.54747350886464e-13\\
83	4.54747350886464e-13\\
84	1.81898940354586e-12\\
85	-2.72848410531878e-12\\
86	-1.36424205265939e-12\\
87	-2.00088834390044e-11\\
88	-4.6429704525508e-10\\
89	-1.04591890703887e-11\\
90	9.09494701772928e-12\\
91	-4.27462509833276e-11\\
92	1.36424205265939e-12\\
93	4.54747350886464e-13\\
94	-1.81898940354586e-12\\
95	-6.86668499838561e-11\\
96	-2.27373675443232e-12\\
97	-4.63796823169105e-09\\
98	1.8007995095104e-10\\
99	-1.33877620100975e-09\\
100	0\\
101	-1.81898940354586e-12\\
102	-2.68300937023014e-11\\
103	0\\
104	-1.63863660418428e-08\\
105	-3.27418092638254e-11\\
106	-1.21008270070888e-09\\
107	1.36424205265939e-12\\
108	-2.45563569478691e-11\\
109	-6.23003870714456e-11\\
110	-1.00044417195022e-10\\
111	5.50244294572622e-11\\
112	-3.05681169265881e-08\\
113	-6.3664629124105e-12\\
114	4.54747350886464e-13\\
115	-4.54747350886464e-13\\
116	-7.36690708436072e-11\\
117	5.30690158484504e-10\\
118	-1.26551185530843e-07\\
119	-4.38258211943321e-08\\
120	-3.44243744621053e-10\\
121	-4.54747350886464e-13\\
122	-7.73070496506989e-12\\
123	-9.09494701772928e-13\\
124	1.08684616861865e-10\\
125	-4.22026460000779e-07\\
126	-2.41107045440003e-09\\
127	-4.27917257184163e-10\\
128	-3.18323145620525e-12\\
129	2.72848410531878e-12\\
130	-1.31876731757075e-11\\
131	-9.27684595808387e-11\\
132	-1.04150785773527e-08\\
133	-5.77529135625809e-11\\
134	-5.91171556152403e-12\\
135	-1.81898940354586e-12\\
136	-1.36424205265939e-12\\
137	-1.93267624126747e-10\\
138	-2.27373675443232e-12\\
139	-1.71849023899995e-09\\
140	-4.39285940956324e-10\\
141	-4.54747350886464e-13\\
142	9.09494701772928e-13\\
143	-7.04858393874019e-11\\
144	-1.09594111563638e-10\\
145	-1.36424205265939e-12\\
146	-2.39260771195404e-08\\
147	-5.50244294572622e-11\\
148	9.09494701772928e-13\\
149	4.54747350886464e-13\\
150	-3.99722921429202e-10\\
151	-2.72848410531878e-12\\
152	-2.33603714150377e-09\\
153	-1.30557964439504e-09\\
154	-8.64019966684282e-12\\
155	1.22054188977927e-09\\
156	-2.65890776063316e-09\\
157	0\\
158	-1.07138475868851e-09\\
159	-2.72848410531878e-12\\
160	-6.11180439591408e-10\\
161	9.09494701772928e-13\\
162	-3.18323145620525e-12\\
163	-1.81898940354586e-12\\
164	0\\
165	4.54747350886464e-13\\
166	-5.54791768081486e-11\\
167	-2.86490831058472e-11\\
168	0\\
169	-9.09494701772928e-13\\
170	1.09139364212751e-11\\
171	-4.65206539956853e-10\\
172	-4.09272615797818e-12\\
173	1.36424205265939e-12\\
174	-4.54747350886464e-13\\
175	-3.2287061912939e-11\\
176	4.54747350886464e-13\\
177	-7.6397554948926e-11\\
178	-1.79429662239272e-08\\
179	-2.46973286266439e-09\\
180	-8.32187652122229e-11\\
181	-2.77395884040743e-09\\
182	-7.24412529962137e-10\\
183	6.82121026329696e-12\\
184	-6.49379217065871e-09\\
185	-4.34238245361485e-09\\
186	-7.27595761418343e-12\\
187	1.09139364212751e-11\\
188	-9.72113411989994e-09\\
189	4.54747350886464e-13\\
190	0\\
191	-6.66659616399556e-09\\
192	-4.50199877377599e-11\\
193	-6.41193764749914e-11\\
194	-2.72848410531878e-12\\
195	-9.09494701772928e-12\\
196	2.72848410531878e-12\\
197	-1.84809323400259e-09\\
198	-3.06590663967654e-09\\
199	-2.34194885706529e-10\\
};
\end{axis}
\end{tikzpicture}%
 \end{tabular}
  \caption{Total energy of the L-shaped body (left) for the energy-momentum scheme (EM) and the midpoint rule (MP), and energy difference (right) for the EM scheme. Evaluated for the GT model.}
  \label{figure:LShapedBodyEnergyGT}
\end{figure}

\begin{figure}[tp]
  \begin{tabular}{cc}
    \setlength{\figH}{0.2\textheight}
    \setlength{\figW}{0.3\textwidth}
    \input{figures/LShape/energyANNMooneyRivlin.tikz}
   &
    \setlength{\figH}{0.2\textheight}
    \setlength{\figW}{0.3\textwidth}
%
%
\definecolor{mycolor1}{rgb}{0.00000,0.44700,0.74100}%
\begin{tikzpicture}

\begin{axis}[%
width=\figW,
height=\figH,
at={(0\figW,0\figH)},
scale only axis,
xmin=0,
xmax=200,
xlabel style={font=\color{white!15!black}},
xlabel={t\,[s]},
ymin=-1e-06,
ymax=1e-06,
ylabel style={font=\color{white!15!black}},
ylabel={\(E_{n+1}-E_{n}\)\,[J]},
axis background/.style={fill=white}
]
\addplot [color=CPSdarkblue, forget plot, line width=1.5pt]
  table[row sep=crcr]{%
6	-1.16415321826935e-10\\
7	-2.27373675443232e-12\\
8	5.00222085975111e-12\\
9	2.72848410531878e-12\\
10	-2.27373675443232e-12\\
11	-1.04591890703887e-11\\
12	-3.18323145620525e-12\\
13	0\\
14	-1.36424205265939e-12\\
15	6.82121026329696e-12\\
16	1.04591890703887e-11\\
17	-1.31876731757075e-11\\
18	-4.36148184235208e-09\\
19	6.82121026329696e-12\\
20	-4.59294824395329e-10\\
21	-1.82353687705472e-10\\
22	-9.09494701772928e-13\\
23	7.27595761418343e-12\\
24	-6.73026079311967e-10\\
25	-2.91038304567337e-11\\
26	-7.91260390542448e-11\\
27	-2.79442247119732e-09\\
28	1.9475146473269e-07\\
29	-3.63797880709171e-12\\
30	1.60071067512035e-10\\
31	-1.34605215862393e-09\\
32	6.7257133196108e-10\\
33	-1.74891283677425e-08\\
34	4.54747350886464e-12\\
35	1.09139364212751e-11\\
36	-2.72848410531878e-12\\
37	-2.17596607399173e-09\\
38	-5.45696821063757e-11\\
39	-1.34286892716773e-09\\
40	-1.00044417195022e-11\\
41	9.09494701772928e-13\\
42	4.54747350886464e-12\\
43	1.09139364212751e-11\\
44	-5.83531800657511e-09\\
45	-1.88720150617883e-10\\
46	-7.36235961085185e-10\\
47	-6.3664629124105e-12\\
48	-6.3664629124105e-12\\
49	-2.27373675443232e-12\\
50	-1.91766957868822e-09\\
51	2.36241248785518e-09\\
52	-1.46580077853287e-07\\
53	-1.81898940354586e-12\\
54	-3.63797880709171e-12\\
55	2.72848410531878e-12\\
56	2.27373675443232e-12\\
57	-1.22054188977927e-09\\
58	-7.36235961085185e-10\\
59	-2.20318270294229e-07\\
60	1.04591890703887e-11\\
61	-1.31876731757075e-11\\
62	2.27373675443232e-12\\
63	-1.04591890703887e-11\\
64	8.18545231595635e-12\\
65	-6.09625203651376e-08\\
66	-1.63709046319127e-11\\
67	-4.86579665448517e-11\\
68	4.59294824395329e-11\\
69	-1.81898940354586e-12\\
70	-9.09494701772928e-13\\
71	-3.30601324094459e-10\\
72	-1.37833922053687e-09\\
73	-2.06728145712987e-09\\
74	9.09494701772928e-13\\
75	4.09272615797818e-12\\
76	5.45696821063757e-12\\
77	9.09494701772928e-12\\
78	-1.37333699967712e-10\\
79	-3.63797880709171e-12\\
80	-1.75259629031643e-08\\
81	9.09494701772928e-12\\
82	9.09494701772928e-13\\
83	7.27595761418343e-12\\
84	-1.81898940354586e-12\\
85	-4.54747350886464e-12\\
86	1.81898940354586e-12\\
87	-9.09494701772928e-12\\
88	-1.8781065591611e-10\\
89	-4.54747350886464e-13\\
90	-1.00044417195022e-11\\
91	-1.45519152283669e-11\\
92	8.18545231595635e-12\\
93	4.54747350886464e-12\\
94	5.45696821063757e-12\\
95	-1.31876731757075e-11\\
96	5.00222085975111e-12\\
97	-1.58115653903224e-09\\
98	7.04858393874019e-11\\
99	-4.6429704525508e-10\\
100	1.54614099301398e-11\\
101	-8.18545231595635e-12\\
102	-8.64019966684282e-12\\
103	-1.45519152283669e-11\\
104	-5.09862729813904e-09\\
105	-1.40971678774804e-11\\
106	-4.7066350816749e-10\\
107	-6.3664629124105e-12\\
108	-9.09494701772928e-12\\
109	-1.45519152283669e-11\\
110	-3.77440301235765e-11\\
111	2.95585778076202e-11\\
112	-1.02554622571915e-08\\
113	-6.82121026329696e-12\\
114	5.91171556152403e-12\\
115	0\\
116	-3.27418092638254e-11\\
117	1.47792889038101e-10\\
118	-3.52097231370863e-08\\
119	-8.48831405164674e-09\\
120	-1.4279066817835e-10\\
121	7.73070496506989e-12\\
122	-1.13686837721616e-11\\
123	4.54747350886464e-13\\
124	2.77395884040743e-11\\
125	-9.91553861240391e-08\\
126	-4.97038854518905e-10\\
127	-1.67801772477105e-10\\
128	1.81898940354586e-12\\
129	-1.36424205265939e-12\\
130	-9.09494701772928e-12\\
131	-9.09494701772928e-12\\
132	-3.10910763801076e-09\\
133	-2.22826201934367e-11\\
134	4.09272615797818e-12\\
135	-3.63797880709171e-12\\
136	1.13686837721616e-11\\
137	-6.63931132294238e-11\\
138	-3.63797880709171e-12\\
139	-5.36147126695141e-10\\
140	-1.1186784831807e-10\\
141	-1.81898940354586e-12\\
142	1.04591890703887e-11\\
143	-6.3664629124105e-12\\
144	-2.09183781407773e-11\\
145	2.72848410531878e-12\\
146	-8.17499312688597e-09\\
147	-3.36513039655983e-11\\
148	5.00222085975111e-12\\
149	4.54747350886464e-12\\
150	-9.09494701772928e-11\\
151	5.45696821063757e-12\\
152	-8.82664608070627e-10\\
153	-4.74301486974582e-10\\
154	-3.18323145620525e-12\\
155	5.30690158484504e-10\\
156	-7.85348674980924e-10\\
157	4.54747350886464e-13\\
158	-4.27462509833276e-10\\
159	1.00044417195022e-11\\
160	-3.18777892971411e-10\\
161	1.36424205265939e-12\\
162	6.82121026329696e-12\\
163	9.09494701772928e-13\\
164	7.73070496506989e-12\\
165	6.82121026329696e-12\\
166	-4.45652403868735e-11\\
167	0\\
168	3.18323145620525e-12\\
169	1.04591890703887e-11\\
170	1.40971678774804e-11\\
171	-1.45519152283669e-10\\
172	3.63797880709171e-12\\
173	-1.81898940354586e-12\\
174	0\\
175	-1.90993887372315e-11\\
176	5.00222085975111e-12\\
177	-2.77395884040743e-11\\
178	-2.44972397922538e-09\\
179	-9.14951669983566e-10\\
180	-1.50066625792533e-11\\
181	-6.92580215400085e-10\\
182	-1.88720150617883e-10\\
183	1.09139364212751e-11\\
184	-1.64163793670014e-09\\
185	-9.85892256721854e-10\\
186	3.63797880709171e-12\\
187	1.36424205265939e-12\\
188	-3.56430973624811e-09\\
189	-1.04591890703887e-11\\
190	2.72848410531878e-12\\
191	-1.27238308778033e-09\\
192	-1.68256519827992e-11\\
193	-1.00044417195022e-11\\
194	1.00044417195022e-11\\
195	-6.3664629124105e-12\\
196	-3.18323145620525e-12\\
197	-7.16681824997067e-10\\
198	-3.95630195271224e-10\\
199	-6.3664629124105e-11\\
};
\end{axis}
\end{tikzpicture}%
 \end{tabular}
  \caption{Total energy of the L-shaped body (left) for the energy-momentum scheme (EM) and the midpoint rule (MP), and energy difference (right) for the EM scheme. Evaluated for the PANN8 model.}
  \label{figure:LShapedBodyEnergyANN}
\end{figure}

\section{Conclusion}\label{sec:conc}
In this contribution, advanced spatial and temporal discretization techniques are tailored to hyperelastic PANN constitutive models, i.e., NN-based constitutive models which fulfill all relevant mechanical conditions of hyperelasticity by construction \cite{linden2023}. One of the benefits of PANN-based constitutive models is their extraordinary flexibility, which is achieved by using NNs as highly flexible functions. 
Actually, the feed-forward neural networks applied for this have a quite simple structure, which is based on only a few ingredients. Furthermore, from a formal point of view, NNs are nothing more but mathematical functions, and thus, constitutive models based on NNs can be applied in numerical methods just like their analytical counterparts.
Still, the explicit mathematical formulation of the NN, and even more its derivatives often required for numerical applications, are quite extensive compared to analytical constitutive models. This suggests to tailor advanced discretization methods for NN-based constitutive models, to arrive at compact mathematical formulations and convenient implementations with superior stability and robustness in static and dynamic analysis. 

\medskip

In the present work, this is achieved by setting the focus of the numerical methods on the scalar-valued strain invariants that hyperelastic stored energy functions depend on, rather than considering the full tensorial strain measures.
Thereby, the evaluation of the hyperelastic potential and its derivatives w.r.t.\ the strain invariants is separated from the derivatives of the strain invariants w.r.t.\ the primal tensorial strain measures. This is a very natural choice, since for NN-based constitutive models, the derivatives of the hyperelastic potential w.r.t.\ the invariants become more complex, while the derivatives of the invariants remain the same as in the analytical theory.
In this regard, an elegant 7-field Hu-Washizu like mixed formulation is proposed, where the additional unknowns are only strain invariants. This novel mixed formulation is applicable for nearly incompressible material behavior, in particular, to avoid locking phenomena.
Furthermore, a novel energy-momentum scheme (EMS) is designed, which is applicable in dynamical simulations to preserve energy and momentum in the discretized version of the weak form. At the same time, the EMS highly improves the numerical stability in transient simulations. For this, a new algorithmic stress formula is proposed, which, again, only depends on strain invariants. Here, the required discrete gradients are formulated in scalar-valued invariants, therefore only consisting of Greenspan formulas. The employed Greenspan formulas are by far less complex for both consistent tangent computation and implementation when compared to the usually employed projection-based discrete gradients. 

\medskip

Both the mixed method and the EMS are applied in finite element analysis. For this, the hyperelastic PANN model proposed by \cite{linden2023} is calibrated to data generated with an analytical Mooney-Rivlin potential.
Unsurprisingly, due to the simple nature of the analytical potential, the results obtained with the PANN model are practically identical to the results obtained with the Mooney-Rivlin model in all simulations.
The newly proposed mixed 7-field Hu-Washizu like formulation showed excellent performance in a scenario where purely displacement-based formulations extensively suffer from locking phenomena. Also, it showed similar results when compared to a state-of-the-art mixed, polyconvex 5-field approach \cite{kraus2019} which considers tensor-valued quantities as additional unknowns. Furthermore, in a benchmark example, it could be shown that with a suitable parallelization strategy, the computation time of the NN-based model is practically the same as for the analytical constitutive model. 
In a dynamical simulation, the tailored EMS showed a remarkable stability and robustness even for coarse time steps when compared to the symplectic midpoint-rule. 
Overall, the numerical examples demonstrate the superior stability and robustness of the proposed numerical methods.

\medskip

Finally, future work should be concerned with the extension of the proposed methods towards multiphysics \cite{klein2022b,zlatic2023}, inelasticity \cite{rosenkranz2023,masi2023,abdolazizi2023}, and structural mechanics. Furthermore, while the present work only considers isotropic material behavior, the extension to further symmetry groups is straightforward \cite{linden2023,Ebbing2010}.

\vspace*{3ex}

\noindent
\textbf{Conflict of interest.} The authors declare that they have no conflict of interest.
\vspace*{1ex}

\noindent
\textbf{Software.}
The source code used for the finite element computations is implemented in \textsc{Matlab} under MIT license and is available at \url{https://github.com/kit-ifm/moofeKIT}.
Version 1.XX of the code, used in this paper, is archived at \cite{moofeKIT}. 
\vspace*{1ex}    

\noindent
\textbf{Acknowledgement.}
M.~Franke acknowledges the financial support provided by the Deutsche Forschungsgemeinschaft (DFG, German Research Foundation, project number 443238377). 
D.K.~Klein and O.~Weeger acknowledge the financial support provided by the Deutsche Forschungsgemeinschaft (DFG, German Research Foundation, project number 492770117), the Graduate School of Computational Engineering at TU Darmstadt, and Hessian.AI. 

\appendix
\numberwithin{equation}{section} 
\section{Continous balance laws}\label{section:balanceLaws}
The verification of the time-continuous balance laws, i.e., balance of angular momentum and balance of total energy for the weak form given in \cref{equation:weak}, will be dealt with in the following. 
\subsection{Balance of angular momentum}
In order to verify the balance of angular momentum, we introduce the angular momentum
\begin{equation}
  \vec{J}_t = \int_{\B}\vec{\varphi}_t\times\rho_0\,\vec{v}_t\,\d{V}\,,
\end{equation}
and the external momentum
\begin{equation}
  \vec{M}^{\text{ext}}_t = \int_{\B}\vec{\varphi}_t\times\vec{B}_t\,\d{V} + \int_{\partial\B^{\subT}}\vec{\varphi}_t\times\vec{T}_t\,\d{A}\,.
\end{equation}
Furthermore, we choose the admissible variations
\begin{equation}\label{equation:variationsAngularMomentum}
\delta\vec{\varphi} = \vec{\xi}\times\vec{\varphi}_t\,,\quad \delta\vec{v} = \vec{\xi}\times\dot{\vec{\varphi}}_t\,,
\end{equation}
where \(\mathds{R}^3\ni\vec{\xi}=\text{const.}\)
For further considerations, we anticipate
\begin{equation}
  \F(\delta\vec{\varphi} = \vec{\xi}\times\vec{\varphi}_t) = \hat{\tens{\xi}}\,\FPhit\,,
\end{equation}
where \(\hat{\tens{\xi}}\) is skew-symmetric, such that
\begin{equation}
  \hat{\tens{\xi}}\,\vec{a} = \vec{\xi}\times\vec{\varphi}_t\quad\forall\vec{a}\in\mathds{R}^3\,.
\end{equation}
Recall \cref{equation:weak}\(_1\) and substitute \cref{equation:variationsAngularMomentum}\(_2\), which finally yields
\begin{equation}\label{equation:weakAngularMomentum1}
  \vec{\xi}\cdot\int_{\B}(\dot{\vec{\varphi}}_t\times\vec{v}_t)\,\rho_0\,\d{V} = 0\,,
\end{equation}
where therein the identity \(\dot{\vec{\varphi}}\times\dot{\vec{\varphi}} = \vec{0}\) has been employed. 
Consider the integrand of the 2nd term in \cref{equation:weak}\(_2\) and substitute \cref{equation:variationsAngularMomentum}\(_1\)
\begin{equation}
  \tens{S}_t:\tfrac{1}{2}\,\delta\CPhit = \tens{S}_t:\frac{1}{2}\,(\FPhit\transp\,(\hat{\tens{\xi}}{\transp}+\hat{\tens{\xi}})\,\FPhit) = 0\,,
\end{equation}
which is due to skew-symmetry of \(\hat{\tens{\xi}}\), i.e.\ \(\hat{\tens{\xi}}{\transp} = -\hat{\tens{\xi}}\).
Recall \cref{equation:weak}\(_2\) and substitute \cref{equation:variationsAngularMomentum}\(_1\)
\begin{equation}
  \vec{\xi}\cdot\int_{\B}\vec{\varphi}_t\times\dot{\vec{v}}_t\,\rho_0\,\d{V} + \Pi^{\text{ext}}(\vec{\xi}\times\vec{\varphi}_t) = 0\,,
\end{equation}
for the above we finally obtain with \cref{equation:weakAngularMomentum1} the desired balance of angular momentum 
\begin{equation}
\begin{aligned}
  \vec{\xi}\cdot\left(\frac{\d}{\d{t}}\,\int_{\B}\vec{\varphi}_t\times\vec{v}_t\,\rho_0\,\d{V} - \int_{\B}\vec{\varphi}_t\times\vec{B}_t\,\d{V}-\int_{\partial\B^{\subT}}\vec{\varphi}_t\times\vec{T}_t\,\d{V}\right) &= 0\nonumber\\
  \Leftrightarrow\quad \vec{\xi}\cdot\left(\frac{\d}{\d{t}}\,\vec{J}_t - \vec{M}^{\text{ext}}_t\right) &= 0\,.
  \end{aligned}
\end{equation}
\subsection{Balance of total energy}
With the definition of the kinetic energy
\begin{equation}
  T_t = \frac{1}{2}\,\int_{\B}\rho_0\,\vec{v}_t\cdot\vec{v}_t\,\d{V}\,,
\end{equation}
at hand, we choose the admissible variations
\begin{equation}\label{equation:variationsTotalEnergy}
  \delta\vec{\varphi} = \dot{\vec{\varphi}}_t,\qquad \delta\vec{v} = \dot{\vec{v}}_t\,,
\end{equation}
and substitute \cref{equation:variationsTotalEnergy}\(_2\) into \cref{equation:weak}\(_1\), such that 
\begin{equation}\label{equation:kineticEnergyBalance}
  \int_{\B}\dot{\vec{\varphi}}_t\cdot\dot{\vec{v}}_t\,\rho_0\,\d{V} = \int_{\B}\vec{v}_t\cdot\dot{\vec{v}}_t\,\rho_0\,\d{V} = \frac{\d}{\d{t}}\,\left(\frac{1}{2}\,\int_{\B}\vec{v}\cdot\vec{v}_t\,\rho_0\,\d{V}\right) = \dot{T}_t\,.
\end{equation}
Consider the derivative w.r.t.\ time of the internal potential \begin{align}\label{equation:potentialEnergyBalance}
  \frac{\d}{\d{t}}\,{}^{\W}\Pi^{\text{int}}_{t} &= \frac{\d}{\d{t}}\,\int_{\B}\W(\IPhit,\IIPhit,\IIIPhit)\,\d{V}\nonumber\\
                                       &= \int_{\B}\partial_{\I}\W\,\tens{I}:\dot{\C}_{t} + \partial_{\II}\W\tens{I}:\dot{\G}_{t} + \partial_{J}\W\,\dot{\JPhit})\,\d{V}\nonumber\\
                                             &= \int_{\B}2\,\left(\partial_{\I}\W\,\tens{I} + \partial_{\II}\W\,\tens{I}\wedge\CPhit + \tfrac{1}{2}\,\partial_{J}\W\,\JPhit^{-1}\,\GPhit\right):\frac{1}{2}\,\dot{\C}_{t}\,\d{V}\nonumber\\
                                             &= \int_{\B}\tens{S}_t:\frac{1}{2}\,\dot{\C}_{t}\,\d{V}\,.
\end{align}
Substitute \cref{equation:variationsTotalEnergy}\(_1\) into \cref{equation:weak}\(_2\) and take \cref{equation:kineticEnergyBalance} and \cref{equation:potentialEnergyBalance} into account, which finally yield the desired balance of total energy 
\begin{equation}
  \frac{\d}{\d{t}}\left(T_t + \Pi^{\text{int}}_t\right) = \dot{E}_t = P^{\text{ext}}_t = -\dot{\Pi}^{\text{ext}}_t\,.
\end{equation}
\section{Time-discrete balance laws}\label{section:semiDiscreteBalanceLaws}
The verification of the time-discrete balance laws, i.e.\ balance of angular momentum and balance of total energy for the time-discrete system given in \cref{equation:semiDiscreteWeak} will be dealt with in the following. 
\subsection{Balance of angular momentum}\label{section:semiDiscreteEnergyBalance}
In order to verify the balance of angular momentum in  the semi-discrete weak form \cref{equation:semiDiscreteWeak}, we choose the admissible variations
\begin{equation}\label{equation:discreteVariationsAngularMomentum}
  \delta\vec{\varphi} = \vec{\xi}\times\vec{\varphi}_{n+\frac{1}{2}},\quad \delta\vec{v} = \vec{\xi}\times\frac{\Delta\vec{\varphi}}{\Delta t}\,,
\end{equation}
where \(\mathds{R}^3\ni\vec{\xi}=\text{const.}\)
Inserting \cref{equation:discreteVariationsAngularMomentum}\(_2\) into \cref{equation:semiDiscreteWeak}\(_1\)
\begin{equation}
  \vec{\xi}\cdot\int_{\B}\frac{1}{\Delta t}\Delta\vec{\varphi}\times\Delta\vec{\varphi}\,\rho_0\,\d{V} = \vec{\xi}\cdot\int_{\B}\frac{\Delta\vec{\varphi}}{\Delta t}\times\vec{v}_{n+\frac{1}{2}}\,\rho_0\,\d{V}\,,
\end{equation}
and exploiting \(\Delta\vec{\varphi}\times\Delta\vec{\varphi}=\vec{0}\), we obtain
\begin{equation}\label{equation:discreteBalanceOfAngularMomentum1}
  \vec{\xi}\cdot\int_{\B}\frac{\Delta\vec{\varphi}}{\Delta t}\times\vec{v}_{n+\frac{1}{2}}\,\rho_0\,\d{V} = 0\,.
\end{equation}
Considering the integrand of the first term of the right-hand side of \cref{equation:semiDiscreteWeak}\(_2\) and plugging in \cref{equation:discreteVariationsAngularMomentum}\(_1\) yields
\begin{equation}
  \tens{S}_{\text{algo}}:\frac{1}{2}\,\F_{{n+\frac{1}{2}}}{\transp}\,(\hat{\tens{\xi}}+\hat{\tens{\xi}}{\transp})\,\F_{{n+\frac{1}{2}}}{\transp} = 0\,.
\end{equation}
With the above in mind, plugging \cref{equation:discreteVariationsAngularMomentum}\(_1\) into the whole \cref{equation:semiDiscreteWeak}\(_2\) and adding \cref{equation:discreteBalanceOfAngularMomentum1}  yields after some algebra the desired semi-discrete balance of angular momentum 
\begin{align}
  \vec{\xi}\cdot\left(\frac{1}{\Delta t}\,\int_{\B}(\Delta\vec{\varphi}\times\vec{v}_{n+\frac{1}{2}} + \vec{\varphi}_{n+\frac{1}{2}}\times\Delta\vec{v})\,\rho_0\,\d{V}\right) &= \vec{\xi}\cdot\left(\int_{\B}\vec{\varphi}_{n+\frac{1}{2}}\times\vec{B}_{{n+\frac{1}{2}}}\,\d{V}\right.\nonumber\\
  &\qquad\quad\left.\: + \int_{\partial\B^{\subT}}\vec{\varphi}_{n+\frac{1}{2}}\times\vec{T}_{{n+\frac{1}{2}}}\,\d{A}\right)\nonumber\\
  \Leftrightarrow\qquad \vec{\xi}\cdot\frac{1}{\Delta t}\,\int_{\B}(\rho_0\,\vec{\varphi}_{n+1}\times\vec{v}_{n+1} - \rho_0\,\vec{\varphi}_{n}\times\vec{v}_{n})\,\d{V} &= \vec{\xi}\cdot\vec{M}_{n,n+1}^{\text{ext}}\nonumber\\
  \Leftrightarrow\qquad \vec{\xi}\cdot\left(\frac{1}{\Delta t}\,\Delta\vec{J} - \vec{M}_{n,n+1}^{\text{ext}}\right) &= 0\,.
\end{align}
Accordingly, the proposed EM scheme consistently reproduces the balance of angular momentum for the time-discrete system which is e.g.\ in contrast to the 2nd order accurate trapezoidal rule. 
\subsection{Balance of total energy}
To verify the balance of energy in the semi-discrete weak form  \cref{equation:semiDiscreteWeak}, we choose admissible variations
\begin{equation}\label{equation:discreteVariationsTotalEnergy}
  \delta\vec{\varphi} = \vec{\varphi}_{n+1} - \vec{\varphi}_n = \Delta\vec{\varphi},\quad \delta\vec{v} = \vec{v}_{n+1} - \vec{v}_n = \Delta\vec{v}\,,
\end{equation}
plug \cref{equation:discreteVariationsTotalEnergy}\(_2\) into \cref{equation:semiDiscreteWeak}\(_1\):
\begin{align}\label{equation:discreteBalanceOfTotalEnergy1}
  \int_{\B} \Delta\vec{\varphi}\,\frac{1}{\Delta t}\,(\vec{\varphi}_{n+1} - \vec{\varphi}_n)\,\rho_0\,\d{V} &= \int_{\B}\Delta\vec{v}\cdot\frac{1}{2}\,(\vec{v}_{n+1}-\vec{v}_n)\,\rho_0\,\d{V}\nonumber\\
                                                                                                            &= \frac{1}{2}\,\int_{\B}\rho_0\,(\vec{v}_{n+1}\cdot\vec{v}_{n+1}-\vec{v}_n\cdot\vec{v}_n)\,\d{V}\nonumber\\
                                                                                                            &= T_{n+1}-T_n = \Delta T\,.
\end{align}
and plug \cref{equation:discreteVariationsTotalEnergy}\(_1\) into the integrand of the first term of the right-hand side of \cref{equation:semiDiscreteWeak}\(_2\): 
\begin{equation}\label{equation:discreteStressPower}
  \tens{S}_{\text{algo}}:\operatorname{sym}(\Delta\F{\transp}\,\F_{{n+\frac{1}{2}}}) =  \tens{S}_{\text{algo}}:(\Delta\F{\transp}\,\F_{{n+\frac{1}{2}}}) = \tens{S}_{\text{algo}}:\frac{1}{2}(\C_{{n+1}} - \C_{{n}}) = \tens{S}_{\text{algo}}:\frac{1}{2}\Delta\C\,.
\end{equation}
Here, the symmetry property \(\mathds{S}^3\cap\mathds{K}^3 = \{0\}\) has been taken into account, where \(\S^3=\{\tens{A}\in\T^3\,\rvert\,\tens{A}=\tens{A}{\transp}\}\) and \(\mathds{K}^3=\{\tens{A}\in\T^3\,\rvert\,\tens{A}=-\tens{A}{\transp}\}\) denote spaces of symmetric and skew-symmetric, quadratic second order tensors, respectively. 
Furthermore, for \cref{equation:discreteStressPower} we insert \cref{equation:algorithmic2ndPKStress}, which yields after some algebra, cf.\ \cref{equation:directionalityProperty}:
\begin{align}
  \tens{S}_{\text{algo}}:\frac{1}{2}\Delta\C &= \left(\D_{\I}\W\,\tens{I} + \D_{\II}\W\,\tens{I}\wedge\C_{\text{algo}} + \tfrac{1}{2}\,\D_{J}\W\,J_{\text{algo}}^{-1}\,\G_{\text{algo}}\right):\Delta\C\nonumber\\
                                                & = \D_{\I}\W\,\tr\Delta\C + \D_{\II}\W\,(\tens{I}\wedge\C_{\text{algo}}):\Delta\C + \tfrac{1}{2}\,\D_{J}\W\,J_{\text{algo}}^{-1}\,\G_{\text{algo}}:\Delta\C\nonumber\\
                                                  & = \D_{\I}\W\,\Delta\I + \D_{\II}\W\,\Delta\II + \D_{J}\W\,\Delta J\,.\label{equation:discreteStressPower2}
\end{align}
Plugging \cref{equation:discreteVariationsTotalEnergy}\(_1\) into the whole \cref{equation:semiDiscreteWeak}\(_2\), considering the above results \cref{equation:discreteBalanceOfTotalEnergy1} and \cref{equation:discreteStressPower2}, and employing the directionality property \cref{equation:directionalityProperty}  yields the desired semi-discrete balance of total energy
\begin{equation}
  \begin{aligned}
    &\Delta T = -\int_{\B}\W(I_{\subC_{{n+1}}},II_{\subC_{{n+1}}},J_{\subC_{{n+1}}})-\W(I_{\subC_{{n}}},II_{\subC_{{n}}},J_{\subC_{{n}}})\,\d{V} + W_{n,n+1}^{\text{ext}}\nonumber\\
    \Leftrightarrow\qquad & \Delta T + \Delta{}^{\W}\Pi^{\text{int}} = W_{n,n+1}^{\text{ext}}\,,
  \end{aligned}
\end{equation}
where
\begin{equation}
  \begin{aligned}
    W_{n,n+1}^{\text{ext}} &= -\left(\Pi_{n+\frac{1}{2}}^{\text{ext}}(\vec{\varphi}_{n+1}) - \Pi_{n+\frac{1}{2}}^{\text{ext}}(\vec{\varphi}_n)\right)\\
    &= -\Pi_{n+\frac{1}{2}}^{\text{ext}}(\Delta\vec{\varphi}) = \int_{\B}\Delta\vec{\varphi}\cdot\vec{B}_{{n+\frac{1}{2}}}\,\d{V} + \int_{\partial\B^{\subT}}\Delta\vec{\varphi}\cdot\vec{T}_{{n+\frac{1}{2}}}\,\d{A}\,.
  \end{aligned}
\end{equation}
As shown above the proposed EM scheme consistently reproduces the balance of total energy for the time-discrete system. 
This achievement can be attributed to the algorithmic stress evaluation, as demonstrated in \cref{equation:discreteStressPower2}. 
In contrast, standard integrators such as the 2nd order accurate midpoint rule fail to maintain this discrete energy balance.    
\section{Partioned discrete gradients of \(\W\)}\label{appendix:discreteGradient}
Introducing the tuple \(\tens{\Pi} = (\Pi^1, \Pi^2, \Pi^3) = (\I, \II, J)\), the partitioned discrete gradients \(\D_{\I}\W,\) \(\D_{\II}\W,\) and \(\D_{J}\W\) with final presentation in \cref{equation:discreteGradientsIIIJ} are computed with 
\begin{equation}\label{equation:formulaDiscreteGradientA}
\begin{aligned}
  D_{\Pi^i}\W = \frac{1}{2}\,(\D_{\Pi^i}\W|_{\Pi^J_{n+1},\Pi_n^K} + \D_{\Pi^i}\W|_{\Pi^J_{n},\Pi_{n+1}^K})\,,\\
  \forall i\in Y,\quad j\in J<Y>K\ni k\: : \: J\ni j<i<k\in K\,,
\end{aligned}
\end{equation}
where in general for tensor-valued quantities \(\Pi^i\) we have 
\begin{equation}\label{equation:formulaDiscreteGradientB}
  \begin{aligned}
    \D_{\Pi^i}\W|_{\Pi^J_{n+1},\Pi_n^K} = &\partial_{\Pi^i}\W(\Pi^i_{n+\frac{1}{2}})|_{\Pi_{n+1}^J, \Pi_{n}^K}\\
    &+ \frac{\W(\Pi_{n+1}^i)|_{\Pi_{n+1}^J, \Pi_{n}^K} - \W(\Pi_{n}^i)|_{\Pi_{n+1}^J, \Pi_{n}^K} - <\partial_{\Pi^i}\W(\Pi^i_{n+\frac{1}{2}})|_{\Pi_{n+1}^J, \Pi_{n}^K}, \Delta\Pi^i>}{<\Delta\Pi^i,\Delta\Pi^i>}\,\Delta\Pi^i\,,
    \end{aligned}
\end{equation}
and
\begin{equation}\label{equation:formulaDiscreteGradientC}
  \begin{aligned}
  \D_{\Pi^i}\W|_{\Pi^J_{n},\Pi_{n+1}^K} = &\partial_{\Pi^i}\W(\Pi^i_{n+\frac{1}{2}})|_{\Pi_{n}^J, \Pi_{n+1}^K}\\
    &+ \frac{\W(\Pi_{n+1}^i)|_{\Pi_{n}^J, \Pi_{n+1}^K} - \W(\Pi_{n}^i)|_{\Pi_{n}^J, \Pi_{n+1}^K} - <\partial_{\Pi^i}\W(\Pi^i_{n+\frac{1}{2}})|_{\Pi_{n}^J, \Pi_{n+1}^K}, \Delta\Pi^i>}{<\Delta\Pi^i,\Delta\Pi^i>}\,\Delta\Pi^i\,,
    \end{aligned}
\end{equation}
whereas for scalar-valued invariants \(\Pi^i\) the above simplifies to the so-called Greenspan formulas (cf.\ \cite{greenspan1984})
\begin{equation}\label{equation:formulaDiscreteGradientD}
  \D_{\Pi^i}\W|_{\Pi^J_{n+1},\Pi_n^K} = \frac{\W(\Pi_{n+1}^i)|_{\Pi_{n+1}^J, \Pi_{n}^K} - \W(\Pi_{n}^i)|_{\Pi_{n+1}^J, \Pi_{n}^K}}{\Delta\Pi^i}\,,
\end{equation}
and
\begin{equation}\label{equation:formulaDiscreteGradientE}
  \D_{\Pi^i}\W|_{\Pi^J_{n},\Pi_{n+1}^K} = \frac{\W(\Pi_{n+1}^i)|_{\Pi_{n}^J, \Pi_{n+1}^K} - \W(\Pi_{n}^i)|_{\Pi_{n}^J, \Pi_{n+1}^K}}{\Delta\Pi^i}\,.
\end{equation}
This finally leads to 
\begin{enumerate}
\item the partioned discrete gradient \(\D_{\I}\W\)
  \begin{equation}
  D_{\I}\W = \frac{1}{2}\,(\D_{\I}\W|_{\Pi^J_{n+1},\Pi_n^K} + \D_{\I}\W|_{\Pi^J_{n},\Pi_{n+1}^K})
  \end{equation}
  with being \(i=1\), \(J=\emptyset\), \(K=\{2,3\}\) we have \(\Pi^1=\I\), \(\Pi^J=\emptyset\), \(\Pi^K = (\II,\III)\) and thus
  \begin{equation}
    \begin{aligned}
      \D_{\I}\W|_{\Pi^J_{n+1},\Pi^K_n} &= \frac{\W(\IPhiNOne,\IIPhiN,\JPhiN)-\W(\IPhiN,\IIPhiN,\JPhiN)}{\Delta\I}\\
      \D_{\I}\W|_{\Pi^J_{n},\Pi^K_{n+1}} &= \frac{\W(\IPhiNOne,\IIPhiNOne,\JPhiNOne)-\W(\IPhiN,\IIPhiNOne,\JPhiNOne)}{\Delta\I}
    \end{aligned}
  \end{equation}
\item the partioned discrete gradient \(\D_{\II}\W\)
  \begin{equation}
  D_{\II}\W = \frac{1}{2}\,(\D_{\II}\W|_{\Pi^J_{n+1},\Pi_n^K} + \D_{\II}\W|_{\Pi^J_{n},\Pi_{n+1}^K})
  \end{equation}
  with being \(i=2\), \(J=1\), \(K=3\) we have \(\Pi^2=\II\), \(\Pi^J=\I\), \(\Pi^K = \III\) and thus
  \begin{equation}
    \begin{aligned}
      \D_{\II}\W|_{\Pi^J_{n+1},\Pi^K_n} &= \frac{\W(\IPhiNOne,\IIPhiNOne,\JPhiN)-\W(\IPhiNOne,\IIPhiN,\JPhiN)}{\Delta\II}\\
      \D_{\II}\W|_{\Pi^J_{n},\Pi^K_{n+1}} &= \frac{\W(\IPhiN,\IIPhiNOne,\JPhiNOne)-\W(\IPhiN,\IIPhiN,\JPhiNOne)}{\Delta\II}
    \end{aligned}
  \end{equation}
\item the partioned discrete gradient \(\D_{J}\W\)
  \begin{equation}
  D_{J}\W = \frac{1}{2}\,(\D_{J}\W|_{\Pi^J_{n+1},\Pi_n^K} + \D_{J}\W|_{\Pi^J_{n},\Pi_{n+1}^K})
  \end{equation}
  with being \(i=3\), \(J=\{1,2\}\), \(K=\emptyset\) we have \(\Pi^3=J\), \(\Pi^J=(\I,\II)\), \(\Pi^K = \emptyset\) and thus
  \begin{equation}
    \begin{aligned}
      \D_{J}\W|_{\Pi^J_{n+1},\Pi^K_n} &= \frac{\W(\IPhiNOne,\IIPhiNOne,\JPhiNOne)-\W(\IPhiNOne,\IIPhiNOne,\JPhiN)}{\Delta J}\\
      \D_{J}\W|_{\Pi^J_{n},\Pi^K_{n+1}} &= \frac{\W(\IPhiN,\IIPhiN,\JPhiNOne)-\W(\IPhiN,\IIPhiN,\JPhiN)}{\Delta J}
    \end{aligned}
  \end{equation}
\end{enumerate}
For the above partioned discrete gradients \((\D_{\I}\W, \D_{\II}\W, \D_{J}\W)\), conveniently shown in \cref{equation:discreteGradientsIIIJ}, the directionality property
\begin{align}
    &\D_{\I}\W\,\Delta\I + \D_{\II}\W\,\Delta\II + \D_{J}\W\,\Delta\III\nonumber\\
    =&\phantom{}\tfrac{1}{2}\,(\W(\IPhiNOne,\IIPhiN,\JPhiN) - \W(\IPhiN,\IIPhiN,\JPhiN)\nonumber\\
    &+ \W(\IPhiNOne,\IIPhiNOne,\JPhiNOne) - \W(\IPhiN,\IIPhiNOne,\JPhiNOne))\\
    &+\tfrac{1}{2}\,(\W(\IPhiNOne,\IIPhiNOne,\JPhiN) - \W(\IPhiNOne,\IIPhiN,\JPhiN)\\
    &+ \W(\IPhiN,\IIPhiNOne,\JPhiNOne) - \W(\IPhiN,\IIPhiN,\JPhiNOne))\\
    &+\tfrac{1}{2}\,(\W(\IPhiNOne,\IIPhiNOne,\JPhiNOne) - \W(\IPhiNOne,\IIPhiNOne,\JPhiN)\\
    &+ \W(\IPhiN,\IIPhiN,\JPhiNOne) - \W(\IPhiN,\IIPhiN,\JPhiN))\\
    =&\phantom{++} \W(\IPhiNOne,\IIPhiNOne,\JPhiNOne) - \W(\IPhiN,\IIPhiN,\JPhiN)\label{equation:directionalityProperty}
\end{align}
is obviously fulfilled.  
\section{Partioned discrete gradients of \(\WCt\)}\label{appendix:CGcDiscreteGradient}
In analogy to \cref{equation:formulaDiscreteGradientA}-\cref{equation:formulaDiscreteGradientE} the partitioned discrete gradients \(\D_{\subC}\WCt\), \(\D_{\subG}\WCt\), and \(\D_{C}\WCt\) of the strain energy \(\WCt(\C,\,\G,\,C)\) eventually are 
\begin{enumerate}
\item the partioned discrete gradient \(\D_{\subC}\WC\)
\begin{equation}
\begin{aligned}
   & (\D_{\subC}\WCt)_{ij} = \frac{1}{2}(\partial_{\subC}\WCt(\C_{n+\frac{1}{2}},\G_n,C_n)
    \\
    &+\partial_{\subC}\WCt(\C_{n+\frac{1}{2}},\G_n,C_n))_{kl}\,(\delta_{ki}\delta_{lj}-\tfrac{(\Delta\C)_{kl}\,(\Delta\C)_{ij}}{(\Delta\C)_{mn}\,(\Delta\C)_{mn}})\\
    &+\tfrac{\WCt(\CPhiNOne,\GPhiN,\cPhiN)+\WCt(\CPhiNOne,\GPhiNOne,\cPhiNOne)-\WCt(\CPhiN,\GPhiN,\cPhiN)-\WCt(\CPhiN,\GPhiNOne,\cPhiNOne)}{2\,(\Delta\C)_{mn}\,(\Delta\C)_{mn}}(\Delta\C)_{ij}
\end{aligned}
\end{equation}
\item the partioned discrete gradient \(\D_{\subG}\WCt\)
\begin{equation}
\begin{aligned}
    &(\D_{\subG}\WCt)_{ij} = \frac{1}{2}(\partial_{\subG}\WCt(\C_{n+1},\G_{n+\frac{1}{2}},C_n)
    \\
    &+\partial_{\subC}\WCt(\C_n,\G_{n+\frac{1}{2}},C_{n+1}))_{kl}\,(\delta_{ki}\delta_{lj}-\tfrac{(\Delta\G)_{ij}\,(\Delta\G)_{kl}}{(\Delta\G)_{mn}\,(\Delta\G)_{mn}})\\
    &+\tfrac{\WCt(\CPhiNOne,\GPhiNOne,\cPhiN)+\WCt(\CPhiN,\GPhiNOne,\cPhiNOne)-\WCt(\CPhiNOne,\GPhiN,\cPhiN)-\WCt(\CPhiN,\GPhiN,\cPhiNOne)}{2\,(\Delta\G)_{mn}\,(\Delta\G)_{mn}}(\Delta\G)_{ij}
    \end{aligned}
\end{equation}
\item the partioned discrete gradient \(\D_{C}\WCt\)
\begin{equation}
    \begin{aligned}
    \D_{C}\WCt =&   \frac{1}{2}\,\left(\tfrac{\WCt(\CPhiNOne,\GPhiNOne,\cPhiNOne)-\WCt(\CPhiNOne,\GPhiNOne,\cPhiN)}{\Delta C}\right.\\
              &+ \left.\tfrac{\WCt(\CPhiN,\GPhiN,\cPhiNOne)-\WCt(\CPhiN,\GPhiN,\cPhiN)}{\Delta C}\right)\,.
    \end{aligned}
\end{equation}
\end{enumerate}
In comparison to an analytical Mooney-Rivlin model as proposed in \cite{betsch2018} the above discrete gradients are pretty challenging to handle for the consistent linearization, since for the PANN constitutive model employed the discrete gradients \(\D_{\subC}\WCt\), \(\D_{\subG}\WCt\) are not constant and furthermore are dependend from all its dependencies \((\C,\G,C)\). 
\section{Discrete gradient of \(\WC\)}\label{appendix:CDiscreteGradient}
The projection-based discrete Gradient for \(\WC(\C)\) is given by 
\begin{equation}
\D_{\subC}\WC = \partial_{\subC}\WC(\C_{n+\frac{1}{2}}) + \frac{\WC(\C_{{n+1}})-\WC(\C_{{n}})-\partial_{\subC}\WC(\C_{n+\frac{1}{2}}):\Delta\C}{\Delta\C:\Delta\C}\,\Delta\C
\end{equation}
where despite the chain rule which is necessary for the PANN constitutive model with its inputs \(\vec{\mathcal{I}}\) the discrete gradient is tensor-valued and therefore is by far complex when compared to the proposed version in \cref{equation:discreteGradientsIIIJ}.

\renewcommand*{\bibfont}{\footnotesize}
\printbibliography
\end{document}